\documentclass[dvipsnames,11pt]{article}
\usepackage{adjustbox}
\usepackage{amssymb}
\usepackage{amsthm}
\usepackage{amsmath}
\usepackage{authblk}
\usepackage{blkarray}
\usepackage{booktabs}
\usepackage{bm}
\usepackage[colorlinks=true, linkcolor=blue, citecolor=blue, urlcolor=blue]{hyperref}
\usepackage{calc}
\usepackage{caption}
\usepackage{cases}
\usepackage{physics}
\usepackage{pbox}
\usepackage{centernot}
\usepackage[utf8]{inputenc}
\usepackage{dcolumn}
\usepackage{float}
\usepackage[margin = 1in]{geometry}
\usepackage{graphicx}
\usepackage{listings}
\usepackage{makecell}
\usepackage{mathtools}
\usepackage[round]{natbib}
\usepackage{pifont}

\usepackage{pgfplots}
\pgfplotsset{compat=newest}
\usepackage{setspace}
\usepackage{soul}
\usepackage{subcaption}
\usepackage{tabularx}
\usepackage{tgpagella}
\usepackage{verbatim}
\usepackage{xcolor}
\usepackage{xfrac}
\usepackage[style=english]{csquotes}
\definecolor{consensus}{RGB}{0,154,229}
\definecolor{gridlock}{RGB}{227,112,70}
\definecolor{zealot1}{RGB}{63,164,77}
\definecolor{zealot2}{RGB}{195,114,209}
\definecolor{vote1}{RGB}{172,142,25}
\usepackage{pgfplots}
\pgfplotsset{compat=1.3}
\MakeOuterQuote{"}
\usepackage{tikz}
\usepackage{marvosym}

\title{Dynamics of voting strategies and public good funding}
\author{Jonathan Engle\thanks{jae23a@fsu.edu} \, \& Bryce Morsky\thanks{bmorsky@fsu.edu}}
\affil{Department of Mathematics, Florida State University, Tallahassee, FL, USA}
\date{\today}
\begin{document}
\maketitle

\begin{abstract}
    We model an electorate voting on the funding of a public good in a two-party system in an evolutionary game theory framework. Voters adopt one of four strategies: Consensus-makers, Gridlockers, Party $1$ Zealots, and Party $2$ Zealots, which they may change via imitation. The public good benefits both individuals locally and those in neighbouring regions due to spillover effects. A system of differential equations governs the spatial movement of individuals and shifts in their voting strategies. Local social interactions drive strategy evolution, while migration occurs toward areas of higher utility, which is a function of both social and economic factors. Our results reveal bistability and significant spatial variations. Locally, populations converge to a politically gridlocked state or a mix of consensus-makers and zealots, determining public good provisioning. We find that public good spillovers generate a free-rider effect and poorly funded regions become spatially tied to, and dependent upon, well-funded ones.
\end{abstract}

\section{Introduction}

The production and maintenance of public goods is a well-studied collective action problem, often formulated as a public goods game \citep{Samuelson54}. As a consequence of the production of a public good, agents can free-ride on the efforts of others \citep{Fischbacher10, Anderoni88, Cornes96}. Many extensions have been analyzed to combat this issue of free-riding and production of public goods. For example, agents that punish free-riders and reward cooperators can be introduced \citep{Szolnoki10,Chen15, Wang21, Shen25}, which are particularly effective in spatial settings. However, space can also create incentives to free-rider through spillovers, where local public goods benefit neighbouring regions in addition to the location that funds them. In the case of centralized and decentralized provisioning of public goods, local public goods can be under provided when there are spillovers \citep{Besley03}. Further, spillovers can render government interventions ineffective \citep{Rojas20}.

Elections and voting play a role in determining levels of public good funding \citep{Hamman11,Ponce-Rodriguez18}. Democratic discussions can overcome the free-rider problem and lead to outcomes that are both efficient and equitable. Or, they can not. In India, for example, state governments have affected the allocations of electricity before an election to tip the electorate in their favour \citep{Baskaran15}. Additionally, politicians who have personal gain in an election may under-provide a public good as its benefits cannot target voters as efficiently as pork-barrel spending \citep{Lizzeri01}. In many developing countries, the allocation of a public good lies in the hands of a broker who is entrusted by the local government to allocate a particular good; as a consequence, brokers often keep the public good for themselves \citep{Shami19}. Outside of legislators actions, voting --- influenced by a variety of social issues such as race, religion, income (including reliance on a public good), and strategic voting \citep{Wani14, Pinto21, Mettler22} --- plays an important role in determining whether or not public goods are funded.

To understand the interplay between the provisioning of public goods and elections, we develop and analyze a system of differential equations to model the provisioning of public goods, inter-regional migration, and strategic voting. The model consists of two parties, one that funds the public good and one that does not, and three voting strategies, Consensus-makers, Gridlockers, and Zealots \cite{Engle24}. Consensus-makers prefer consensus, voting for the majority whether that is to fund the public good or not. Gridlockers vote to maintain intermediate funding of the public good, and Zealots always vote for their party. We explore how these strategies affect public good provisioning in both non-spatial and spatial settings.

\section{Methods}

\subsection{Social dynamics}

Here, we are interested in modeling voting strategies in a two party system, which determines the funding for a public good. We assume that Party $1$ will fund the public good, while Party $2$ will not. And, the public good benefits all voters regardless of how they voted. Voters adopt a voting strategy, Consensus-maker, Gridlocker, Party $1$ Zealot, and Party $2$ Zealot. Consensus-makers are voters who prefer that a majority is obtained by either party in their location. Consensus-makers least desire gridlock, where the vote is evenly split between the two parties. Consensus-makers value social conformity. Conversely, Gridlockers, as the name suggests, prefer gridlock over either party forming a majority. Lastly, Zealots always vote for their affiliated party, since they simply want it to earn the most votes. These voting strategy types have been previously included in an agent-based model \citep{Engle24}.

The population can be divided into proportions of each voting strategy, Consensus-makers $c$, Gridlockers $g$, Party $1$ Zealots $z_1$, and Party $2$ Zealots $z_2$, where $c+g+z_1+z_2=1$. From $z_2=1-c-g-z_1$, we reduce to the variables $c$, $g$, and $z_1$. To model how these voting strategies vote, we let $v_c$, $v_g$, $v_{z_1}=1$, and $v_{z_2}=0$ be the intended votes of Consensus-makers, Gridlockers, Party $1$ Zealots, and Party $2$ Zealots for Party $1$. Note that Zealots always vote their party. The votes for Consensus-makers and Gridlockers, however, will depend on the intended votes of others. $v = cv_{c}+gv_{g}+z_1$ is the proportion of the population voting for Party $1$. Thus, the vote for Party $2$ is $1-v$. Table \ref{tbl:vars} summarizes these variables.

We consider the following voting dynamics for Consensus-makers and Gridlockers:
\begin{subequations}
\begin{align}
    \dot{v}_c&=(1-v_c)v^2-v_c(1-v)^2,\label{eq:v_c_dot}\\
    \dot{v}_g&=(1-v_g)(1-v)^2-v_gv^2. \label{eq:v_g_dot}
\end{align}
\end{subequations}
Since $1-v_c$ represents Consensus-makers voting for party $2$, these Consensus-makers then interact with Party $1$ voters at a rate of $v$ and have a probability of $v$ switching to vote to Party $2$. Similarly, Party $1$ voting Consensus-makers ($v_c$) interact with Party $2$ voters at a rate $1-v$. They then imitating them with have probability $1-v$, thus becoming Party $2$ voters. We use the same type of logic for constructing $\dot{v}_g$, where $1-v_g$ are Gridlockers voting for Party $2$. These voters interact with Party $2$ individuals at a rate of $1-v$ and have a probability of $1-v$ of switching their vote to Party $1$ (since they prefer gridlock). Similarly, Party $1$ voting Gridlockers interact with Party $1$ voters at a rate $v$ and switch their vote with probability $v$.

Next, we consider the dynamics of individuals changing their strategies, which occurs through imitation. We assume that these dynamics occur relatively slowly compared to the voting dynamics. Let $f_i(v)$ be the fitness of the voting strategy $i$, representing how content with their strategy and thus persuasive these voters are. The fitnesses for each strategy are:
\begin{subequations}
\begin{align}
    f_c(v) &=  \frac{1+\cos{(2\pi v)}}{2}, \label{eq:fc} \\
    f_g(v) &= \frac{1-\cos{(2\pi v)}}{2},\label{eq:fg} \\
    f_{z_1}(v) &= \frac{1-\cos(\pi v)}{2},\label{eq:fz1} \\
    f_{z_2}(v) &= \frac{1+\cos(\pi v)}{2}\label{eq:fz2}.
\end{align}
\end{subequations}
These fitness functions are constructed in such a way that they continuously reflect the preferences of each voting strategy. The Consensus-makers receive the highest payoff when a consensus result is obtained, either when Party $1$ obtains a total majority ($v=1$) or Party $2$ obtains the total majority ($v=0$). Gridlockers, however, have the highest fitness at Gridlock and lowest when either party has a total majority. As for Zealots, they obtain the maximum payoff when the vote is in favour of their party, $v=1$ for Party $1$ Zealots and $v=0$ for Party $2$ Zealots. Additionally, we incorporate diminishing returns for all strategies as the vote approaches their desired result. For example, as the vote approaches a total majority for Party $1$, the slopes of the fitness functions for Consensus-makers and Party $1$ Zealots level off. Additionally, to avoid giving any voting strategy an advantage, all fitness functions satisfy $\int_0^1 f_i(v)dv=1/2$.

\begin{table}[h!]
\centering
\begin{tabular}{lll}
\toprule
Variable & Definition \\
\midrule
$c$ & proportion of Consensus-makers  \\
$g$ & proportion of Gridlockers  \\
$z_1$ &proportion of Party $1$ Zealots\\
$z_2=1-c-g-z_1$& proportion of Party $2$ Zealots\\
$v_{c}$ & Consensus-makers voting for Party $1$ \\
$v_{g}$ & Gridlockers voting for Party $1$  \\
$v_{z_1}=1$ & Party 1 Zealots voting for for Party $1$  \\
$v_{z_2}=0$ & Party 2 Zealots voting for for Party $1$  \\
$v=cv_c+gv_g+z$ & total votes for Party $1$  \\
\bottomrule
\end{tabular}
\caption{Parameter and variable definitions.} 
\label{tbl:vars}
\end{table}

We use the replicator equation to model imitation dynamics, where the rate of imitation is proportional to fitness differences. Except, we assume that Zealots from the opposing parties do not imitate each other. For Zealot of one party to become a Zealot of the other party, they must first imitate a Consensus-maker or Gridlocker. With this caveat, the imitation dynamics are:
\begin{subequations}
\begin{align}
    \dot{c}& = cg(f_c-f_g)+cz_1(f_c-f_{z_1})+c(1-c-g-z_1)(f_c-f_{z_2}), \label{eq:imitation_c}\\
    \dot{g}&=gc(f_g-f_c)+gz_1(f_g-f_{z_1})+g(1-c-g-z_1)(f_g-f_{z_2}),\label{eq:imitation_g} \\
    \dot{z_1}&=z_1c(f_{z_1}-f_c)+z_1g(f_{z_1}-f_g).\label{eq:imitation_z}
\end{align}
\end{subequations}
As before, we may reduce dimensions since $z_2 = 1-c-g-z_1$.

\subsection{Spatial interactions}

The previous section assumes that the populations are well-mixed. Here, we detail a spatial model in which there are nodes of well-mixed voters (representing neigbhourhoods, cities, or space more generally). Social dynamics occur within these nodes, factoring in only the local votes and populations within. However, individuals can move to neighbouring nodes. Consider an $L \times L$ grid of nodes, which we assume has periodic boundaries so that we have no edge effects (i.e.,\ it is a torus). For our simulations, we let $L=10$ and, unless otherwise stated, initial conditions are drawn from a uniform distribution. Each node is represented by the position $(x,y)$ for $x,y \in \{0, \ldots, L-1\}$. Equations \ref{eq:imitation_c}-\ref{eq:imitation_z} are then modified such that each state variable and fitness is relative to the location $(x,y)$. Further, we add a spatial flux term $MF_i(x,y)$ to each equation, representing the flow between nodes. For example, the combined social and spatial dynamics for Party $1$ Zealots is
\begin{multline}
        \dot{z}_1(x,y) = z_1(x,y)c(x,y)(f_{z_1}(v(x,y))-f_c(v(x,y)) \\ + z_1(x,y)g(x,y)(f_{z_1}(v(x,y))-f_g(v(x,y))) + MF_{z_1}(x,y).
\end{multline}
The parameter $M>0$ represents the rate of movement relative to the social dynamics. If $M$ is sufficiently high, the populations mix rapidly and the model is approximately equivalent to the ODE model. Therefore, we consider a low $M=10^{-2}$ for our simulations.

We consider both undirected and directed movement represented by the choice of $F_i$ for $i=c,g,z_1,z_2$. For undirected movement, individuals are equally likely to move to other nodes, and do so at the rate $M$. Thus,
\begin{equation}
    F_i(x,y) = \frac{1}{L^2}\sum_{(j,k) \neq (x,y)} i(j,k) - i(x,y),
\end{equation}
Under this type of movement, individuals have no preference to where they move.

Directed movement considers the case where individuals move to regions with higher utility. Utility is defined as a function of both the social utility (i.e.,\ the fitness $f_i(x,y)$) and economic utility, which is the amount of public good available in that location. Since Party $1$ funds the public good, economic utility is determined locally by $v(x,y)$. However, public goods often have spillover effects, where they benefit neighbouring regions. Infrastructure, such as highways, is an example of this effect. Therefore, economic utility is also partially determined by the votes of neighbouring nodes. Utility for strategy $i$ at node $(x,y)$ is thus
\begin{multline}
        u_i(v(x,y)) = \lambda \left((1-s)v(x,y)+\frac{s}{4}\sum_{(j,k) \in N} v(\bmod(x+j,L),\bmod(y+k,L)) \right) \\ + (1-\lambda)f_i(v(x,y)). \label{eq:utility_spillover}
\end{multline}
$N = \{(1,0),(-1,0),(0,1),(0,-1) \}$ gives the von Neumann neighbourhood of the focal node $(x,y)$. $\lambda \in [0,1]$ represents the weighting of economic vs.\ social utility, while $s \in [0,1]$ is the degree of spillover.

Under directed movement, the rates of movement are modified by the differences in utility between the focal node and the node they are moving to. Where utility is relatively higher, individuals move there more quickly. However, even if utility is lower, we still assume individuals may move, just at a lower rate. This effect is modeled by taking the exponential of the difference in utilities between nodes for a given voting strategy. Therefore, the flux for directed movement is
\begin{equation}
        F_i(x,y) = \frac{1}{L^2} \sum_{(j,k) \neq (x,y)} i(j,k) e^{\Delta u} - i(x,y)e^{-\Delta u},
\end{equation}
where
\begin{equation}
    \Delta u = u_i(v(\bmod(x+j,L),\bmod(y+k,L))) - u_i(x,y).
\end{equation}
Note that directed movement could lead to the accumulation of voters in particular nodes. Thus, we cannot reduce dimensions and must explicitly include $\dot{z}_2$, which has a similar form to Equation \ref{eq:imitation_z}. Therefore, to track the abundance of the population across space, we track the population size at each node, $p(x,y) = c(x,y) + g(x,y) + z_1(x,y) + z_2(x,y)$, as well as the population size relative to largest population, $\tilde{p}(x,y) = p(x,y)/\max\{p(x,y)\}$.

\section{Results}

\subsection{ODE model}

Here, we analyze the two different types of social dynamics. The first being the fast voting dynamics, Equations \ref{eq:v_c_dot} and \ref{eq:v_g_dot}, which allow us to understand the equilibrium behaviour of each strategy given a particular composition of the electorate. The second type of social dynamics we study are the slow strategy imitation dynamics, Equations \ref{eq:imitation_c}, \ref{eq:imitation_g}, and \ref{eq:imitation_z}, which model the change in voting strategies.

\begin{figure}[h!]
\captionsetup[subfigure]{justification=centering}
    \centering
    \begin{subfigure}[]{0.3\columnwidth}
        \caption{$z_1=0$}\label{fig:bifurcation_diagram_z(0.0)_g(1-c-z)}
        \includegraphics[width=\textwidth]{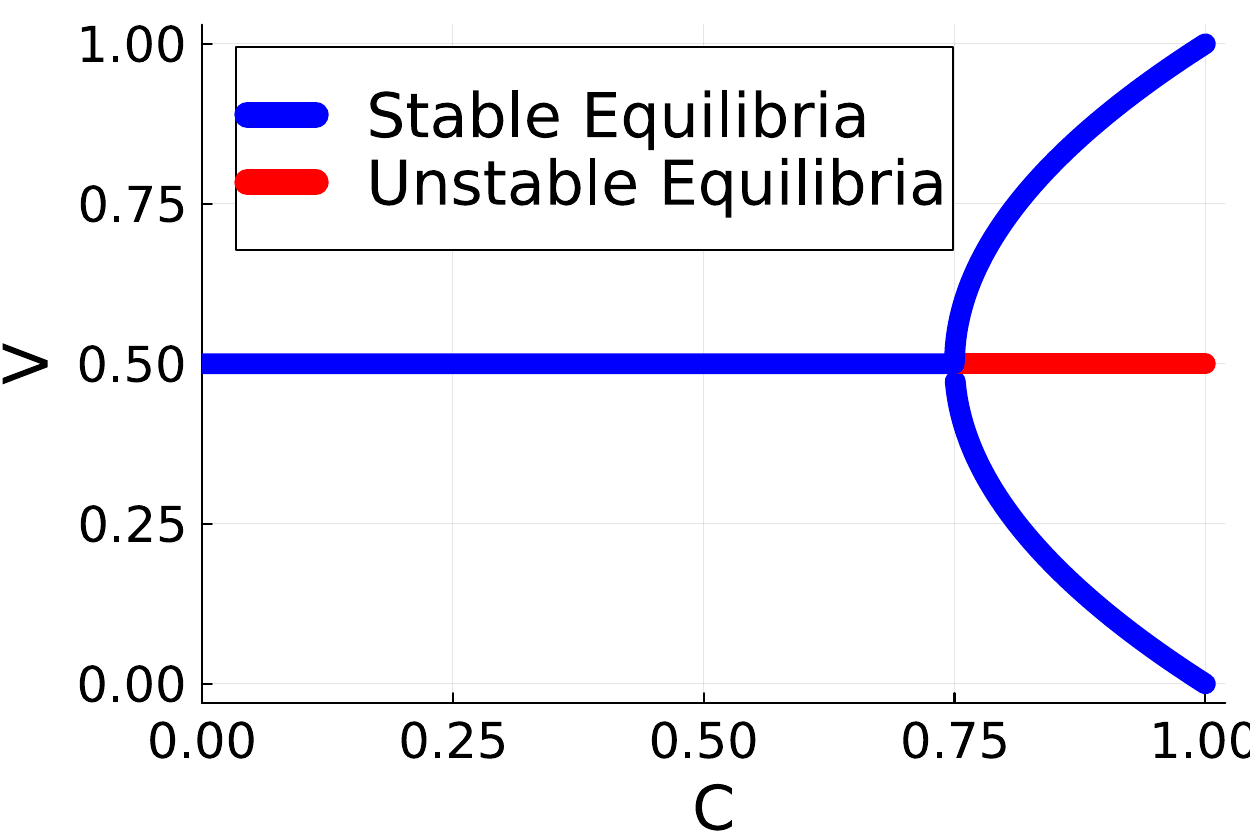}
    \end{subfigure}
    \begin{subfigure}[]{0.3\columnwidth}
        \caption{$z_1=0.01$}\label{fig:bifurcation_diagram_z(0.01)_g(1-c-z)}
        \includegraphics[width=\textwidth]{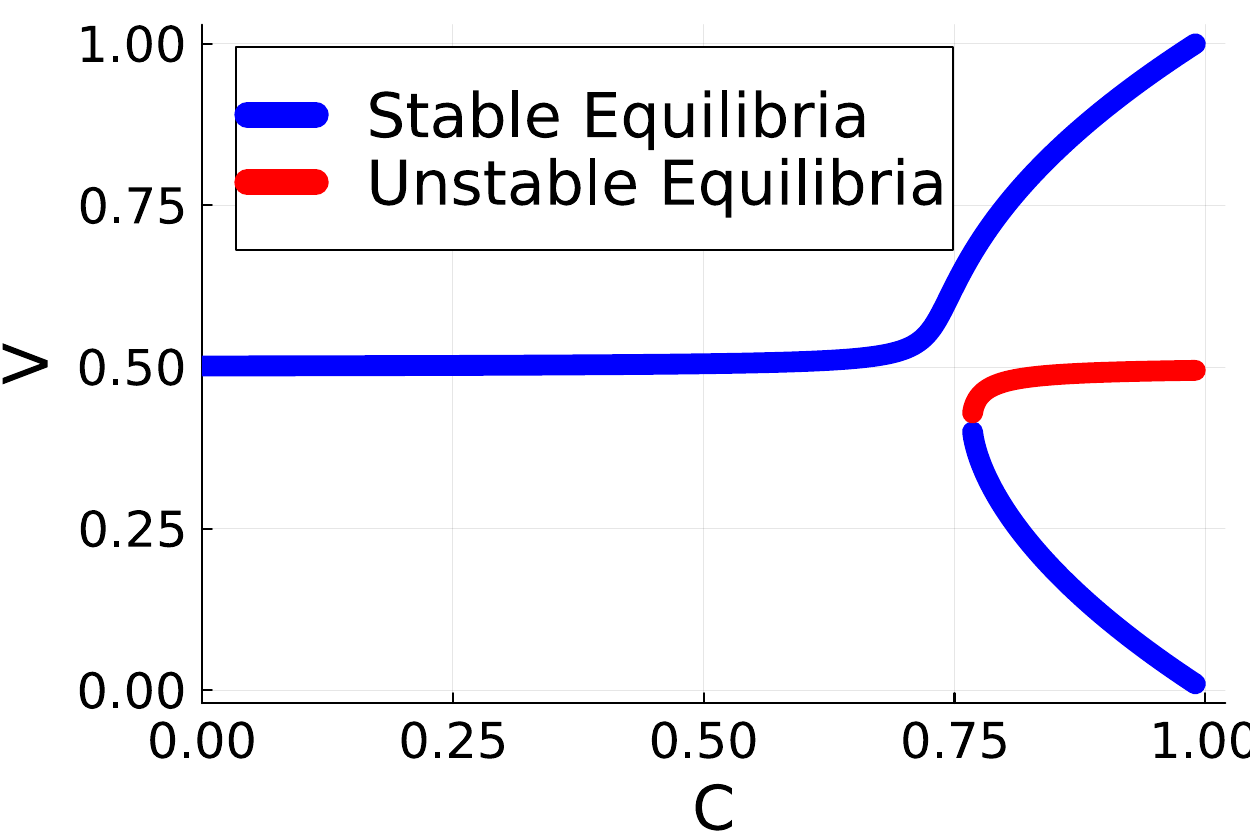}
    \end{subfigure}  
    \begin{subfigure}[]{0.3\columnwidth}
        \caption{ $z_1=0.2$}\label{fig:bifurcation_diagram_z(0.2)_g(1-c-z)}
        \includegraphics[width=\textwidth]{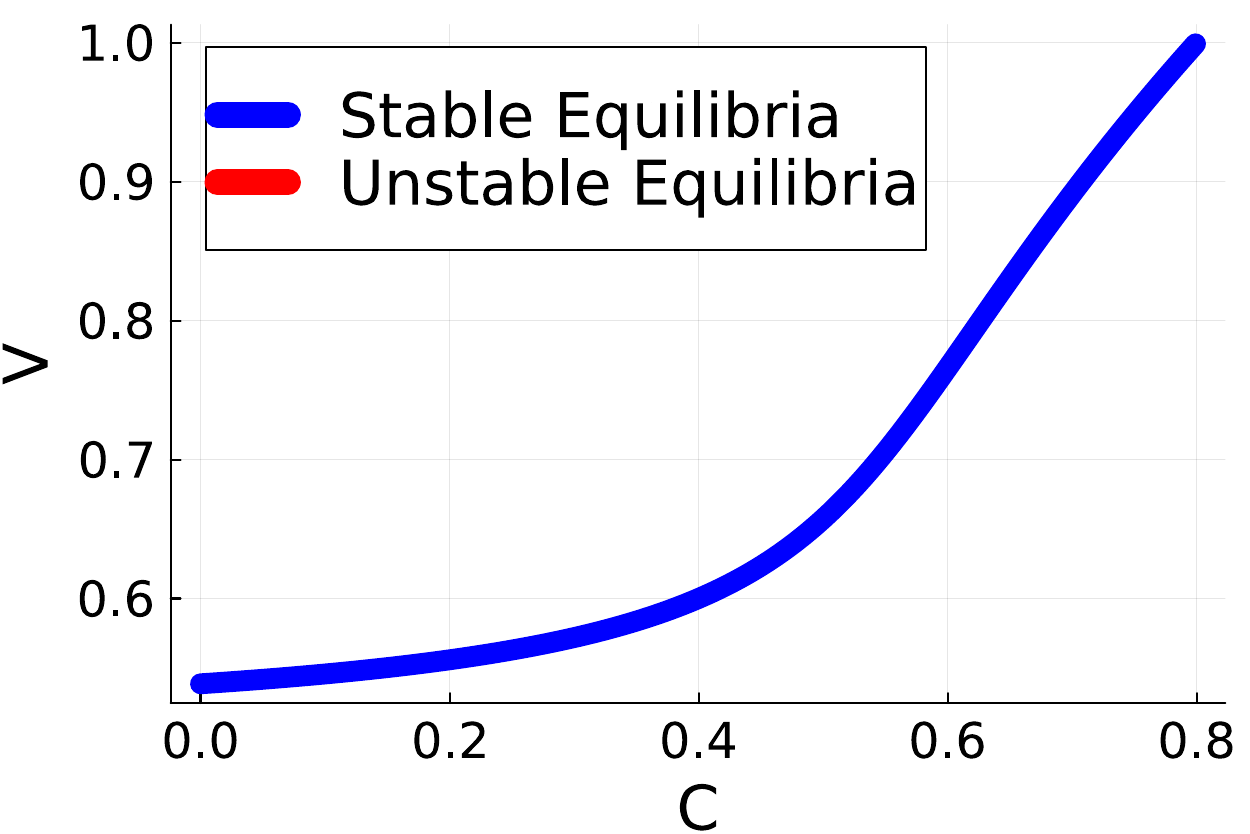}
    \end{subfigure}
    \caption{An assortment of bifurcation diagrams in the $cv$-plane for $g=1-v-c$, $z_2=0$, and varying $z_1$. The bifurcation diagrams show bistability in the system. As we increase the number of party 1 Zealots in the system, the number of unstable equilibria decreases, and fewer Consensus-makers are needed to obtain a consensus result.  }
    \label{fig:bifurcation_plots}
\end{figure}

First, we will perform linear stability analysis on the fast voting dynamics. Equations \ref{eq:v_c_dot} and \ref{eq:v_g_dot} have the following equilibria:
\begin{subequations}
\begin{align}
    \bar{v}_c&=\frac{\bar{v}^2}{2\bar{v}^2-2\bar{v}+1},\label{eq:v_c}\\
    \bar{v}_g&=\frac{(1-\bar{v})^2}{2\bar{v}^2-2\bar{v}+1}.\label{eq:v_g}
\end{align}
\end{subequations}
Substituting these into $\bar{v} = c\bar{v}_{c}+g\bar{v}_{g}+z_1$ and simplifying gives us
\begin{equation}
     \bar{v}^2c+g(1-\bar{v})^2+(z_1-\bar{v})(2\bar{v}^2-2\bar{v}+1) = 0. \label{eq:v2}
\end{equation}
The corresponding Jacobian matrix is
\begin{equation}
J=\begin{bmatrix}
    2c[(1-v_c)v+v_c(1-v)]-v^2-(1-v)^2 & 2g[(1-v_c)v+v_c(1-v)] \\
    -2c[(1-v_g)(1-v)+v_gv] & -2g[(1-v_g)(1-v)+v_gv]-v^2-(1-v)^2\end{bmatrix}.\label{eq:JacobianVotingDynamics}
\end{equation}
We utilize a root finding function in Julia (\cite{Roots.jl}) to numerically solve for the equilibria and then find the eigenvalues of the Jacobian. Figure \ref{fig:bifurcation_plots} depicts bifurcation diagrams in the $cv$-plane. We fix $z_2=0$ (thus $g=1-c-z_1$), and we vary the proportion of Party $1$ Zealots $z_1$. In a Zealot absent environment (Figure \ref{fig:bifurcation_diagram_z(0.0)_g(1-c-z)}), a pitchfork bifurcation emerges when the Consensus-maker population is $3/4$ of the total population, and the vote is split at $1/2$ for each party. For $c<3/4$, Gridlockers successfully create gridlock. However, for $c>3/4$, the equilibrium $\var{v}=1/2$ is unstable. If the vote tips in one direction or the other, then we obtain some consensus. As $c$ increases, consensus grows. Introducing Party $1$ Zealots promotes consensus on Party $1$, diminishing outcomes of Party $2$ reaching a majority (Figure \ref{fig:bifurcation_diagram_z(0.01)_g(1-c-z)}). For a sufficient proportion of $z_1$, only consensus on Party $1$ occurs (Figure \ref{fig:bifurcation_diagram_z(0.2)_g(1-c-z)}).

\begin{figure}[h!]
\centering
    \includegraphics[scale=0.7]{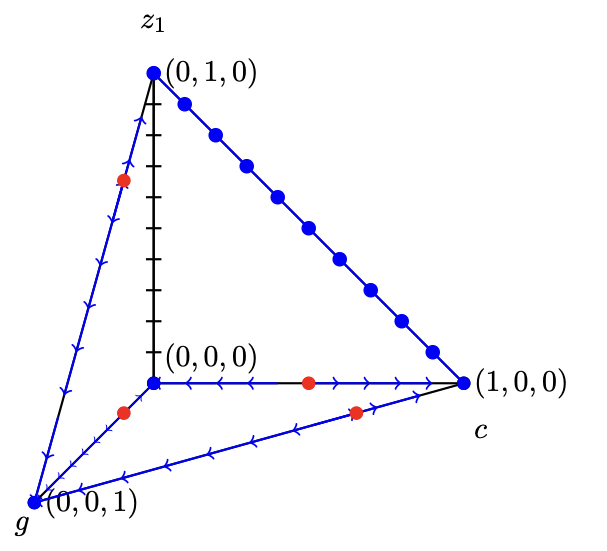}
   \caption{Stability analysis of the boundary of the $cgz_1$-simplex. Where \textcolor{blue}{\textbullet} is a stable equilibrium and \textcolor{red}{\textbullet} is a saddle equilibrium. The arrows indicate the direction of strategy change.} 
    \label{fig:Ternary}
\end{figure}

Next, we consider the slow imitation dynamics (Equations \ref{eq:imitation_c}-\ref{eq:imitation_z}).  With these equations, we can linearize and analyze the stability of equilibria on the boundary of the simplex. The corresponding Jacobian entries can be found in Appendix \ref{eq:JacobianVotingDynamics}. We depict the results for the edge cases of the simplex in Figure \ref{fig:Ternary}. Where the voting dynamics are bistable, we assume that we are at the stable equilibria in which Party $1$ has a higher proportion of votes. On the edge that connects $(0,0,1)$ and $(1,0,0)$ (i.e.\ $g=1-c$), we obtain the same saddle equilibrium as Figure \ref{fig:bifurcation_diagram_z(0.0)_g(1-c-z)} when $g=1/4$ and $c=3/4$. Additionally, when there is a mix of Gridlockers and Party $1$ Zealots (on the $z_1g$-plane), we observe a similar behaviour: there are saddle equilibria at $z_1=3/4$ and $g=1/4$, and at $z_2=3/4$ and $g=1/4$. On the edge that connects the origin to $(1,0,0)$ (where all voters are Consensus-makers), we obtain a saddle equilibrium at $(0,1/2,0)$. Here, Consensus-makers are in gridlock, with the population split voting for each party. Any small perturbation towards either side will shift the system towards the corresponding vertex. Along the $z_1$-axis, there is no change due to the voting behaviour of Zealots. And, on the edge that connects $(0,0,1)$ to $(1,0,0)$ (i.e.\ a population of Consensus-makers and Zealots), we obtain a line of stable equilibria by which consensus is obtained.

\subsection{Undirected movement}

\begin{figure}[h!]
\captionsetup[subfigure]{justification=centering}
    \centering
    \begin{subfigure}[]{0.3\columnwidth}
        \caption{Uniform random IC}\label{fig:TS_diffusion_gridlock}
        \includegraphics[width=\textwidth]{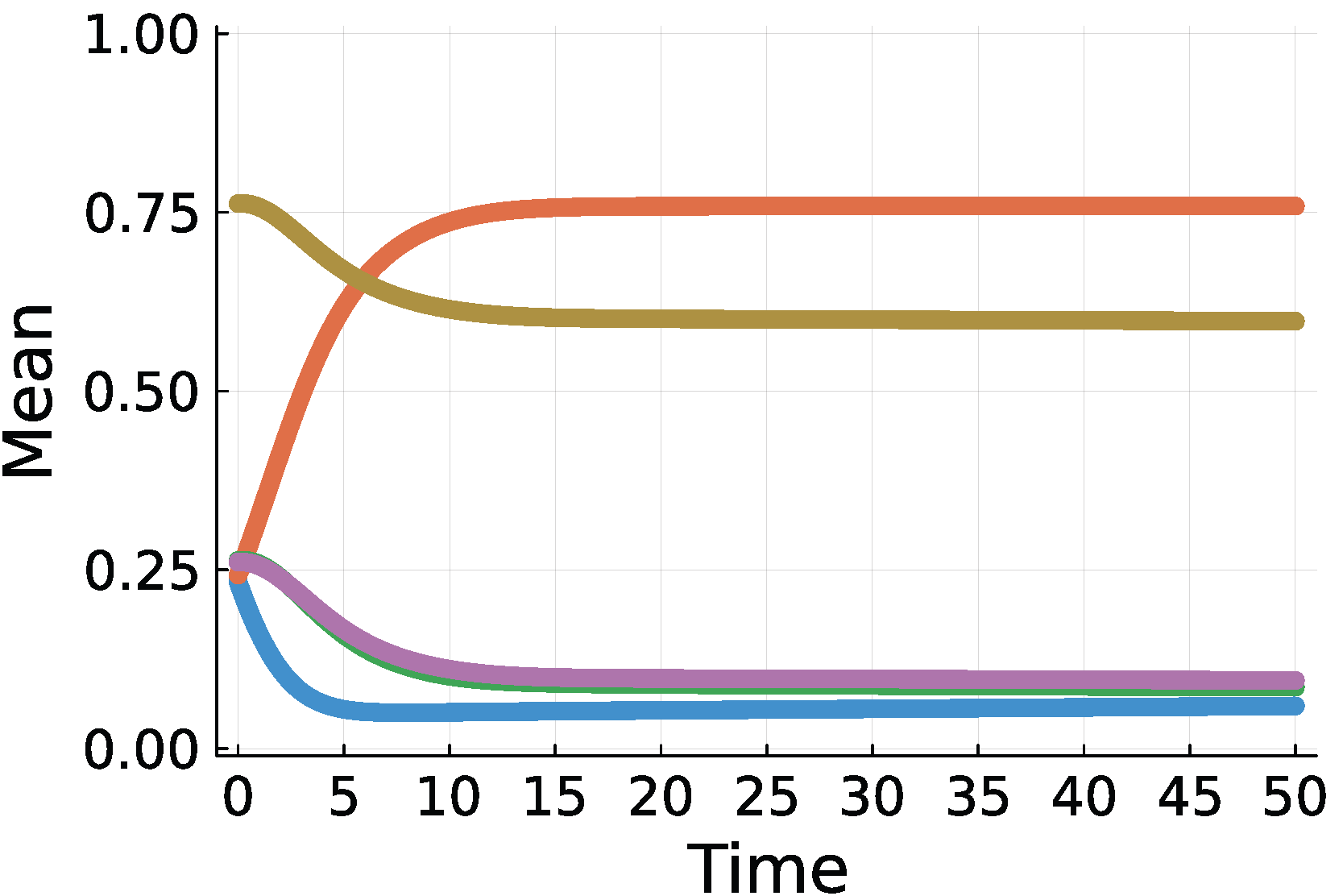}
    \end{subfigure}
    \begin{subfigure}[]{0.3\columnwidth}
        \caption{$c$ and $z_1$ biased IC}\label{fig:TS_diffusion_consensus}
        \includegraphics[width=\textwidth]{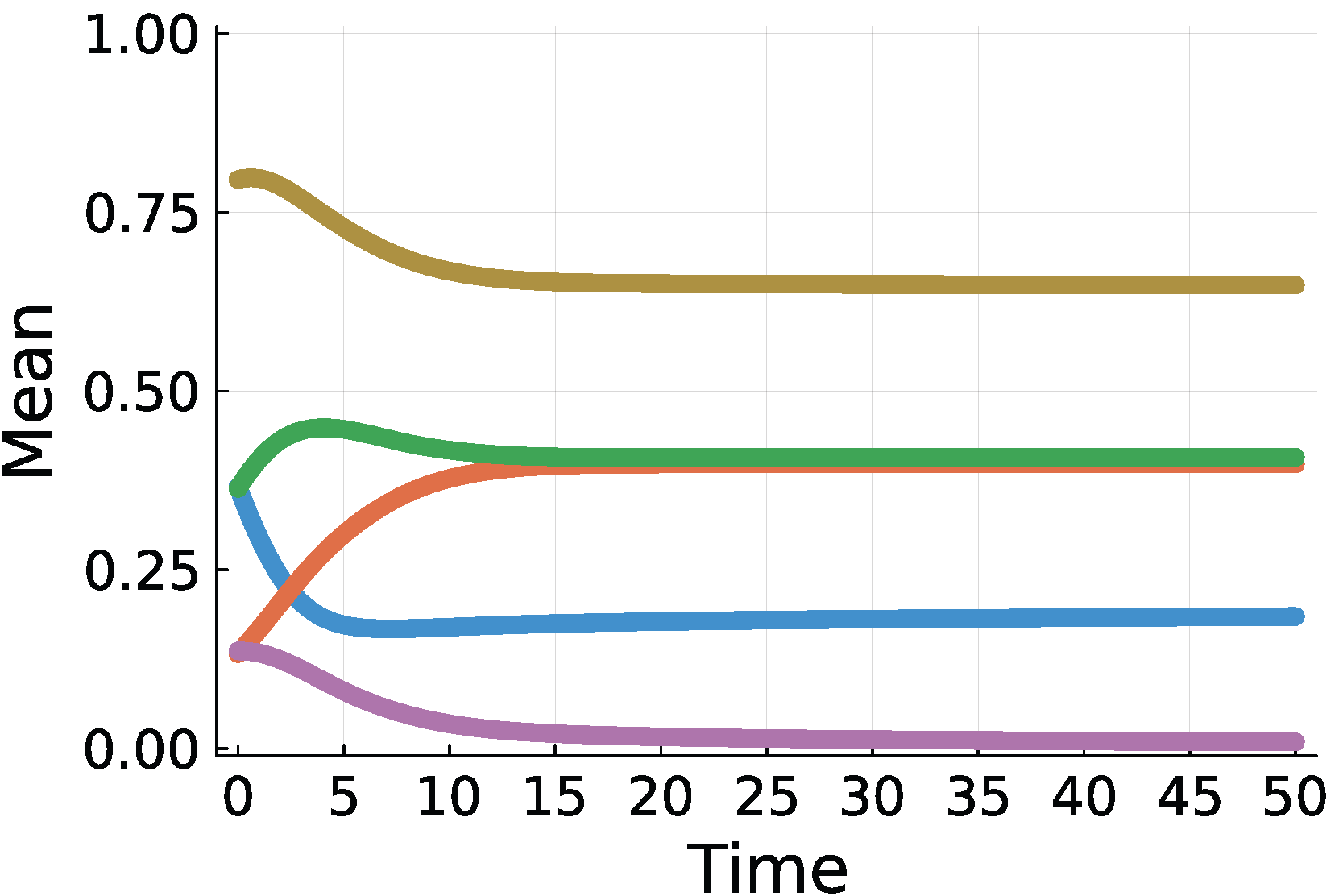}
    \end{subfigure}
    \begin{tikzpicture}
        \matrix (legend){
            \draw[consensus, line width=3pt] (0,0) -- (0.6,0) node[right, black]{Mean Consensus-makers}; &
            \draw[gridlock, line width=3pt] (0,0) -- (0.6,0) node[right, black]{Mean Gridlockers}; &
            \draw[zealot1,line width=3pt] (0,0) -- (0.6,0) node[right, black]{Party $1$ Zealots}; \\
            \draw[zealot2, line width=3pt] (0,0) -- (0.6,0) node[right, black]{Mean Party $2$ Zealots}; &
            \draw[vote1, line width=3pt] (0,0) -- (0.6,0) node[right, black]{Mean Vote for Party $1$};\\};
    \end{tikzpicture}
        \caption{Representative time series for the undirected movement scenario with and different initial conditions. Panel (a) depicts a representative example of convergence to a high prevalence of Gridlockers (and thus gridlock). Initial conditions were drawn from a uniform distribution. For panel (b), the densities of Consensus-makers and Party $1$ Zealots are initially high, resulting in convergence to a high vote for Party $1$. Note that movement conserves population size and thus $z_2$ is not plotted.}
    \label{fig:TS_diffusion}
\end{figure}
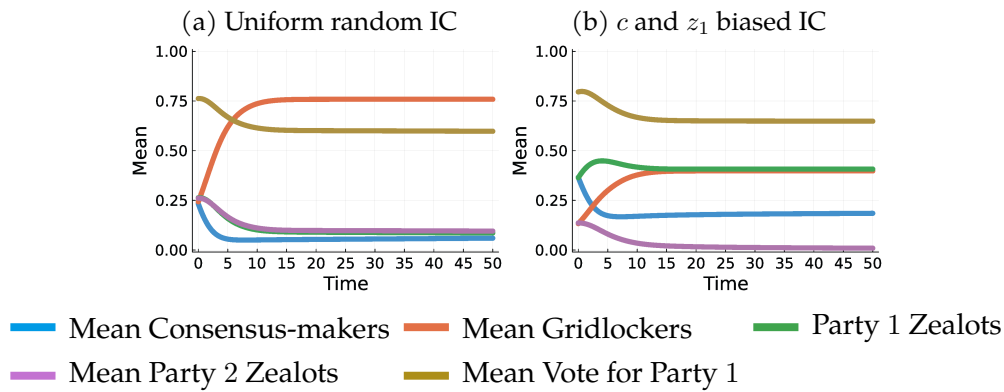

\begin{figure}[h!]
\captionsetup[subfigure]{justification=centering}
    \centering
    \begin{subfigure}[]{0.3\columnwidth}
        \caption*{Population Colourbar}
        \includegraphics[width=\textwidth]{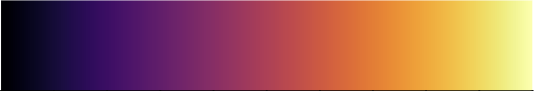}
            \vspace{2pt}
    {\small 0 \hfill 1}
    \end{subfigure} \hspace{2cm}
    \begin{subfigure}[]{0.3\columnwidth}
        \caption*{Party Colourbar}
        \includegraphics[width=\textwidth]{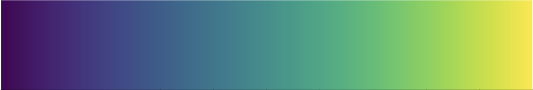}
            \vspace{2pt}
    {\small 0 \hfill 1}
    \end{subfigure}\\
    \begin{subfigure}[]{0.15\columnwidth}
        \caption{$c$}\label{fig:HM_diffusion_Consensusmakers}
        \includegraphics[width=\textwidth]{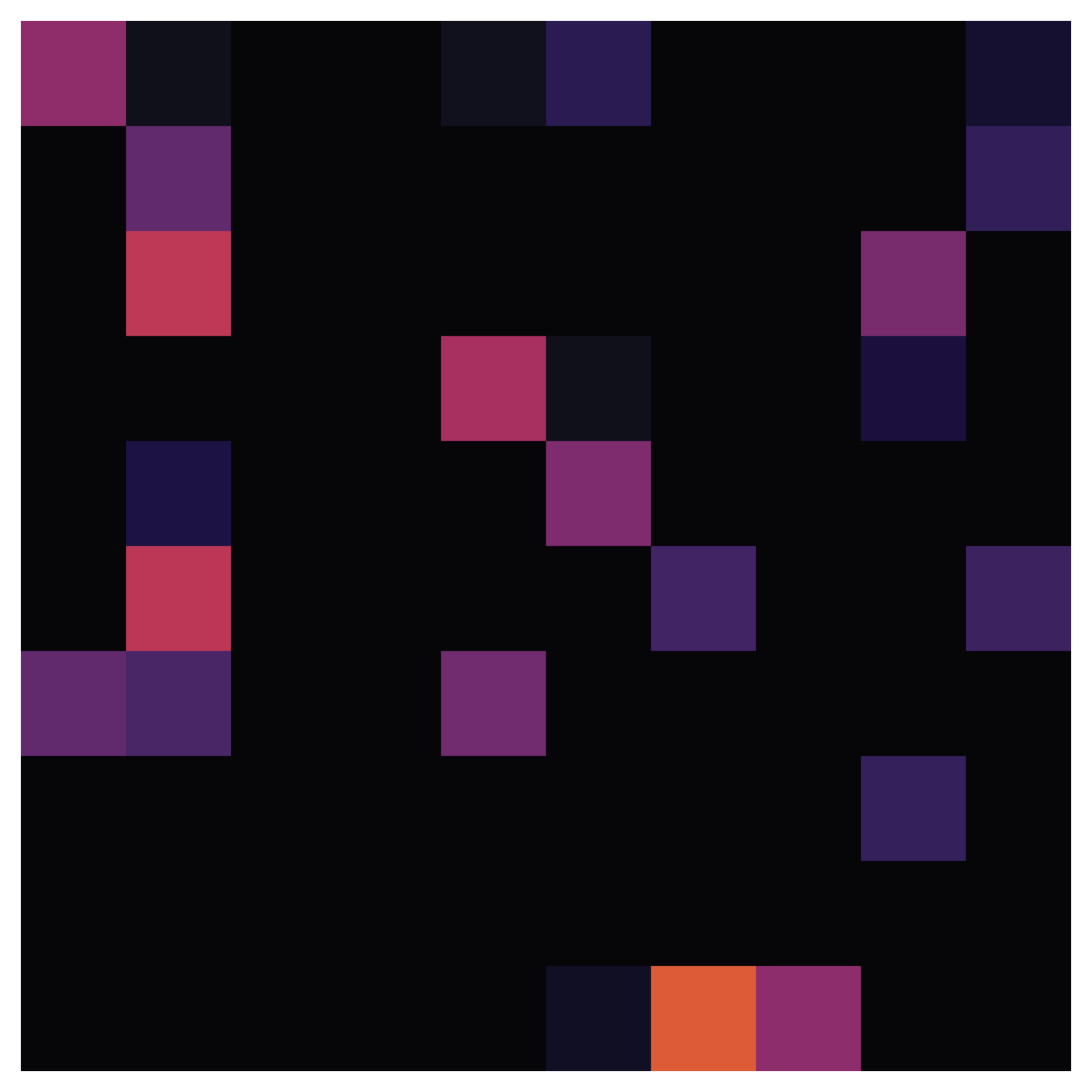}
    \end{subfigure}
    \begin{subfigure}[]{0.15\columnwidth}
        \caption{$g$}\label{fig:HM_diffusion_Gridlockers}
        \includegraphics[width=\textwidth]{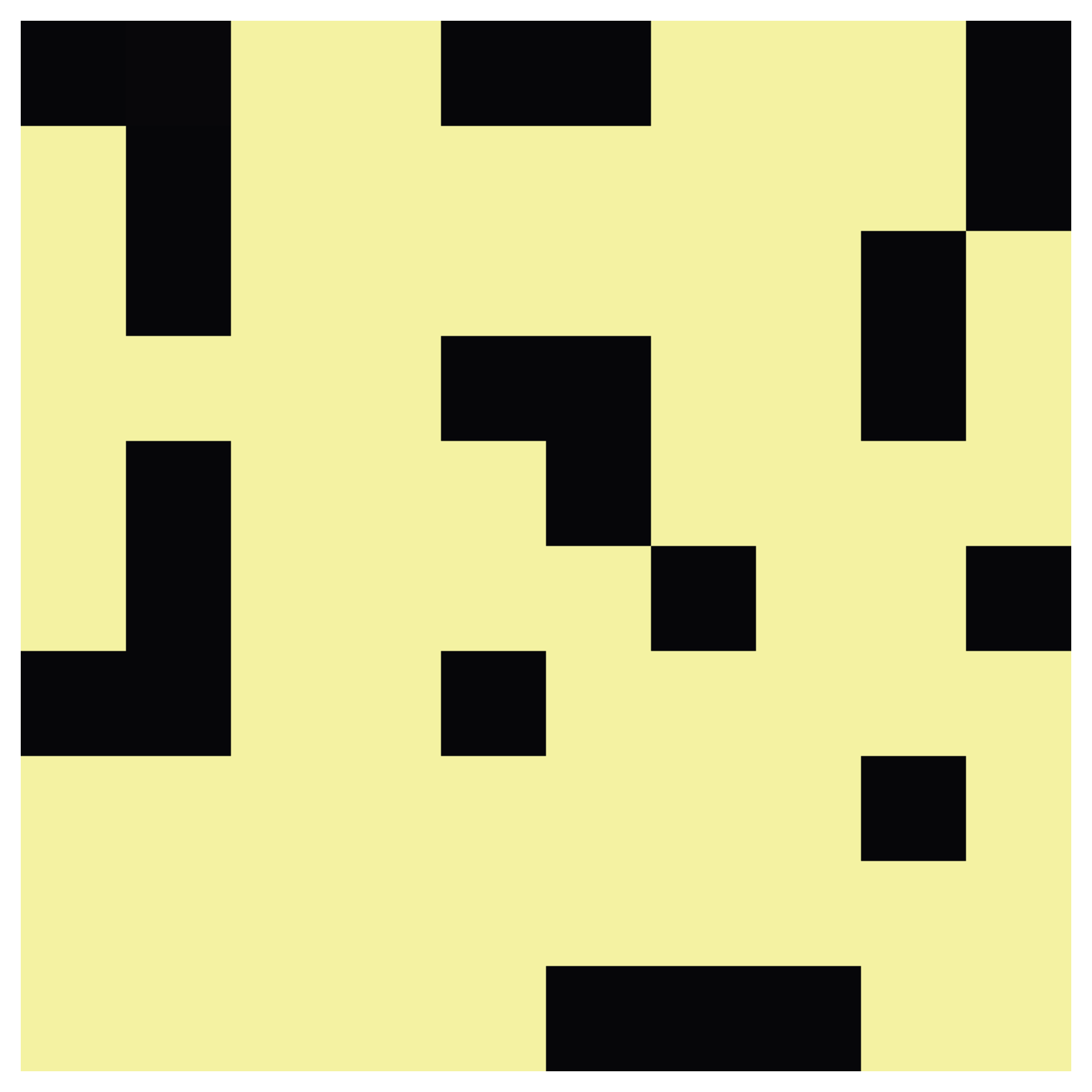}
    \end{subfigure}
        \begin{subfigure}[]{0.15\columnwidth}
        \caption{$z_1$}\label{fig:HM_diffusion_Zealots1}
        \includegraphics[width=\textwidth]{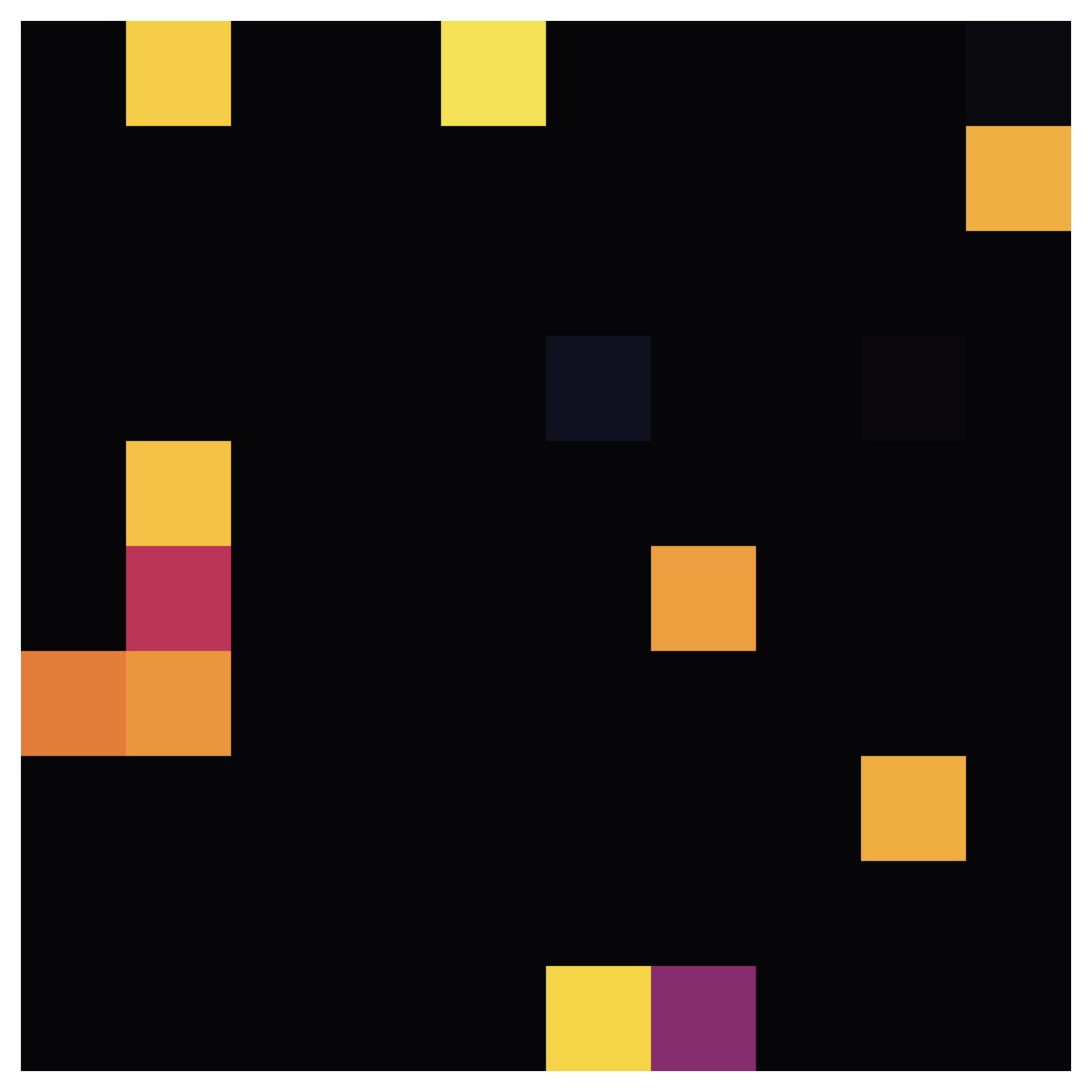}
    \end{subfigure}
        \begin{subfigure}[]{0.15\columnwidth}
        \caption{$v$}\label{fig:HM_diffusion_vote}
        \includegraphics[width=\textwidth]{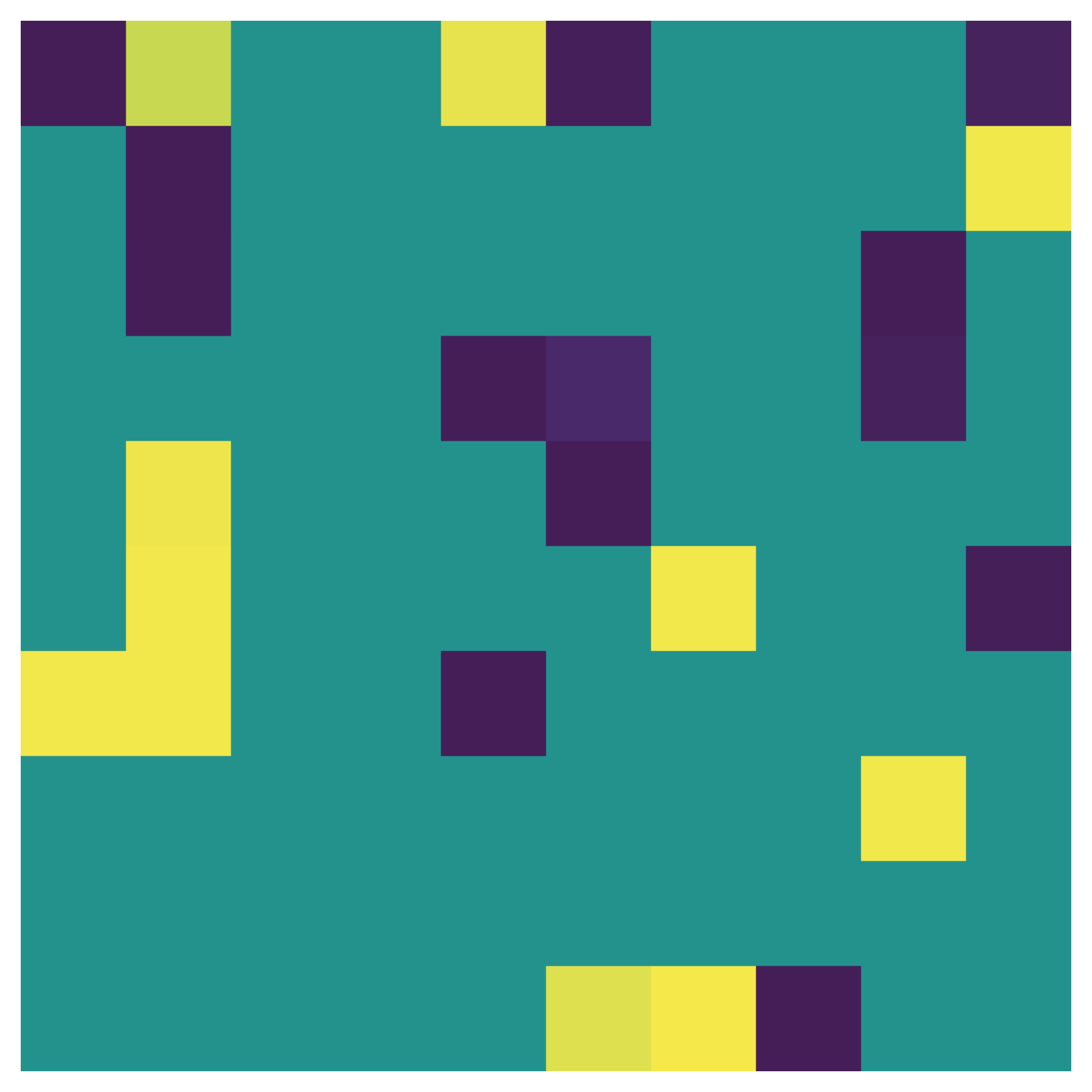}
    \end{subfigure}\\
    \caption{Heatmaps for the undirected movement scenario at time $t=50$. Pockets of Consensus-makers and Zealots emerge, though the majority of the population are Gridlockers. Note that movement conserves population size and thus $z_2$ is not plotted.}
    \label{fig:HM_diffusion}
\end{figure}

We begin by considering the case of local undirected movement, where voters spread to neighbouring nodes regardless of utility. Figure \ref{fig:TS_diffusion} shows representative time series for different initial conditions. If the prevalence of each strategy is initially equal on average, then ultimately gridlock is prevalent (Figure \ref{fig:TS_diffusion_gridlock}). A sizable majority of the population (approximately $75\%$) are Gridlockers, and the remainder are a mix of the remaining strategies. Note that in this simulation, the initial densities of each type of Zealot are marginally higher than Consensus-makers, which is due to the randomness of the initial conditions. This difference persists through to the final time. In Figure \ref{fig:TS_diffusion_consensus}, we constrain the initial conditions for Gridlockers and Party $2$ Zealots to be less than $1/4$ of the total population, resulting in a high prevalence of Party $1$ Zealots and votes for Party $1$ by Consensus-makers. Gridlockers are still sizable, between $25-30\%$ of the population by the end of the simulation, while Party $2$ Zealots vanish and Consensus-makers stabilize at similar levels to Gridlockers.

Figure \ref{fig:HM_diffusion} depicts the spatial distribution of the results of Figure \ref{fig:TS_diffusion_gridlock} at time $t=50$. Consensus-makers and Zealots form pockets where they are highly prevalent and vote for one party or the other. Since Consensus-makers are indifferent to which party is a majority, they align with the party of dominant Zealots in their node. Notice how the random favourable initial condition for Party $2$ Zealots creates more areas voting for Party $2$ than $1$. Outside of these enclaves, Gridlockers are widespread. Thus, space and diffusion can generate population heterogeneity unobserved in the ODE model.

\subsection{Directed movement without spillovers}

\begin{figure}[h!]
\captionsetup[subfigure]{justification=centering}
    \centering
    \begin{subfigure}[]{0.3\columnwidth}
        \caption{$\lambda=0$}
         \includegraphics[width=\textwidth]{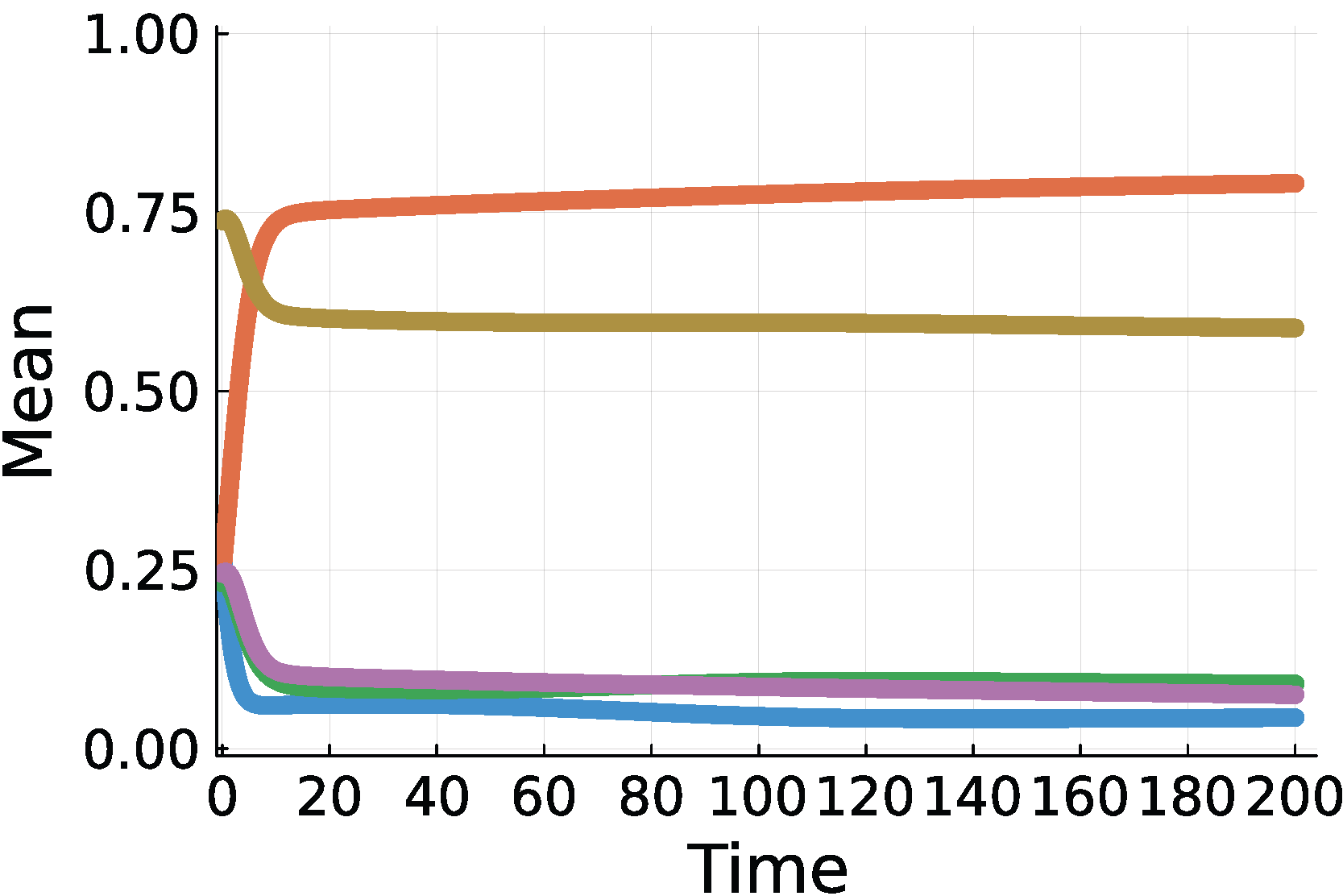}
    \end{subfigure}
    \begin{subfigure}[]{0.3\columnwidth}
        \caption{$\lambda=\tfrac{1}{2}$}
        \includegraphics[width=\textwidth]{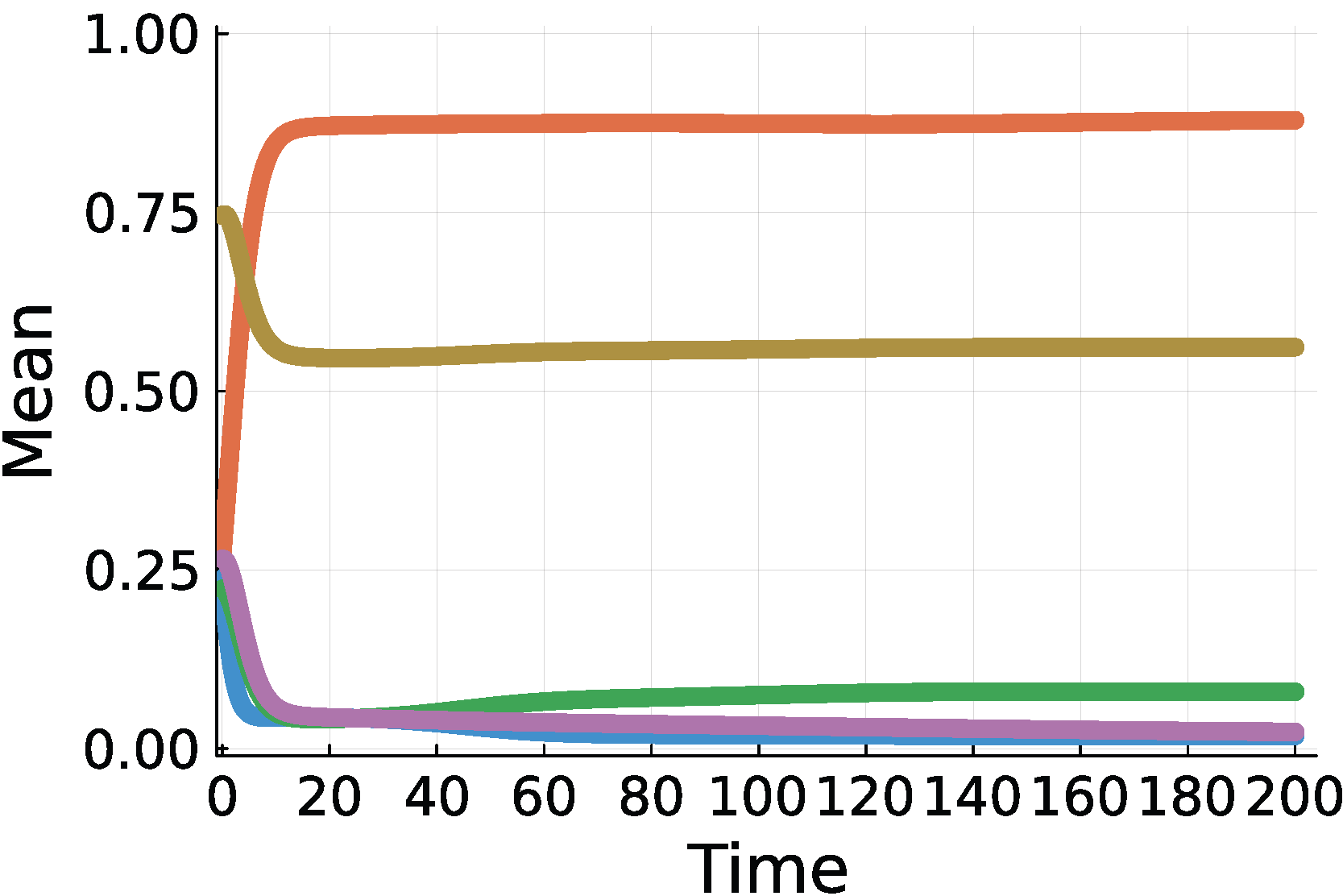}
    \end{subfigure}
        \begin{subfigure}[]{0.3\columnwidth}
        \caption{$\lambda=1$}
        \includegraphics[width=\textwidth]{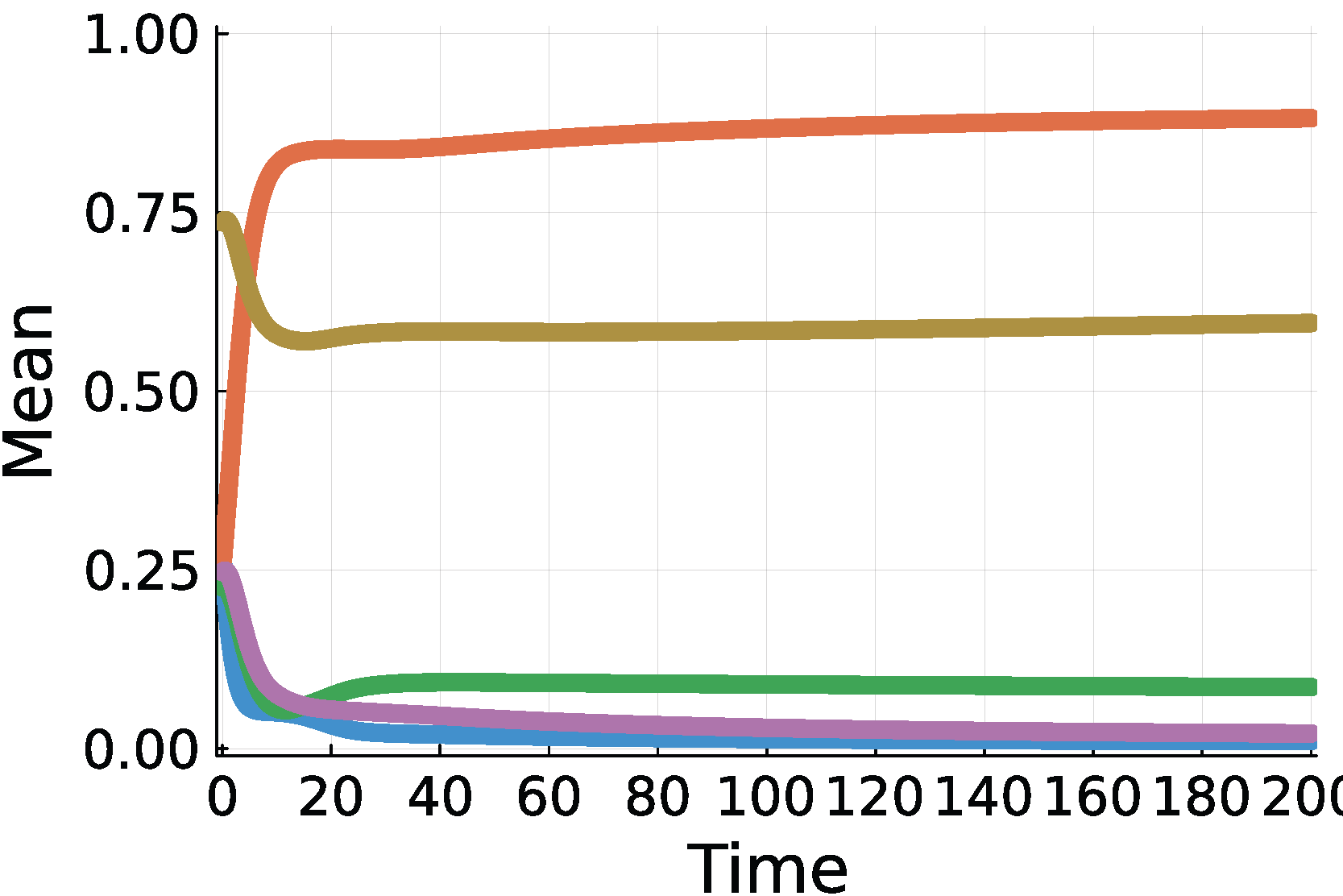}
    \end{subfigure}\\
    \begin{tikzpicture}
        \matrix (legend) {
            \draw[consensus, line width=3pt] (0,0) -- (0.6,0) node[right, black]{Mean Consensus-makers}; &
            \draw[gridlock, line width=3pt] (0,0) -- (0.6,0) node[right, black]{Mean Gridlockers}; &
            \draw[zealot1,line width=3pt] (0,0) -- (0.6,0) node[right, black]{Mean  Party $1$ Zealots}; \\
            \draw[zealot2, line width=3pt] (0,0) -- (0.6,0) node[right, black]{Mean Party $2$ Zealots}; &
            \draw[vote1, line width=3pt] (0,0) -- (0.6,0) node[right, black]{Mean Vote for Party $1$}; \\};
    \end{tikzpicture}
        \caption{Representative time series for the directed movement scenario without spillovers. The preferences for the public good are $\lambda=0,0.5,$ and $1$ for the columns left to right.}
    \label{fig:TS_directed_movement}
\end{figure}
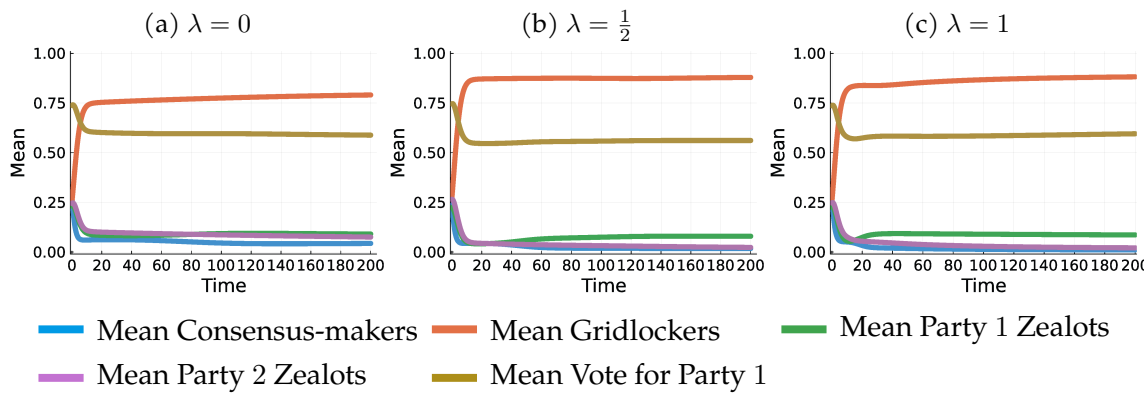

Here we consider directed movement with no spillovers ($s=0$). Figure \ref{fig:TS_directed_movement} depicts the time series for this scenario. Like the undirected movement scenario (Figure \ref{fig:TS_diffusion}), Gridlockers and gridlock are common, and this is exacerbated when economic utility drives movement ($\lambda=1$). Party $1$ Zealots are also more common, relative to the non-Gridlocker strategies, the higher $\lambda$. When movement is driver by economic utility, individuals relocate to regions with high $v$, which are where Party $1$ Zealots are more frequent. As Gridlockers and Party $2$ Zealots flock to these locations, they can adapt their strategies, becoming Consensus-makers or Party $1$ Zealots. As such, $v$ remains unchanged by migration. On the other hand, if their influx is sufficiently high, they can undermine votes for Party $1$ and bring about gridlock.

\begin{figure}[ht!]
\captionsetup[subfigure]{justification=centering}
    \centering
    \begin{subfigure}[]{0.3\columnwidth}
        \caption*{Population Colorbar}
        \includegraphics[width=\textwidth]{figs_png/Population_Colorbar.png}
            \vspace{2pt}
    {\small 0 \hfill 1}
    \end{subfigure} \hspace{2cm}
    \begin{subfigure}[]{0.3\columnwidth}
        \caption*{Party Colorbar}
        \includegraphics[width=\textwidth]{figs_png/PartyColorbar.png}
            \vspace{2pt}
    {\small 0 \hfill 1}
    \end{subfigure}\\
    \begin{subfigure}[]{0.15\columnwidth}
        \caption{$c, \lambda=0$}
        \includegraphics[width=\textwidth]{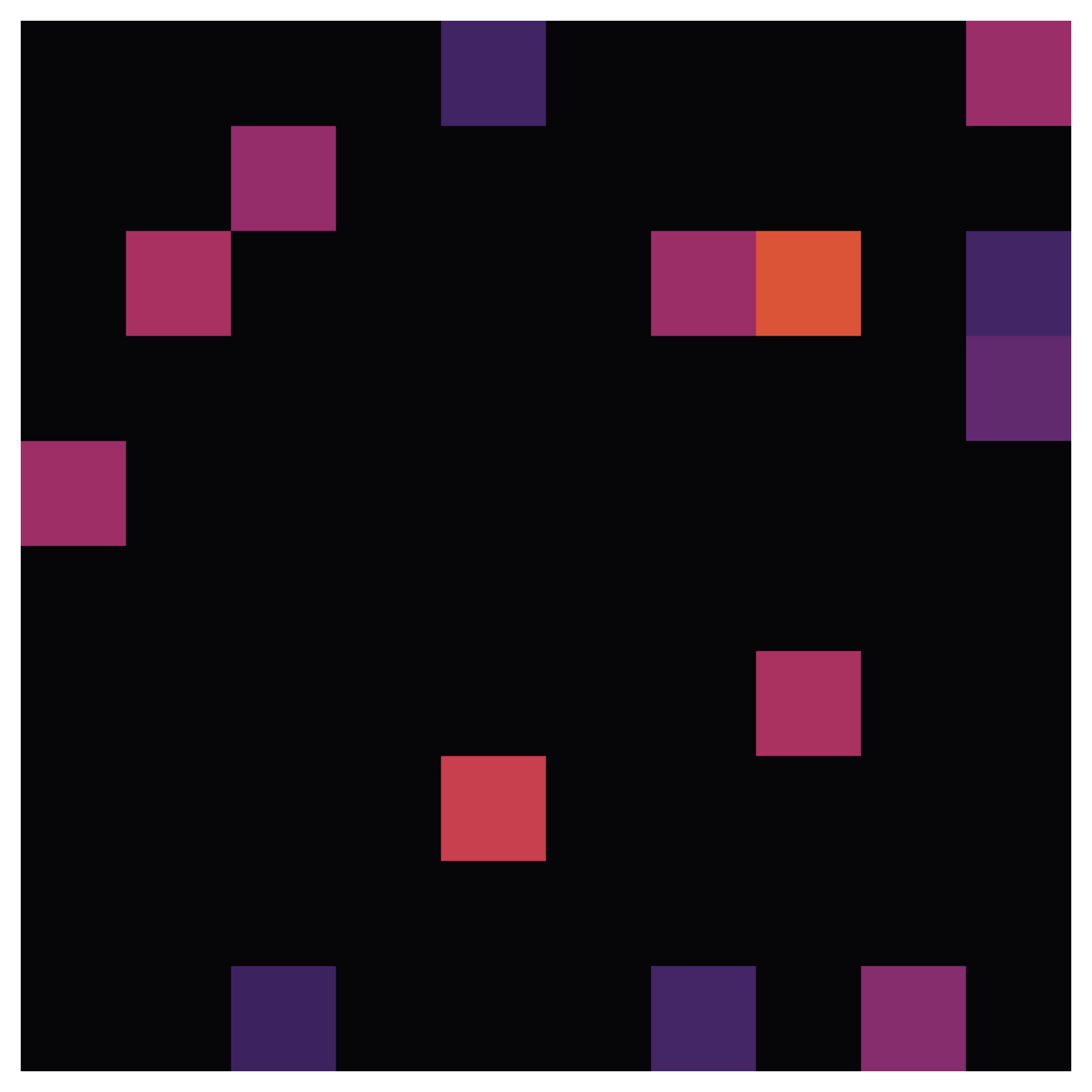}
    \end{subfigure}
    \begin{subfigure}[]{0.15\columnwidth}
        \caption{$g, \lambda=0$}
        \includegraphics[width=\textwidth]{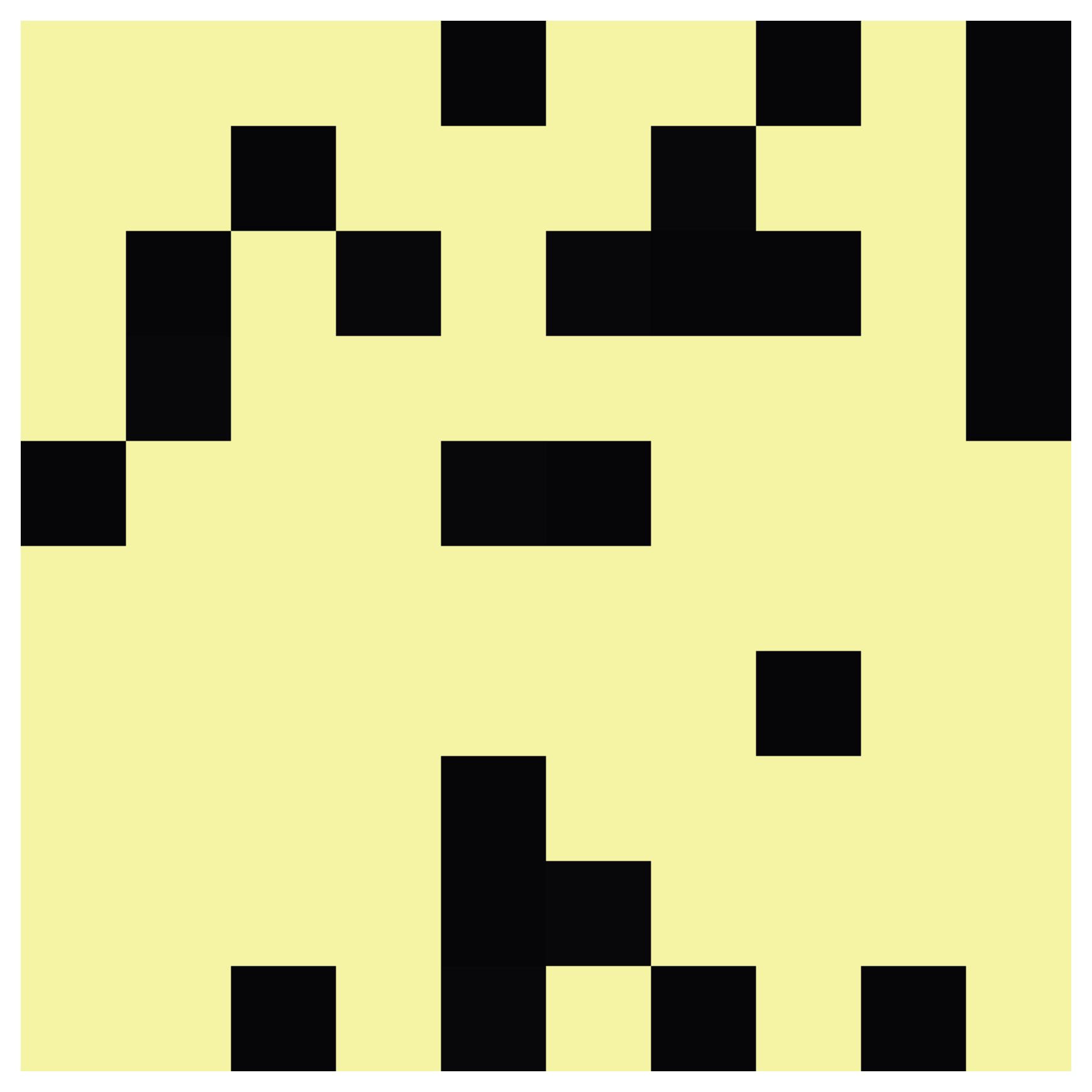}
    \end{subfigure}
        \begin{subfigure}[]{0.15\columnwidth}
        \caption{$z_1, \lambda=0$}
        \includegraphics[width=\textwidth]{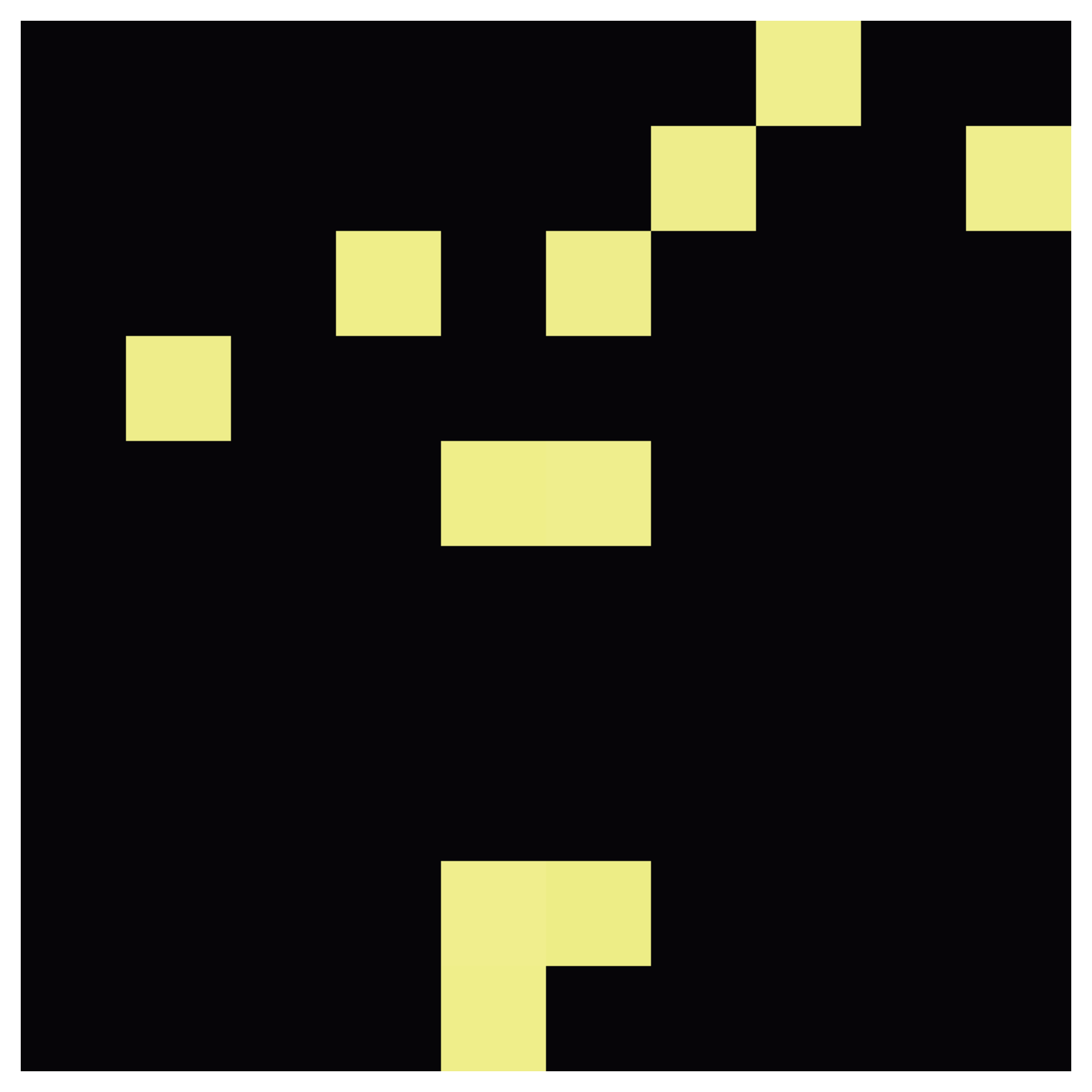}
    \end{subfigure}
        \begin{subfigure}[]{0.15\columnwidth}
        \caption{$z_2, \lambda=0$}
        \includegraphics[width=\textwidth]{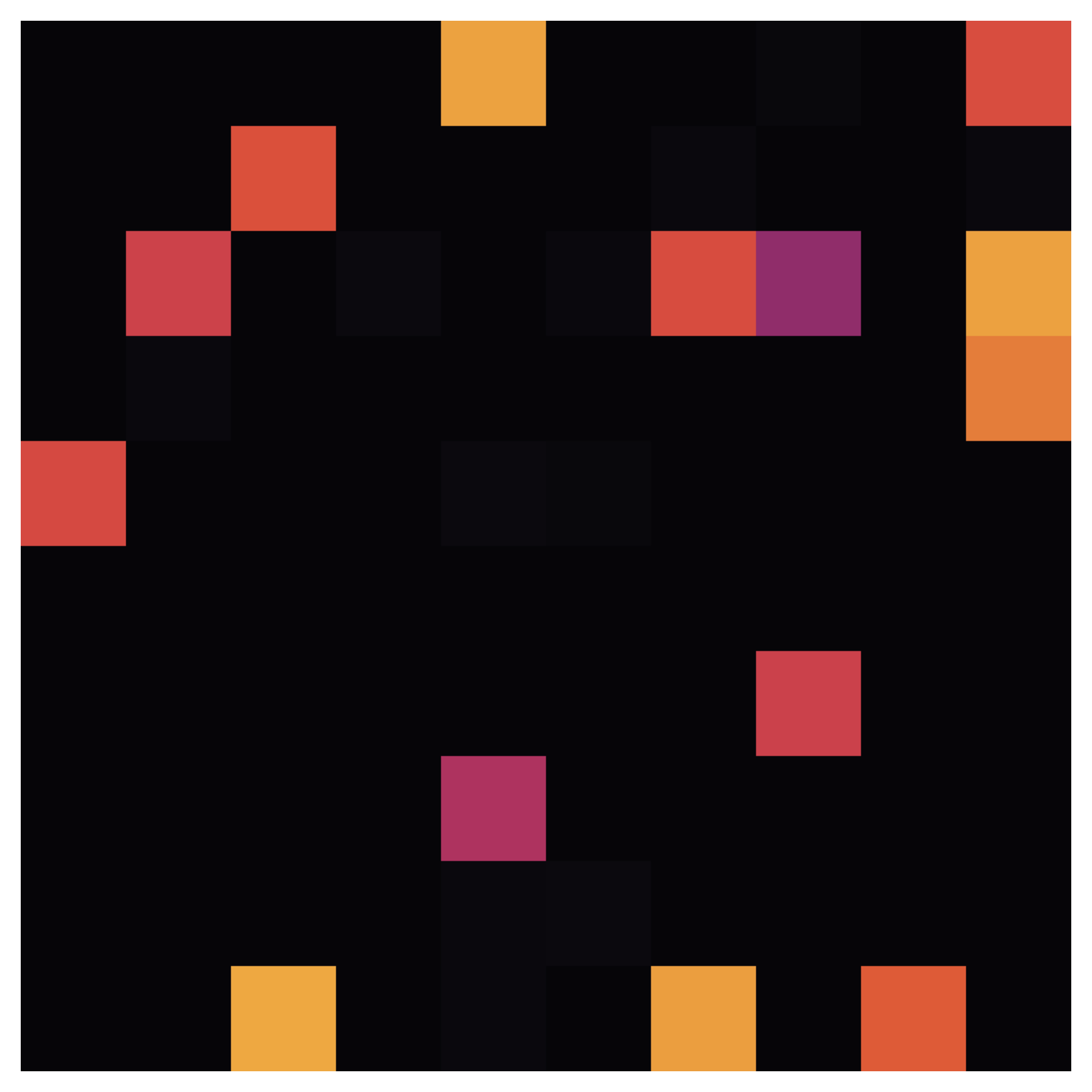}
    \end{subfigure}
         \begin{subfigure}[]{0.15\columnwidth}
         \caption{$\tilde{p}, \lambda=0$}
         \includegraphics[width=\textwidth]{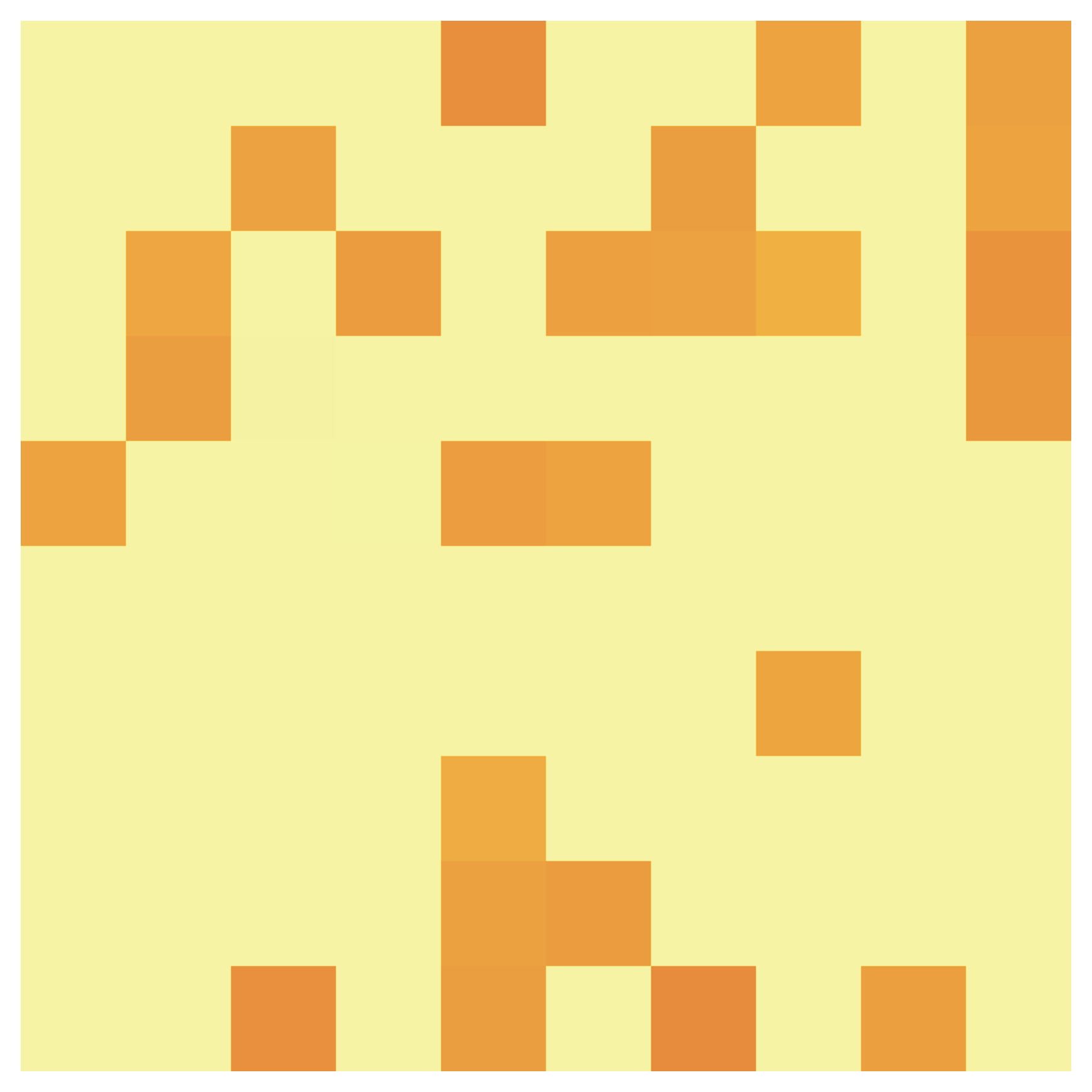}
     \end{subfigure}
        \begin{subfigure}[]{0.15\columnwidth}
        \caption{$v, \lambda=0$}
        \includegraphics[width=\textwidth]{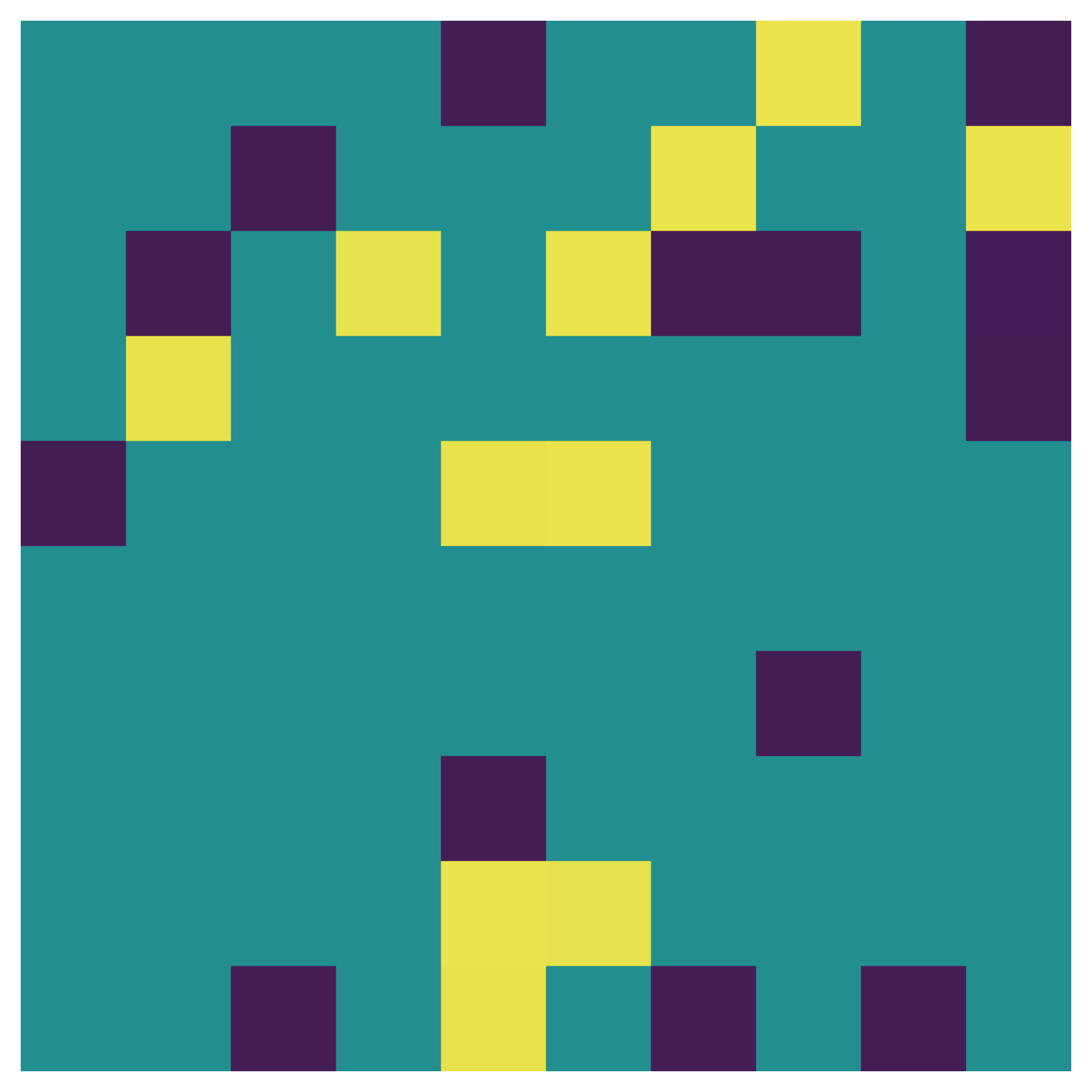}
    \end{subfigure}\\
    \begin{subfigure}[]{0.15\columnwidth}
        \caption{$c, \lambda=\tfrac{1}{2}$}\label{fig:HM_directed_c_lambda0.5}
        \includegraphics[width=\textwidth]{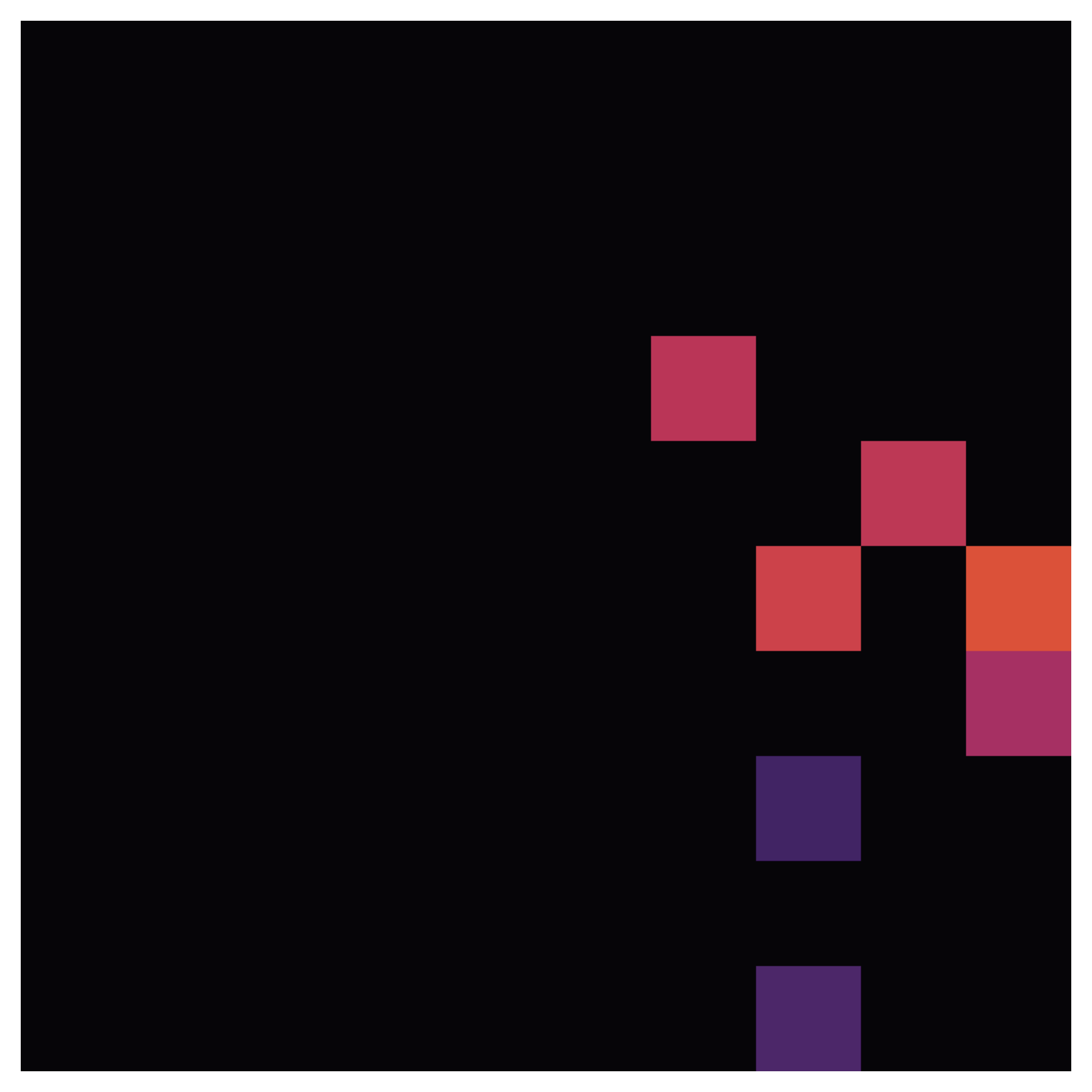}
    \end{subfigure}
    \begin{subfigure}[]{0.15\columnwidth}
        \caption{$g, \lambda=\tfrac{1}{2}$}
        \includegraphics[width=\textwidth]{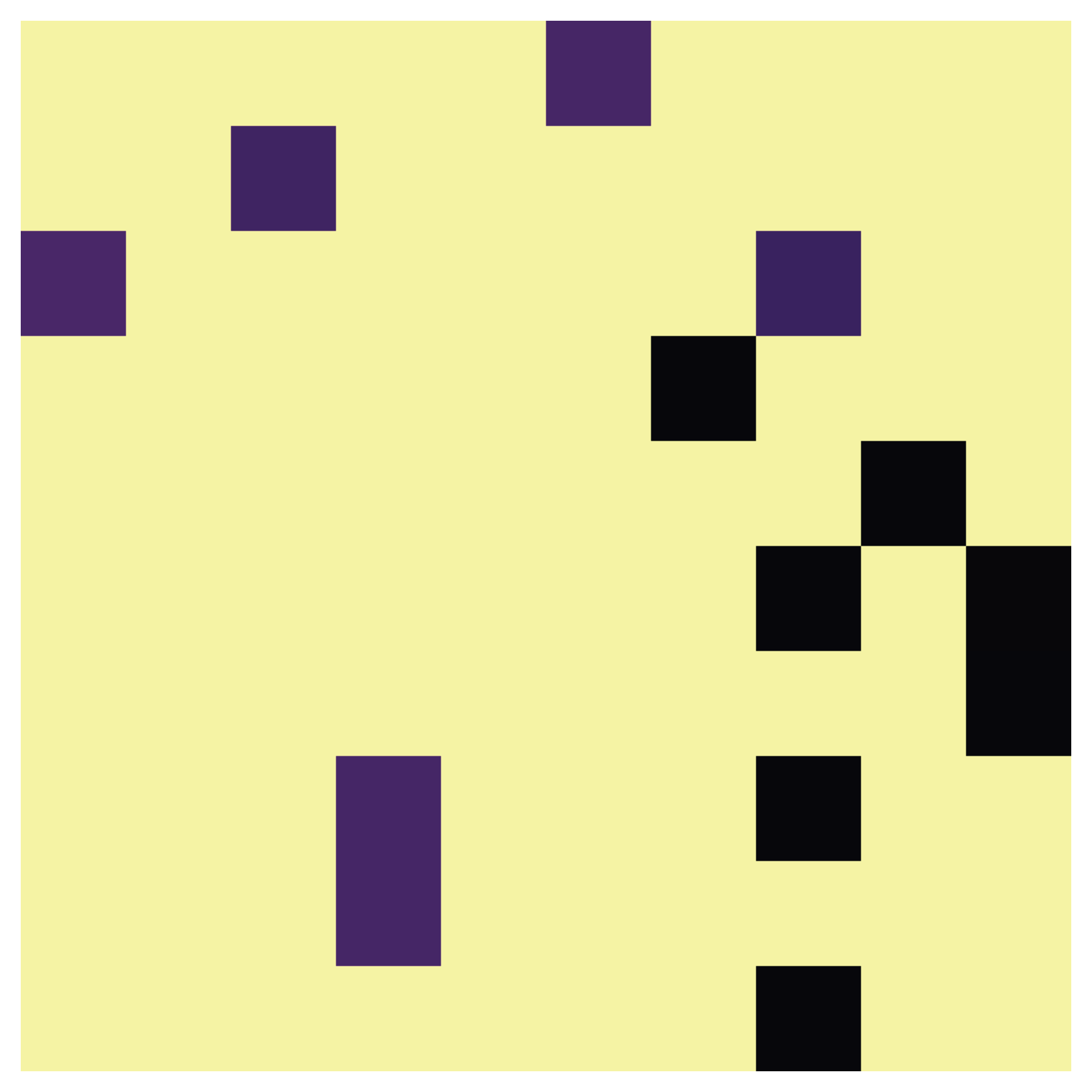}
    \end{subfigure}
        \begin{subfigure}[]{0.15\columnwidth}
        \caption{$z_1, \lambda=\tfrac{1}{2}$}
        \includegraphics[width=\textwidth]{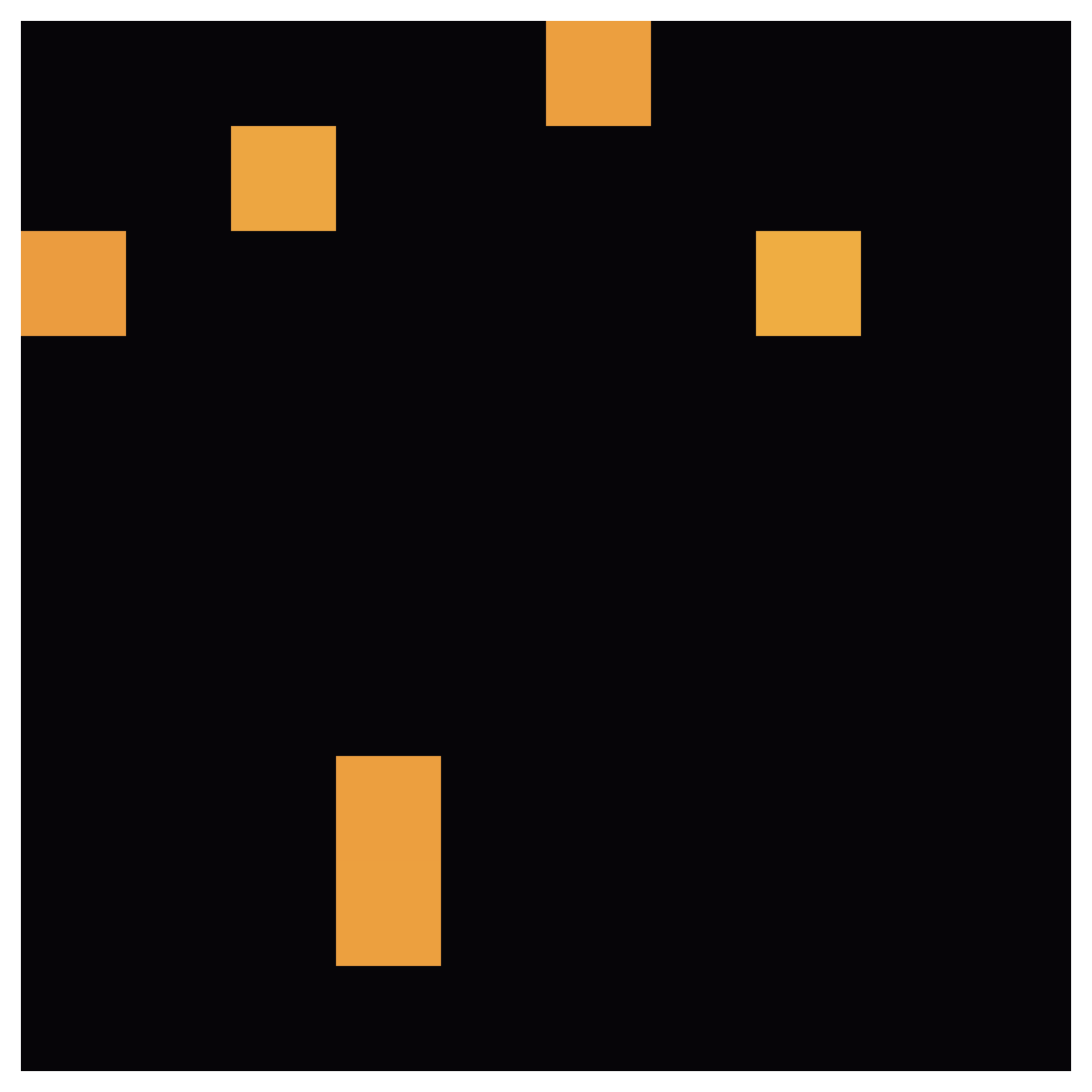}
    \end{subfigure}
        \begin{subfigure}[]{0.15\columnwidth}
        \caption{$z_2, \lambda=\tfrac{1}{2}$}
        \includegraphics[width=\textwidth]{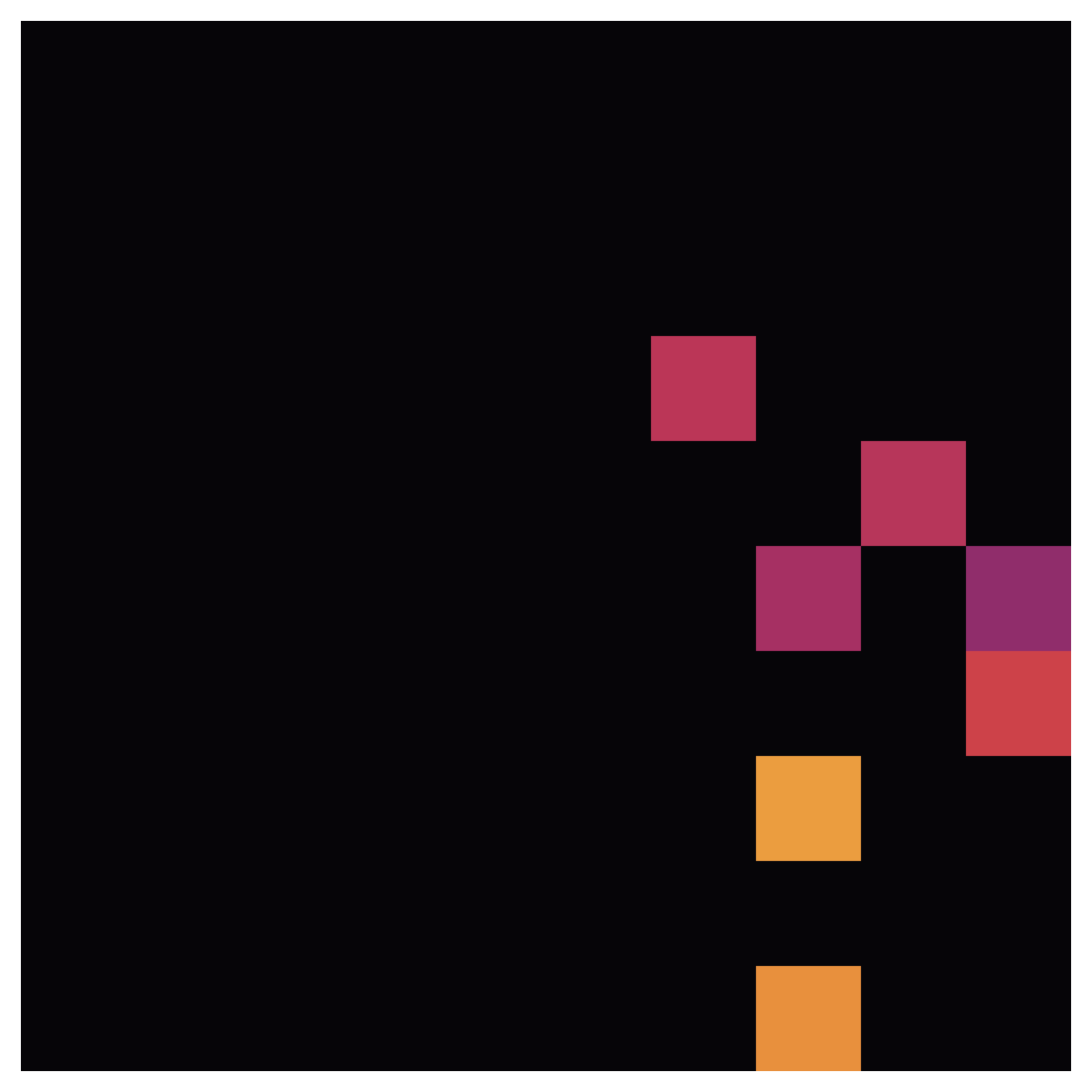}
    \end{subfigure}
         \begin{subfigure}[]{0.15\columnwidth}
         \caption{$\tilde{p}, \lambda=\tfrac{1}{2}$}
         \includegraphics[width=\textwidth]{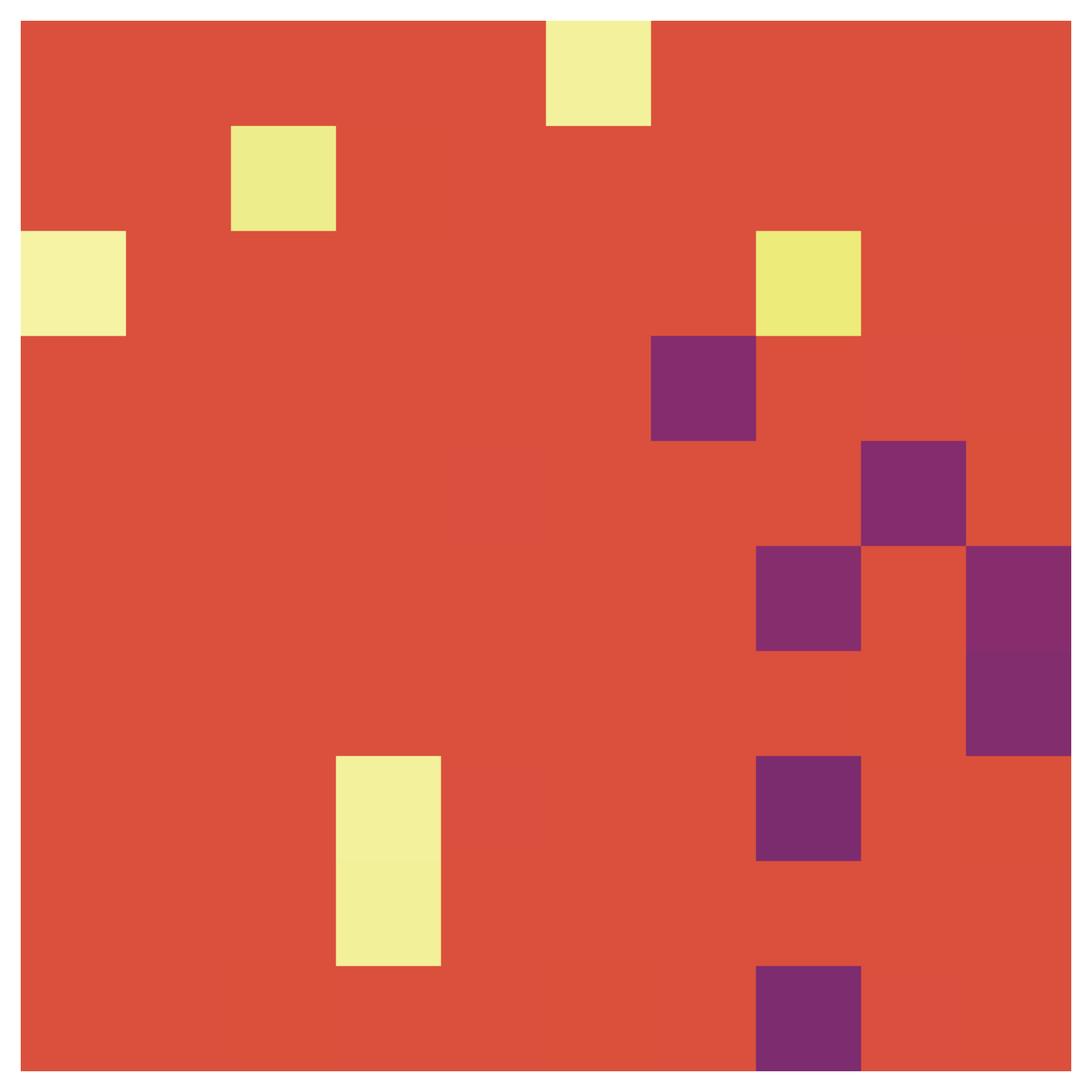}
     \end{subfigure}
        \begin{subfigure}[]{0.15\columnwidth}
        \caption{$v, \lambda=\tfrac{1}{2}$}\label{fig:HM_directed_v_lambda0.5}
        \includegraphics[width=\textwidth]{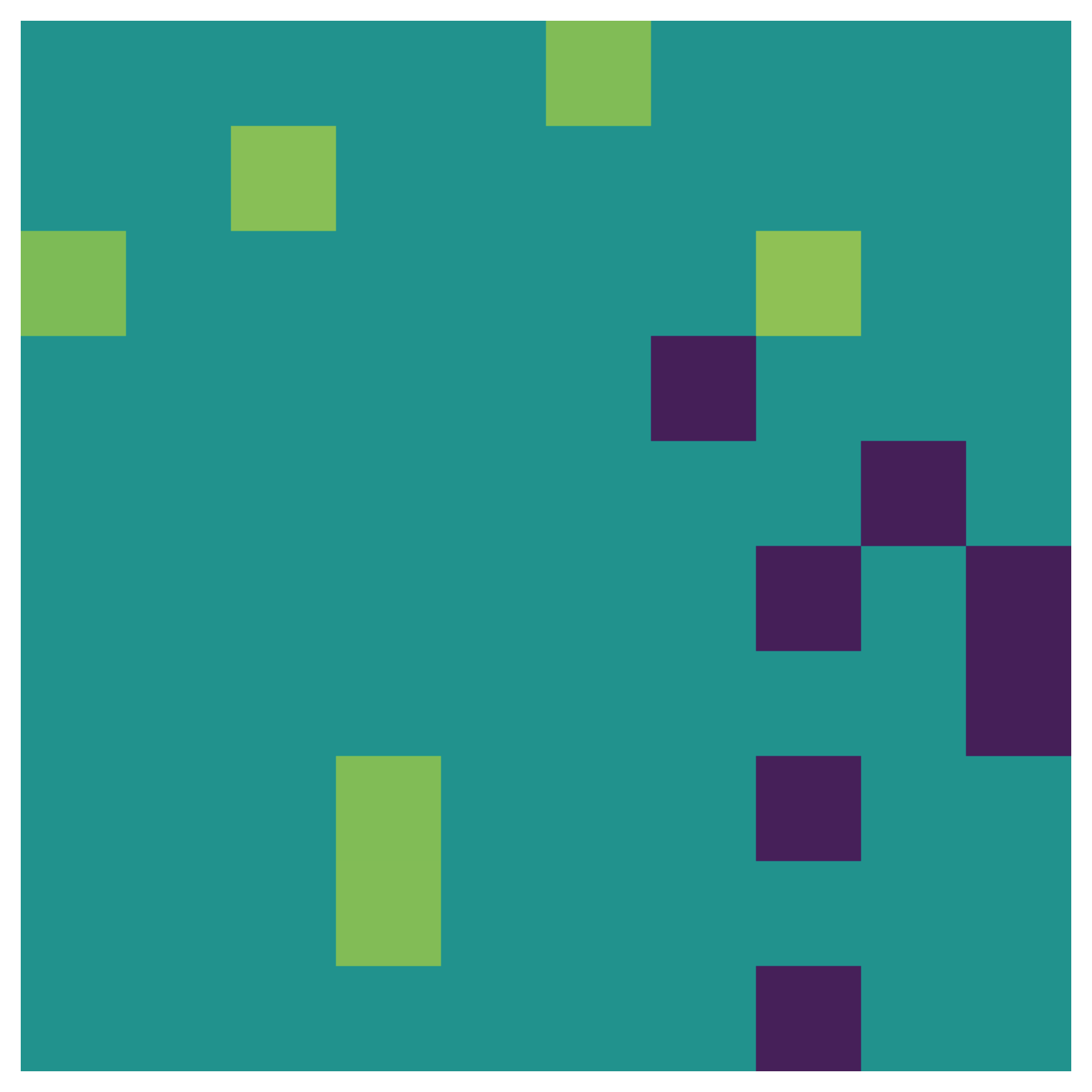}
    \end{subfigure}\\
    \begin{subfigure}[]{0.15\columnwidth}
        \caption{$c, \lambda=1$}\label{fig:HM_directed_c_lambda1}
        \includegraphics[width=\textwidth]{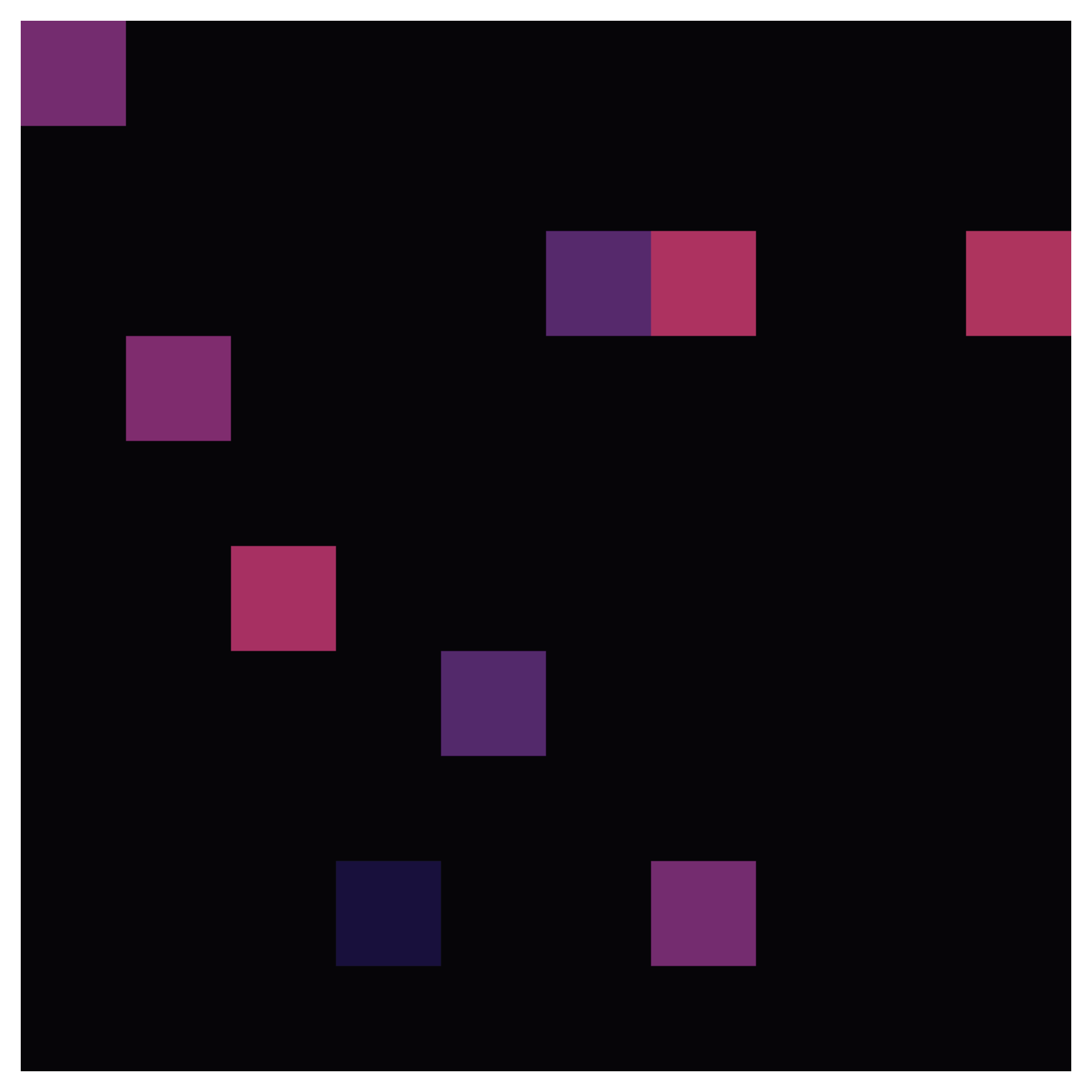}
    \end{subfigure}
    \begin{subfigure}[]{0.15\columnwidth}
        \caption{$g, \lambda=1$}
        \includegraphics[width=\textwidth]{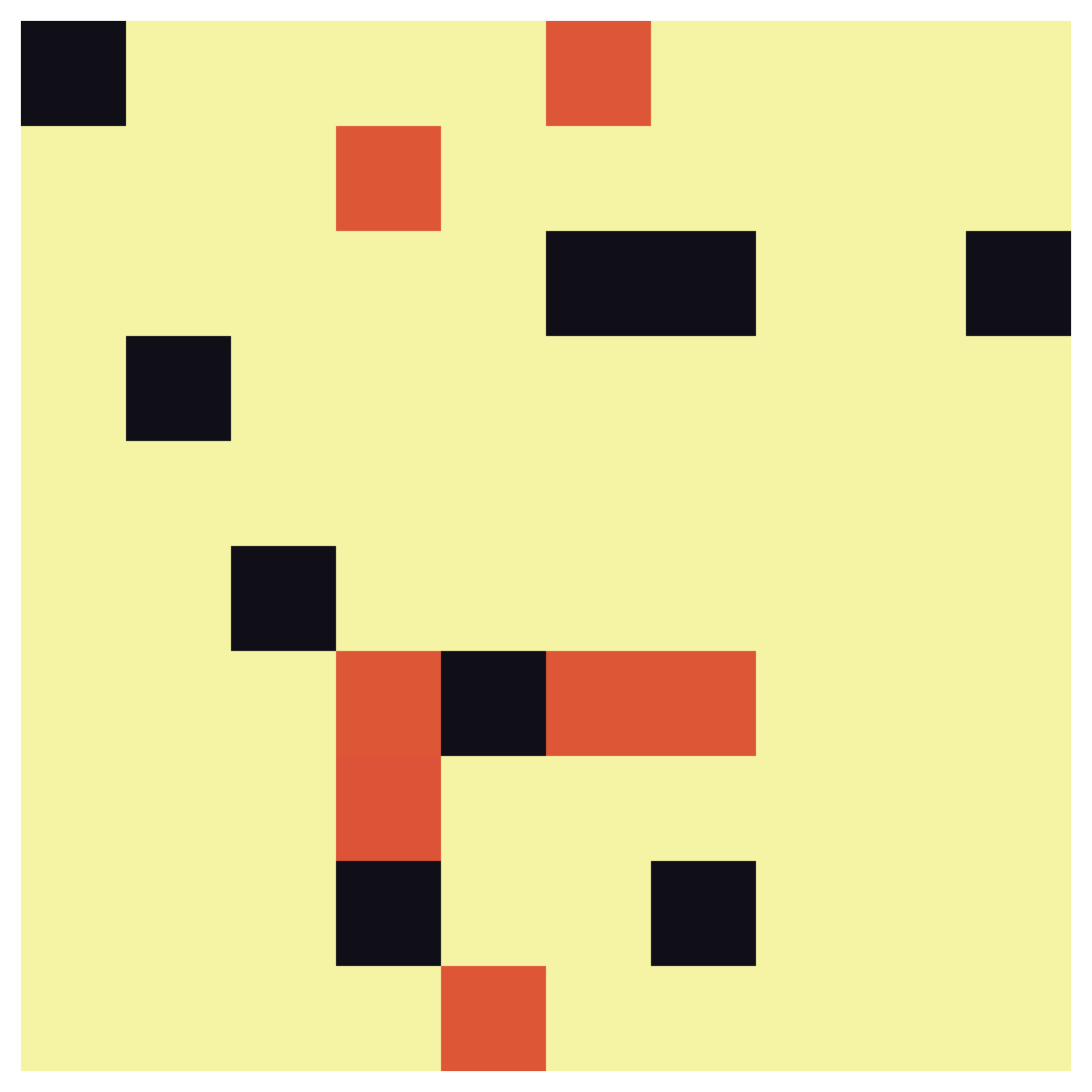}
    \end{subfigure}
        \begin{subfigure}[]{0.15\columnwidth}
        \caption{$z_1, \lambda=1$}
        \includegraphics[width=\textwidth]{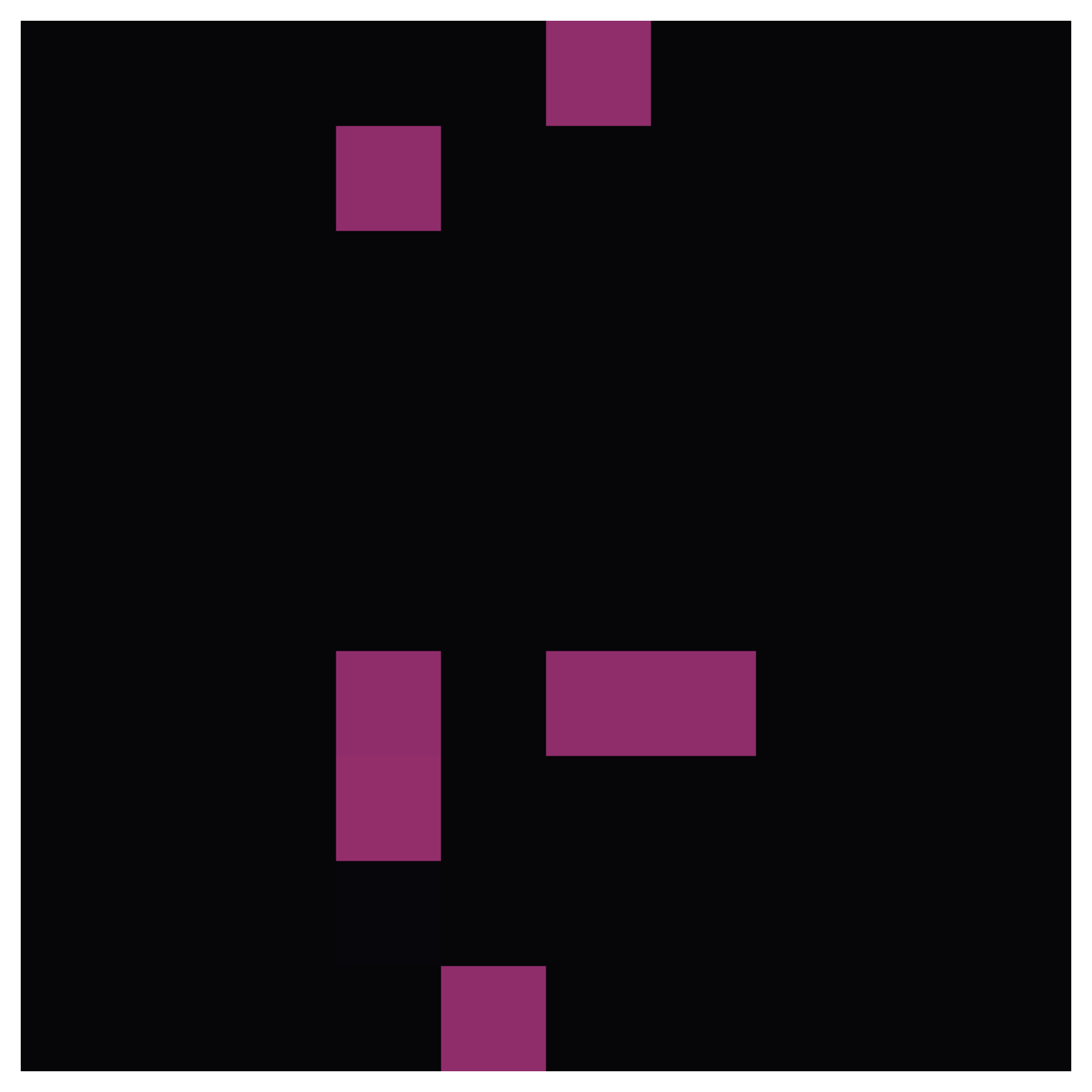}
    \end{subfigure}
        \begin{subfigure}[]{0.15\columnwidth}
        \caption{$z_2, \lambda=1$}
        \includegraphics[width=\textwidth]{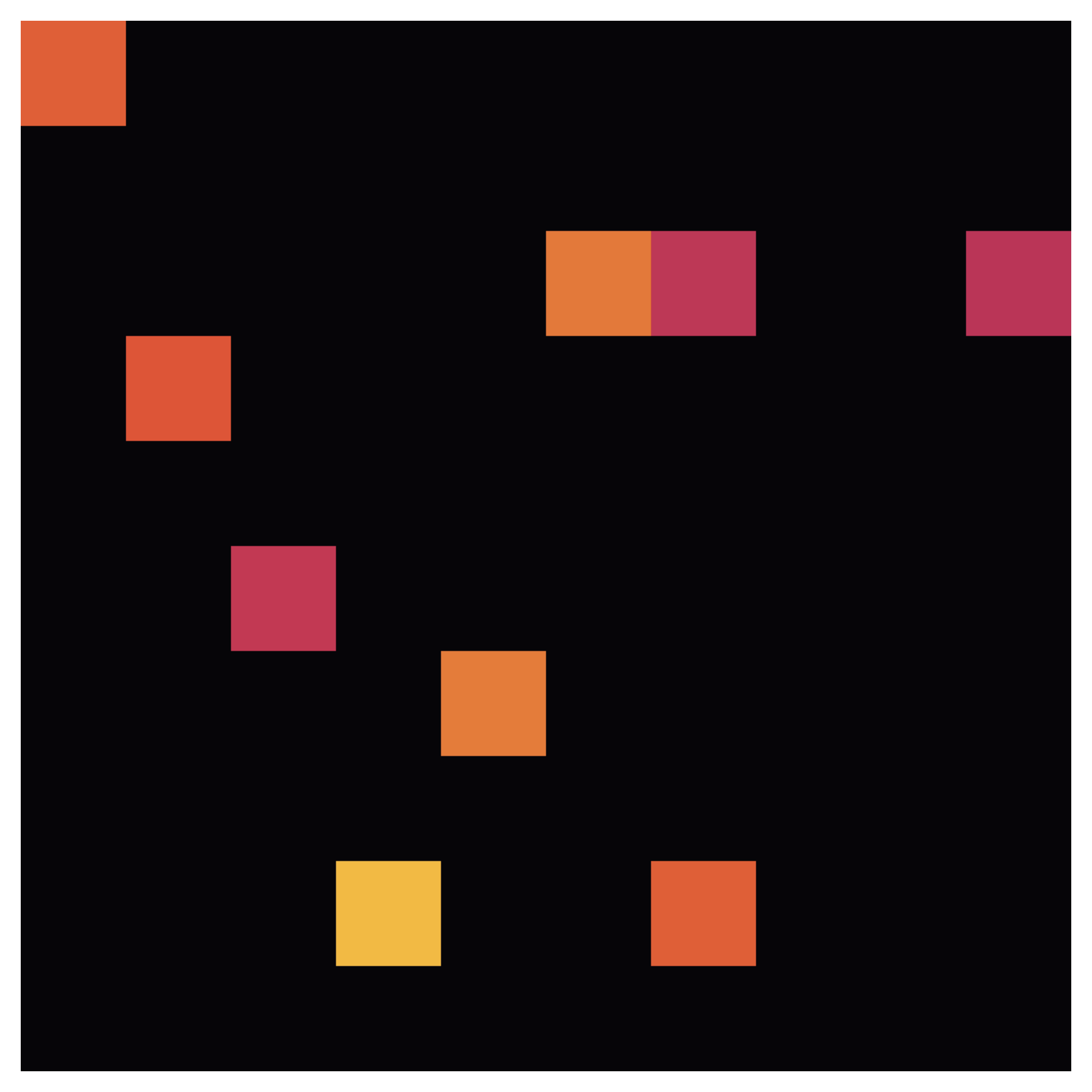}
    \end{subfigure}
         \begin{subfigure}[]{0.15\columnwidth}
         \caption{$\tilde{p}, \lambda=1$}
         \includegraphics[width=\textwidth]{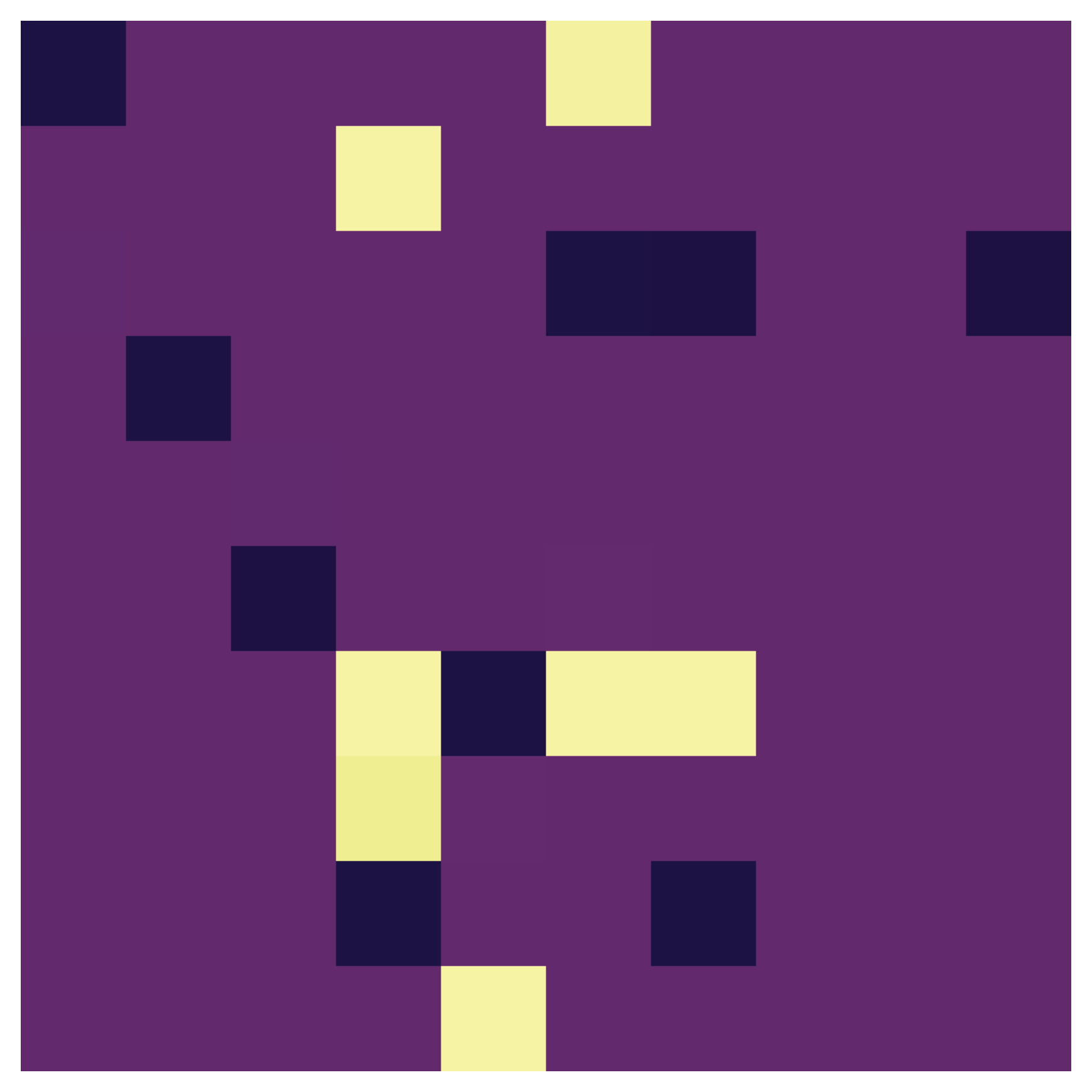}
     \end{subfigure}
        \begin{subfigure}[]{0.15\columnwidth}
        \caption{$v, \lambda=1$}\label{fig:HM_directed_v_lambda1}
        \includegraphics[width=\textwidth]{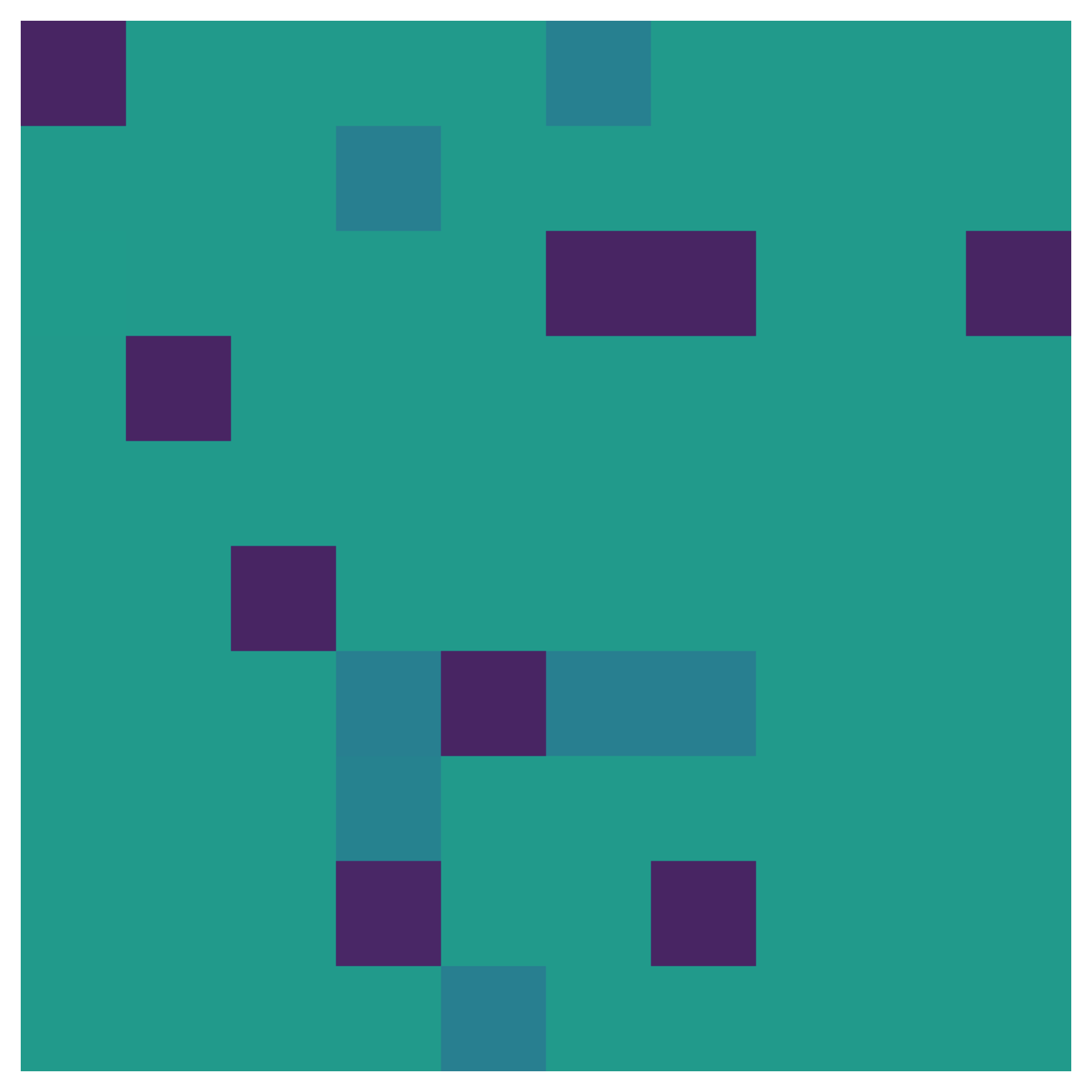}
    \end{subfigure}
    \caption{Heatmaps for the directed movement scenario with no spillovers at time $t=200$. Here, we vary the preference utility.}
    \label{fig:HM_directed_movement}
\end{figure}

Figure \ref{fig:HM_directed_movement} depicts representative heatmaps for the directed movement scenario. The first row depicts the case where movement is strictly based on the utility accrued from their voting preference ($\lambda=0$). In the middle row, there is a balance between economic and social utility driving movement ($\lambda=1/2$). And, in the bottom row, movement is strictly determined by the utility gained from the public good ($\lambda=1$). As in the undirected movement scenario, Gridlockers are prevalent across space, and there are pockets of other strategies. The results for $\lambda=0$ are most similar to the undirected movement scenario. The small non-Gridlock pockets totally vote for one party or the other ($v=0$ or $v=1$). However, the total population at these nodes is lower than the Gridlocker ones, which is caused by the higher prevalence of Gridlockers. Note that the heatmaps for the strategy concentrations show the proportion of the population at each node that is that strategy, not their abundance (which is reflected in the $\tilde{p}$ plots).

When movement is determined in part by economic utility ($\lambda>0$), Gridlockers and other strategies can coexist at nodes. For the case of $\lambda=1/2$, Gridlockers and Party $1$ Zealots can coexist, resulting in a vote $v$ intermediate between $v=1/2$ and $v=1$. Though Gridlockers suppress the vote by voting for Party $2$, the public good is relatively well funding. Thus, these nodes have a relatively high economic utility that attracts migrants, resulting in relatively large populations. Party $2$ Zealots and Consensus-makers constitute the remaining pockets. The populations at these nodes are small, and the public good is unfunded. For $\lambda=1$, economic utility alone drives migration. Regions of Gridlockers are still prevalent. However, their population densities are much smaller than the $\lambda=1/2$ and $\lambda=0$ cases. Strategies coexist at the other nodes. Where Party $1$ Zealots and Gridlockers coexist, the populations are large. $v>1/2$, though less than the case when $\lambda = 1/2$. The pockets of Part $2$ Zealots and Consensus-makers are very small and $v$ very low. As before, poor funding of the public good attracts few migrants.

\subsection{Directed movement with spillovers}

\begin{figure}[htbp!]
\captionsetup[subfigure]{justification=centering}
    \centering
    \begin{subfigure}[]{0.3\columnwidth}
            \caption{$\lambda=\tfrac{1}{2}, s=\tfrac{4}{5}$}
        \includegraphics[width=\textwidth]{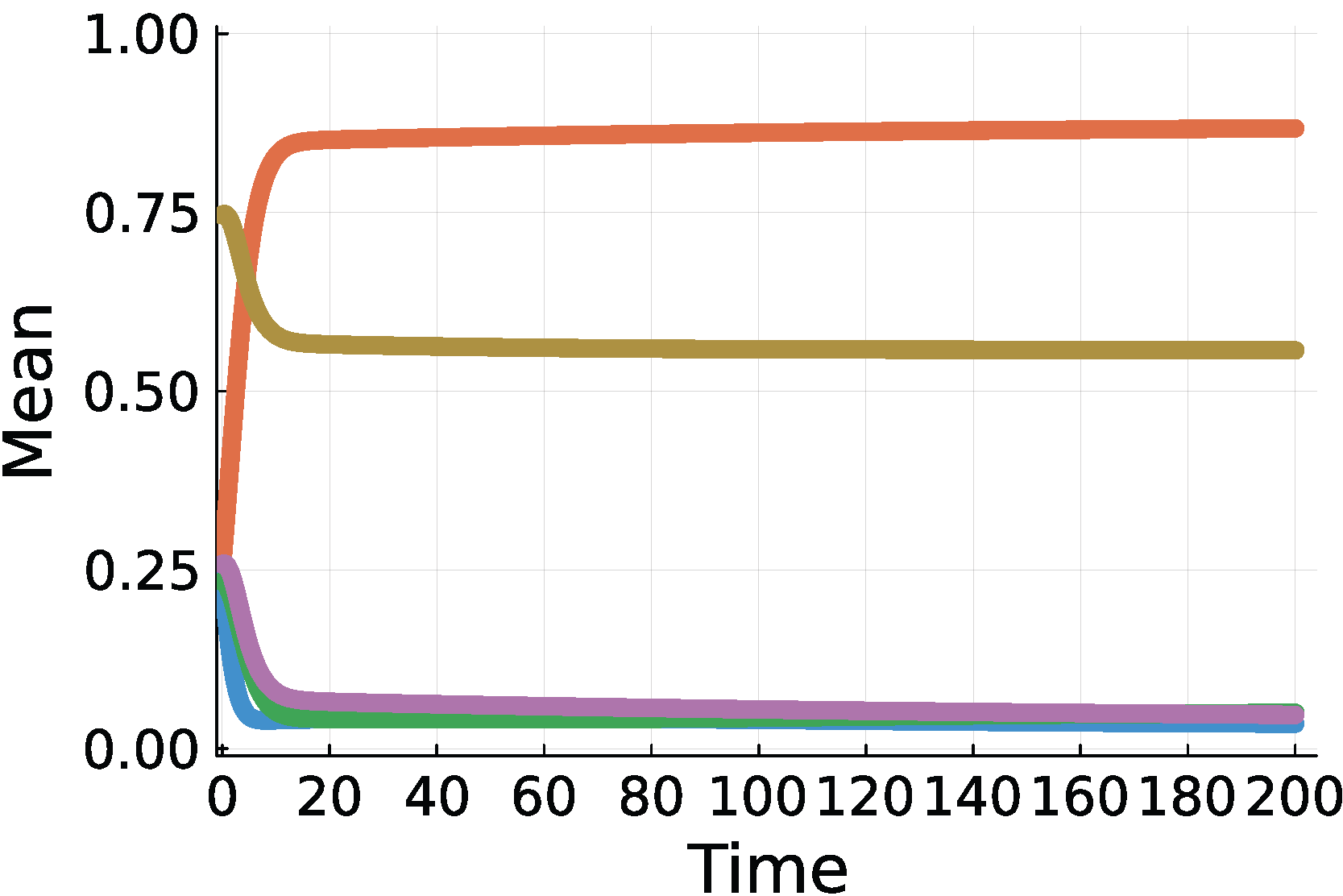}
    \end{subfigure}
        \begin{subfigure}[]{0.3\columnwidth}
            \caption{$\lambda=\tfrac{1}{2}, s=1$}
        \includegraphics[width=\textwidth]{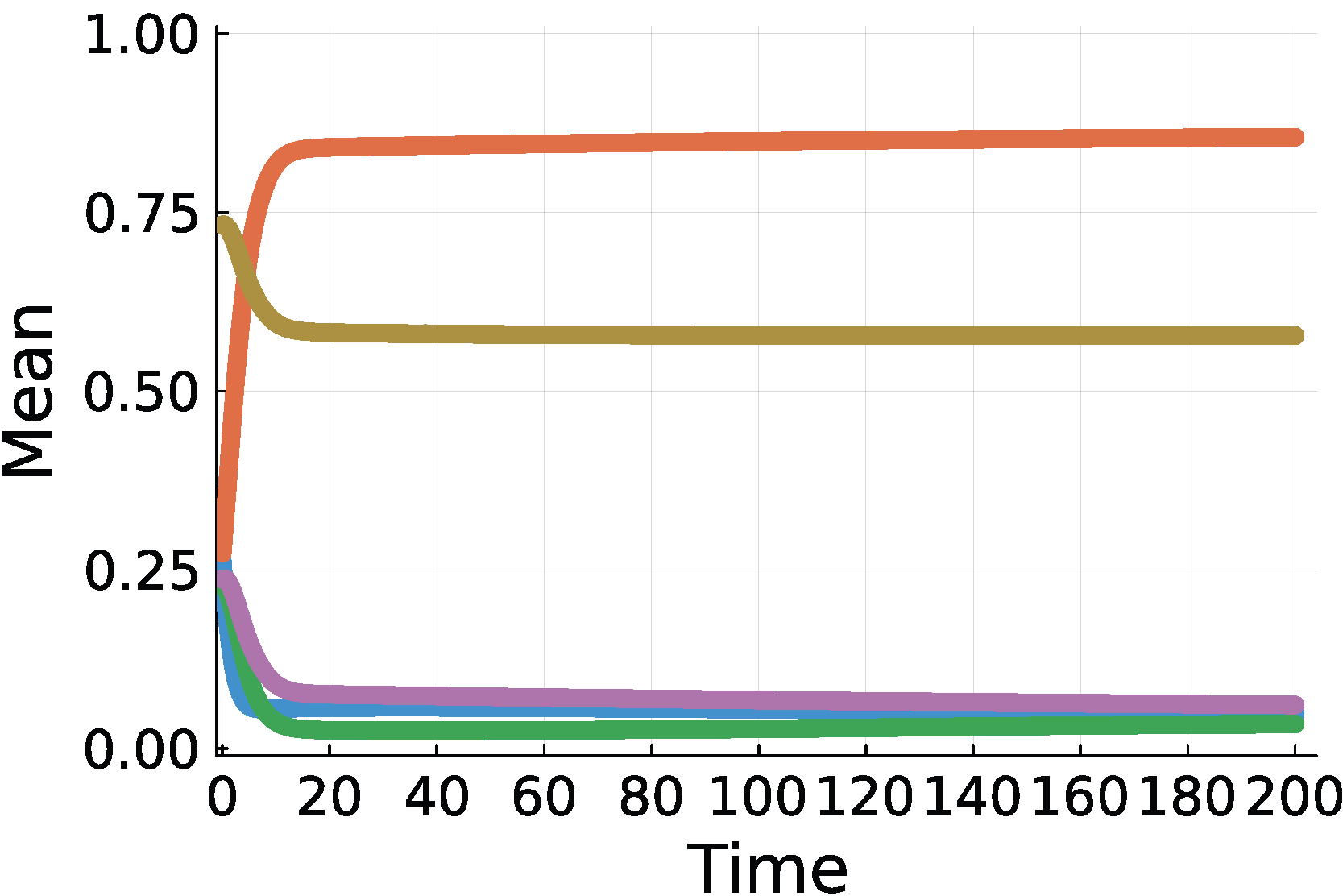}
    \end{subfigure}\\
    \begin{subfigure}[]{0.3\columnwidth}
            \caption{$\lambda=1, s=\tfrac{4}{5}$}
        \includegraphics[width=\textwidth]{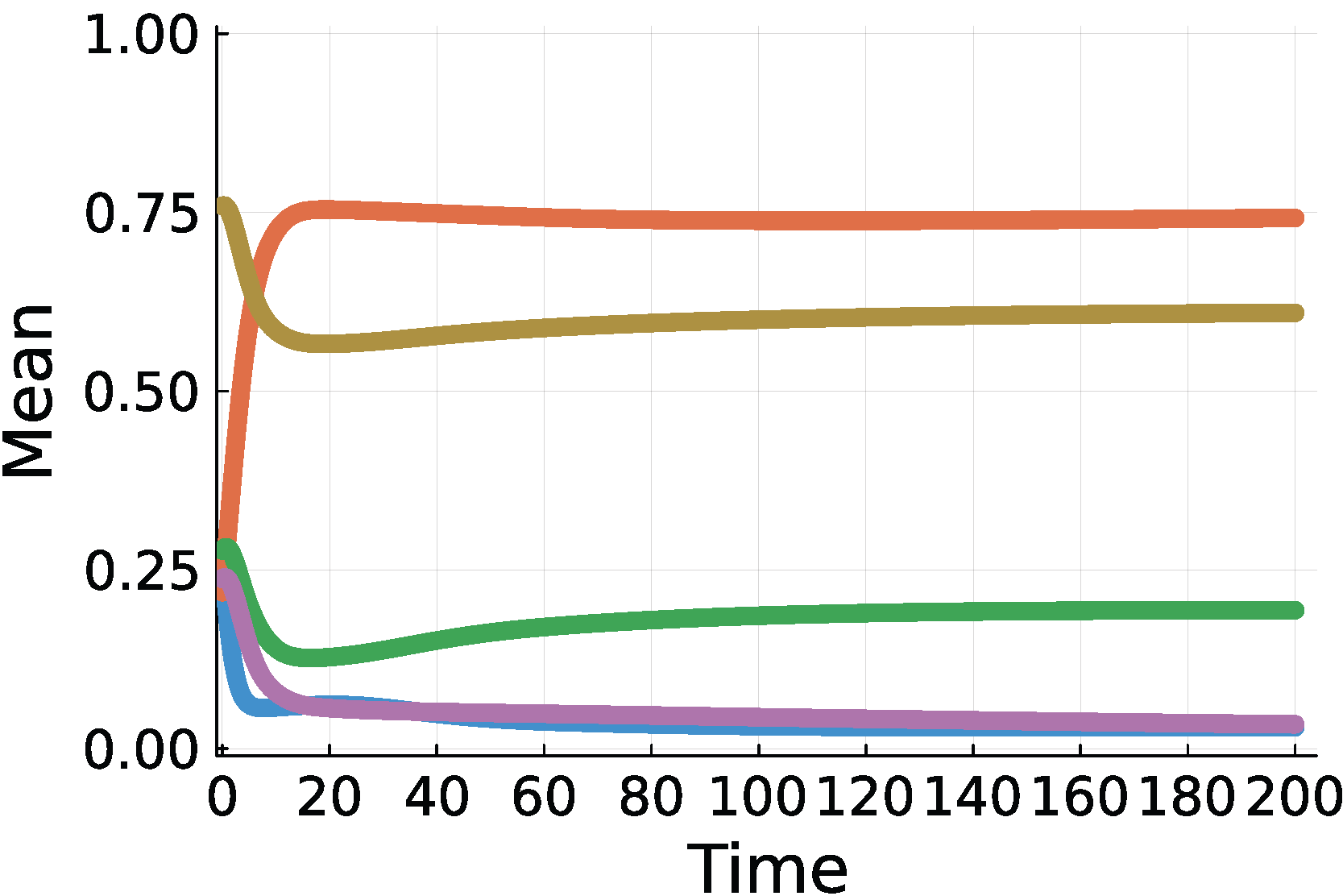}
    \end{subfigure}
    \begin{subfigure}[]{0.3\columnwidth}
            \caption{$\lambda=1, s=1$}
        \includegraphics[width=\textwidth]{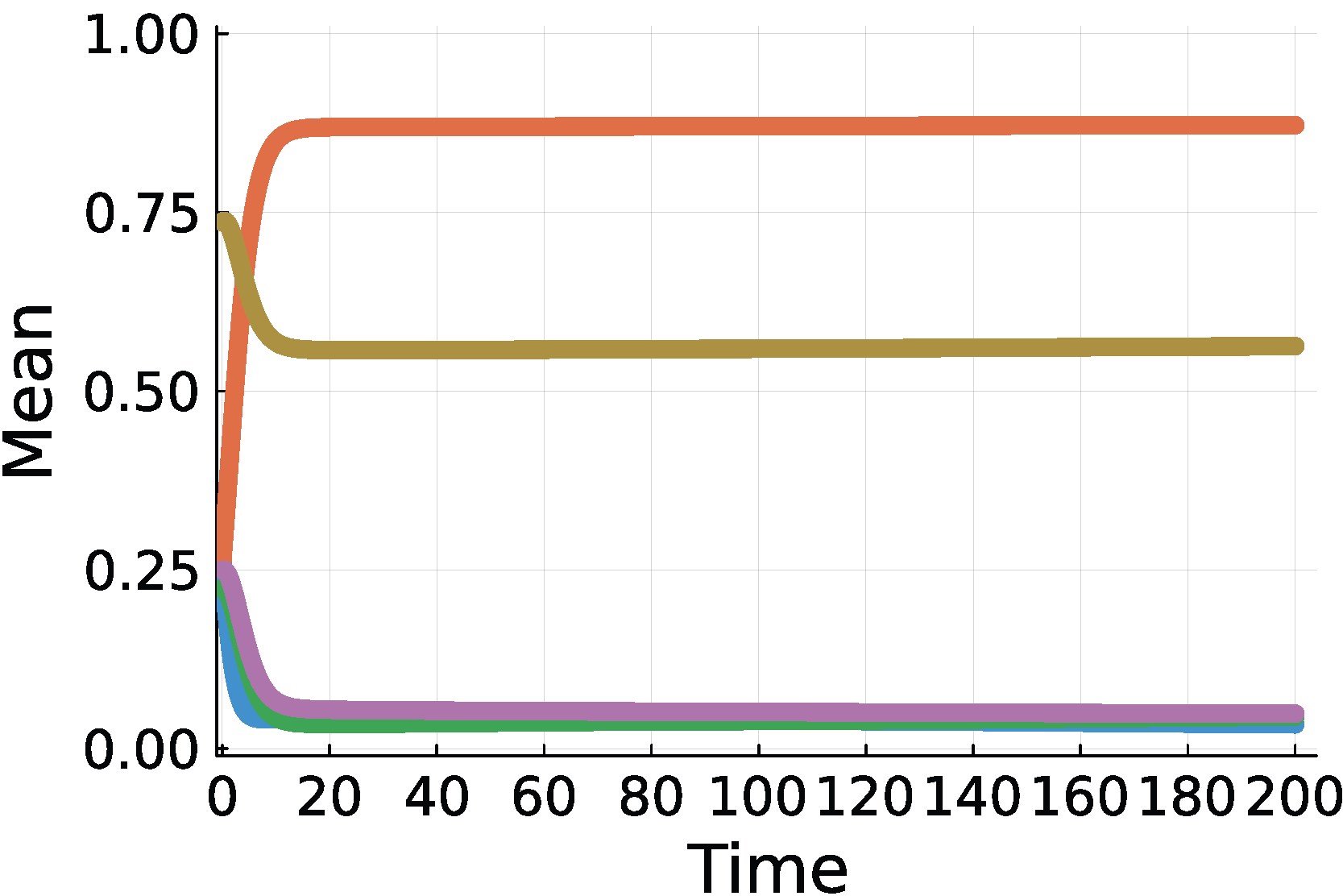}
    \end{subfigure}
    \begin{tikzpicture}
        \matrix (legend) {
            \draw[consensus, line width=3pt] (0,0) -- (0.6,0) node[right, black]{Mean Consensus Makers}; &
            \draw[gridlock, line width=3pt] (0,0) -- (0.6,0) node[right, black]{Mean Gridlockers}; &
            \draw[zealot1,line width=3pt] (0,0) -- (0.6,0) node[right, black]{Mean Party $1$ Zealots}; \\
            \draw[zealot2, line width=3pt] (0,0) -- (0.6,0) node[right, black]{Mean Party $2$ Zealots}; &
            \draw[vote1, line width=3pt] (0,0) -- (0.6,0) node[right, black]{Mean Vote for Party $1$}; \\
        };
    \end{tikzpicture}
        \caption{Representative time series for the directed movement scenario with spillovers. The preferences for the economic utility are $\lambda=1/2$ and $1$, and the spillovers are $s=4/5$ and $1$.}
    \label{fig:TS_spillovers}
\end{figure}
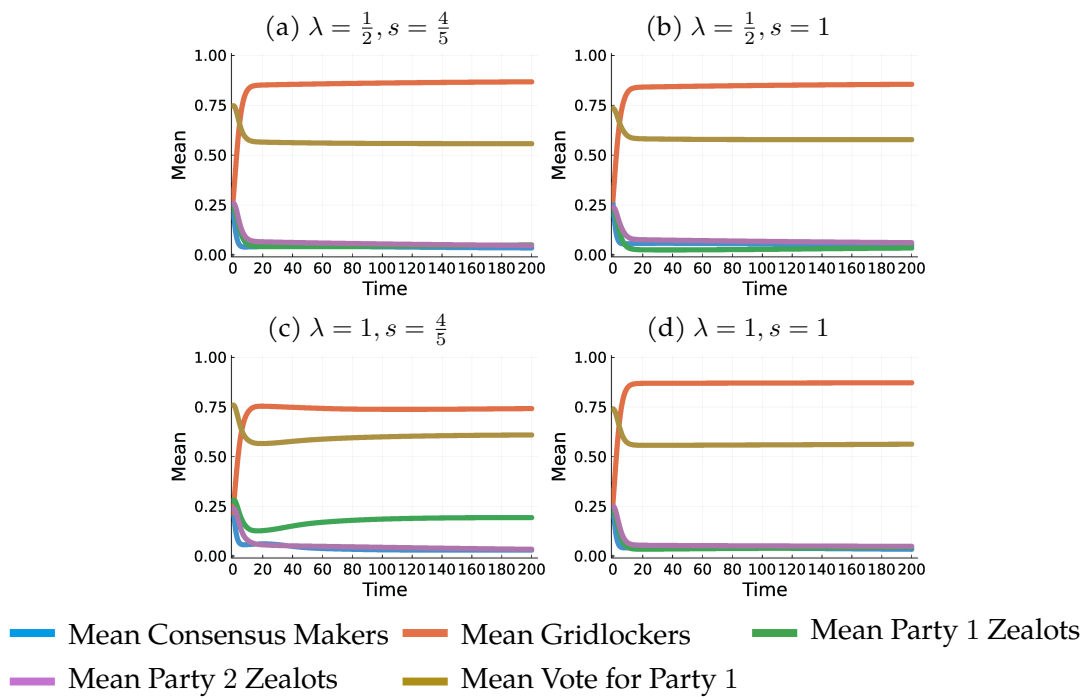

In this section, we consider the spillover effects of the public good. We consider the spillovers rate $s=4/5$, where the focal node and all neighbours equally reap the benefits of the public good neighbour, and $s=1$, where only the neighbours receive the benefit. This latter case could be generated due to taxes the focal node must pay to earn the public good \citep{Uler11}. Further, it results in a public goods game with neighbouring nodes. Though $s=1$ is an extreme case, it helps clarify and illustrate the effects of spillovers. In addition to considering different values of $s$, we also consider $\lambda=1/2$ and $1$ (note the public good, and thus spillovers, are irrelevant for $\lambda=0$).

Figure \ref{fig:TS_spillovers} depicts a representative time series. Gridlockers are prevalent. However, Party $1$ Zealots are approximately $20\%$ of the population when $\lambda=1$ and $s=4/5$. Increasing the spillover rate for $\lambda=1/2$ has a marginal effect, since migrants are still sorting partly based on voting strategy. This preference for like-minded individuals diminishes the incentives from the public goods game to free-ride on public goods funded by neighbouring cities. However, when economic utility is the sole driver of movement, increasing spillovers increases and decreases the densities of Gridlockers and Party $1$ Zealots, respectively.

\begin{figure}[ht!]
\captionsetup[subfigure]{justification=centering}
    \centering
    \begin{subfigure}[]{0.3\columnwidth}
        \caption*{Population Colorbar}
        \includegraphics[width=\textwidth]{figs_png/Population_Colorbar.png}
            \vspace{2pt}
    {\small 0 \hfill 1}
    \end{subfigure} \hspace{2cm}
    \begin{subfigure}[]{0.3\columnwidth}
        \caption*{Party Colorbar}
        \includegraphics[width=\textwidth]{figs_png/PartyColorbar.png}
            \vspace{2pt}
    {\small 0 \hfill 1}
    \end{subfigure}\\
    \begin{subfigure}[]{0.15\columnwidth}
        \caption{$c,\lambda=\tfrac{1}{2},s=\tfrac{4}{5}$}
        \includegraphics[width=\textwidth]{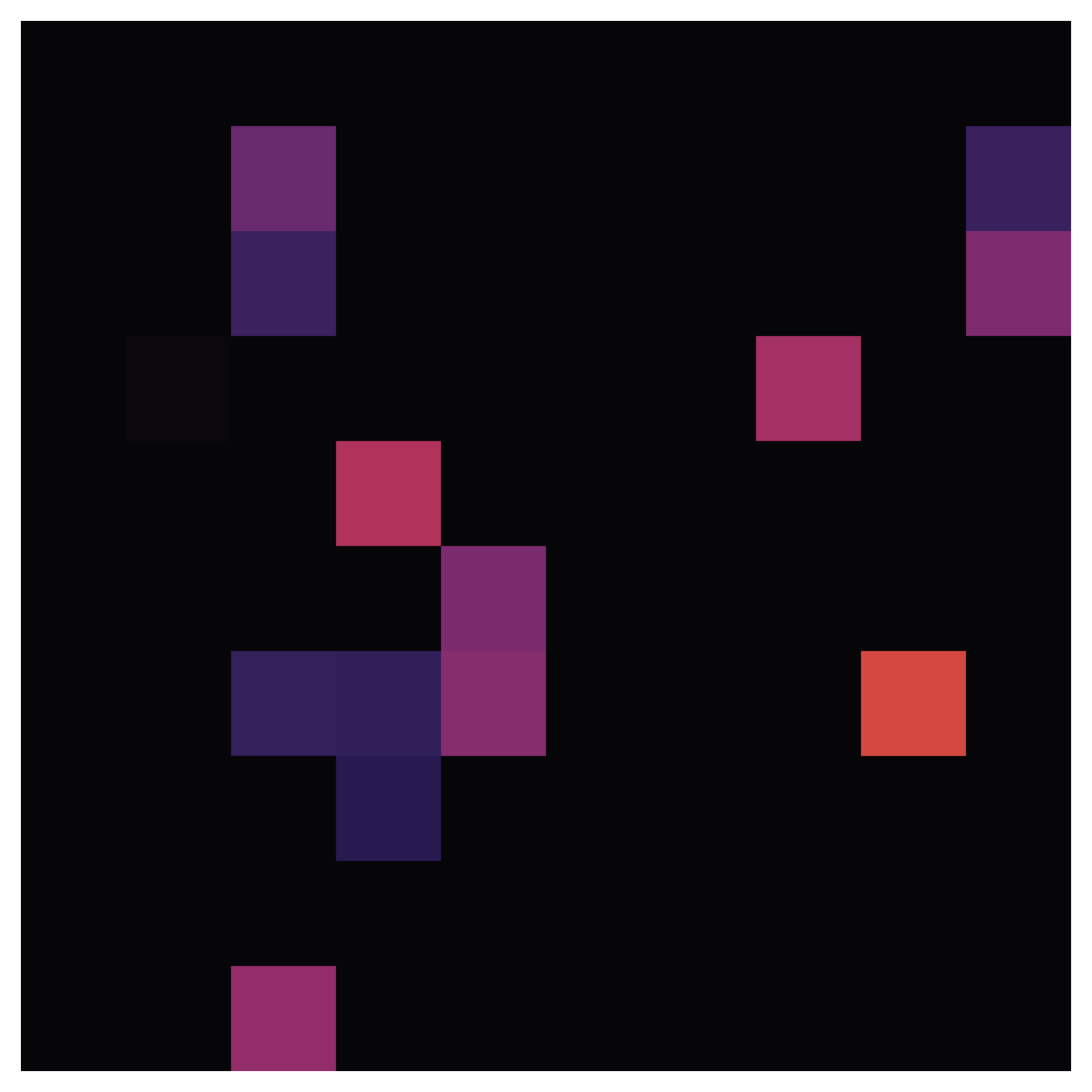}
    \end{subfigure}
    \begin{subfigure}[]{0.15\columnwidth}
        \caption{$g,\lambda=\tfrac{1}{2},s=\tfrac{4}{5}$}
        \includegraphics[width=\textwidth]{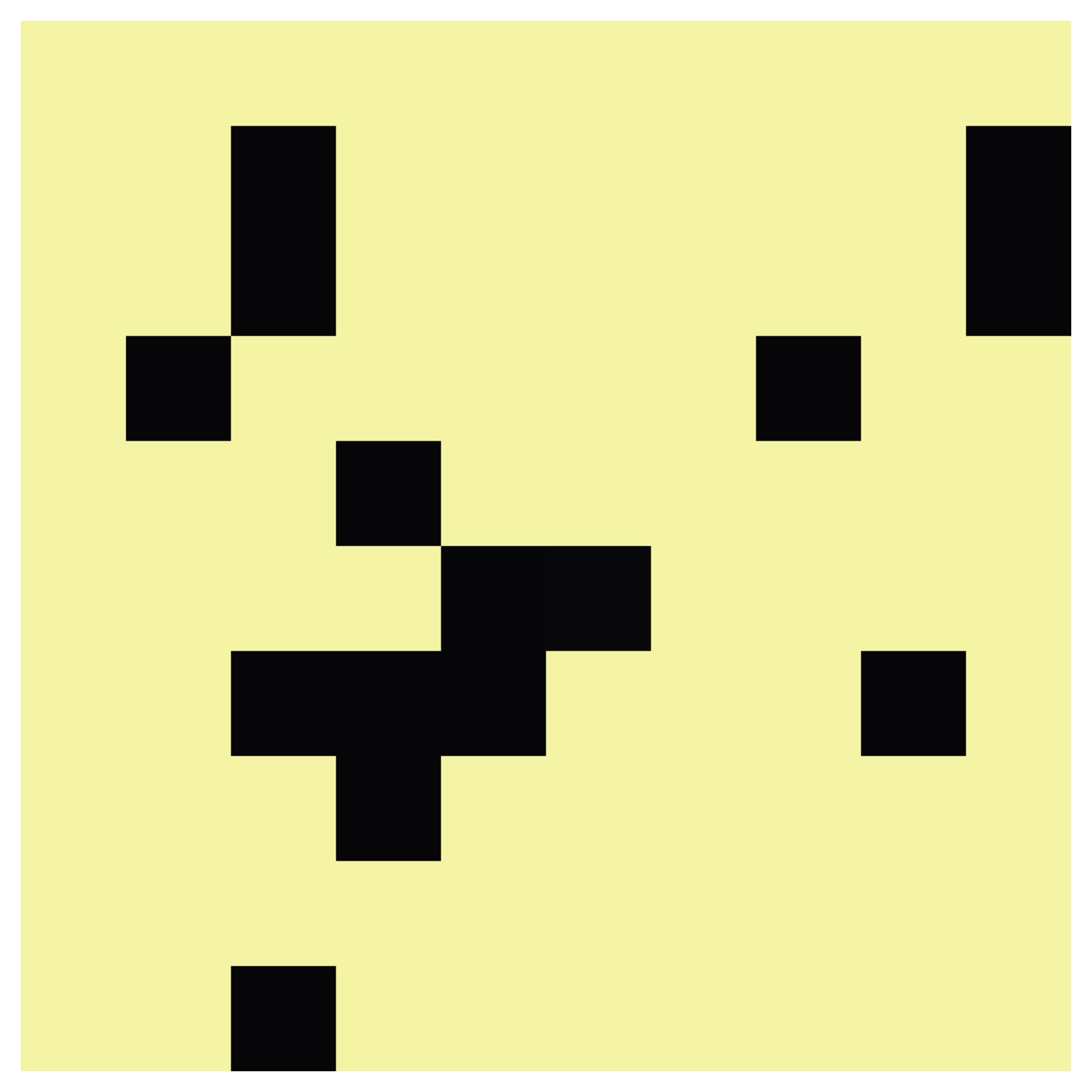}
    \end{subfigure}
        \begin{subfigure}[]{0.15\columnwidth}
        \caption{$z_1,\lambda=\tfrac{1}{2},s=\tfrac{4}{5}$}
        \includegraphics[width=\textwidth]{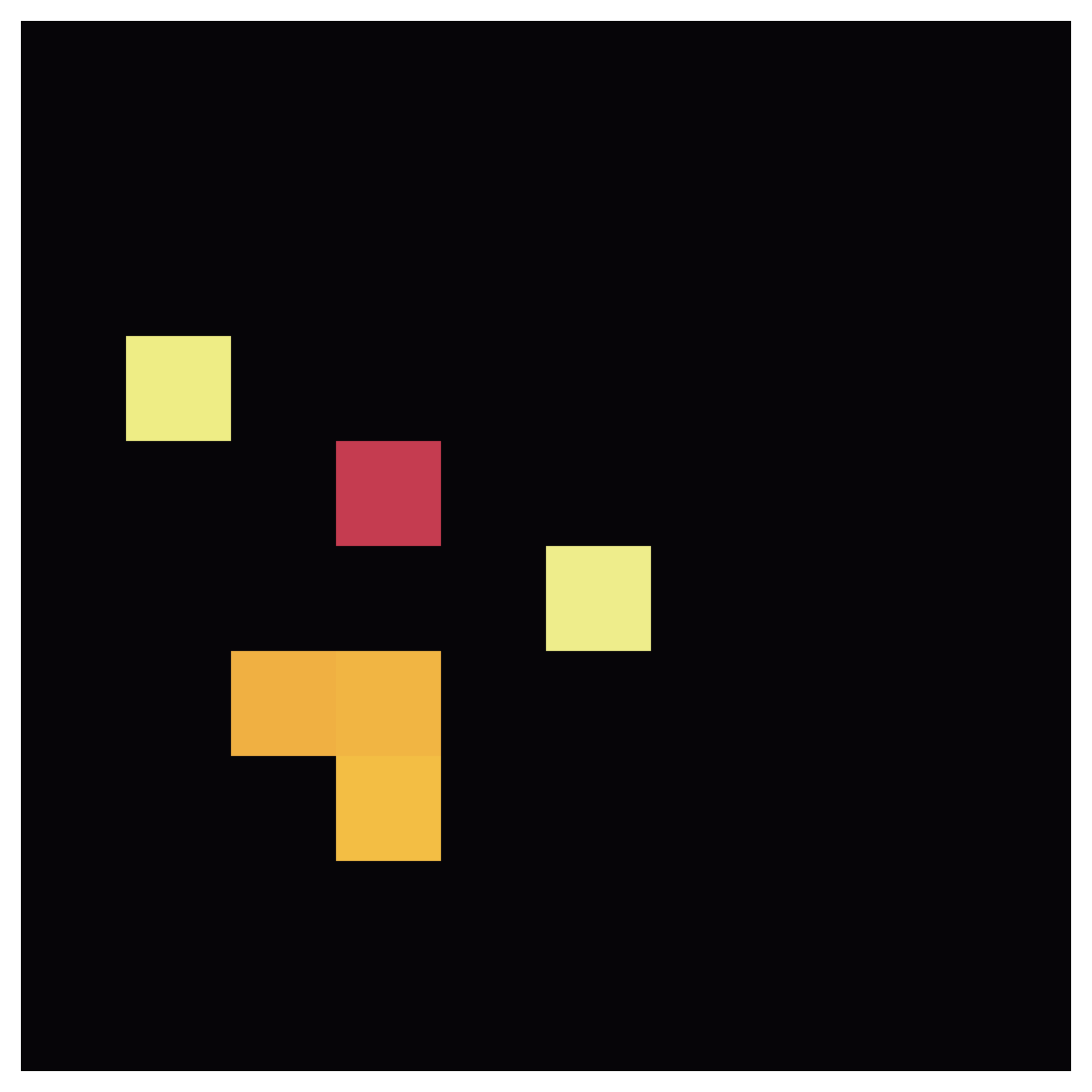}
    \end{subfigure}
        \begin{subfigure}[]{0.15\columnwidth}
        \caption{$z_2,\lambda=\tfrac{1}{2},s=\tfrac{4}{5}$}
        \includegraphics[width=\textwidth]{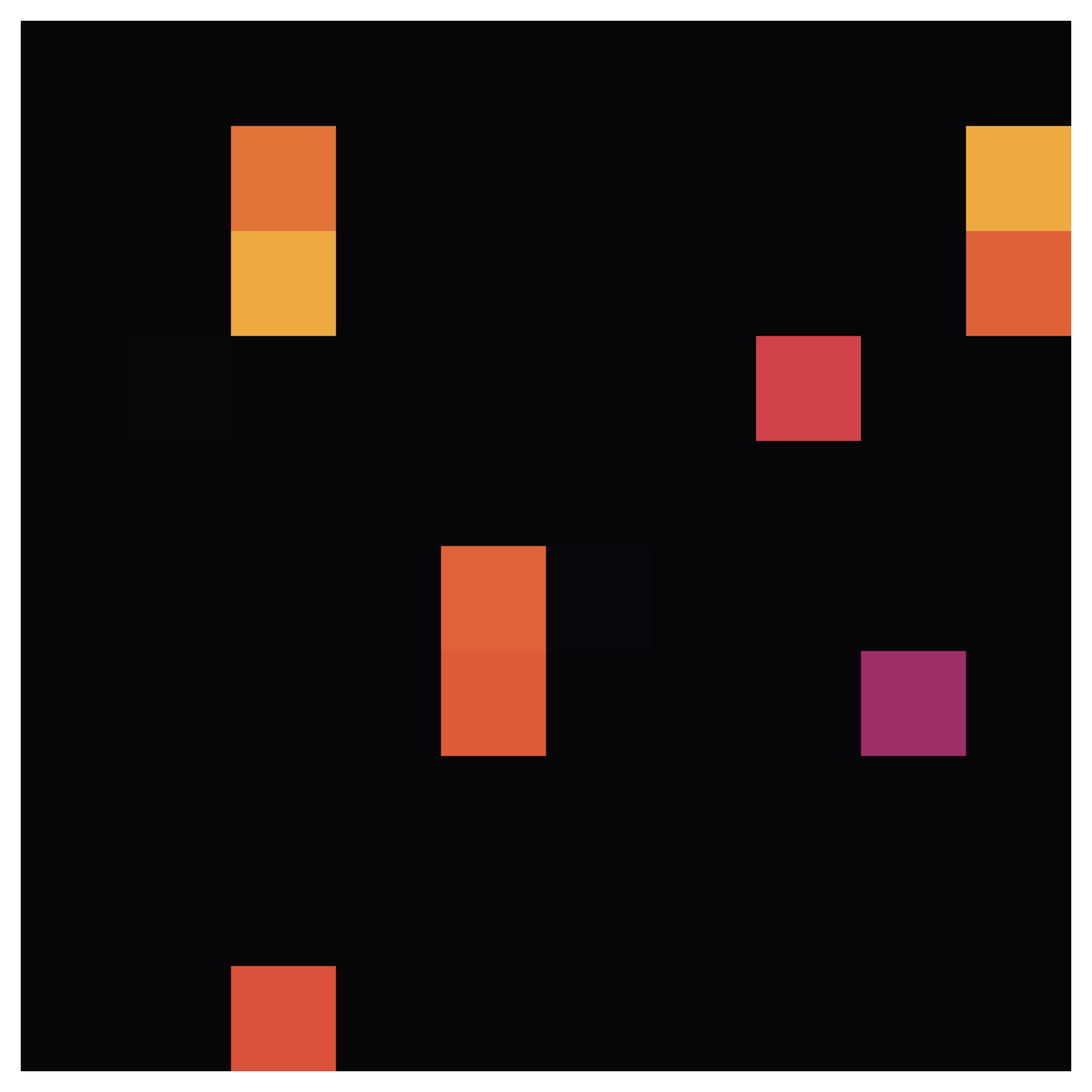}
    \end{subfigure}
         \begin{subfigure}[]{0.15\columnwidth}
         \caption{$\tilde{p},\lambda=\tfrac{1}{2},s=\tfrac{4}{5}$}
         \includegraphics[width=\textwidth]{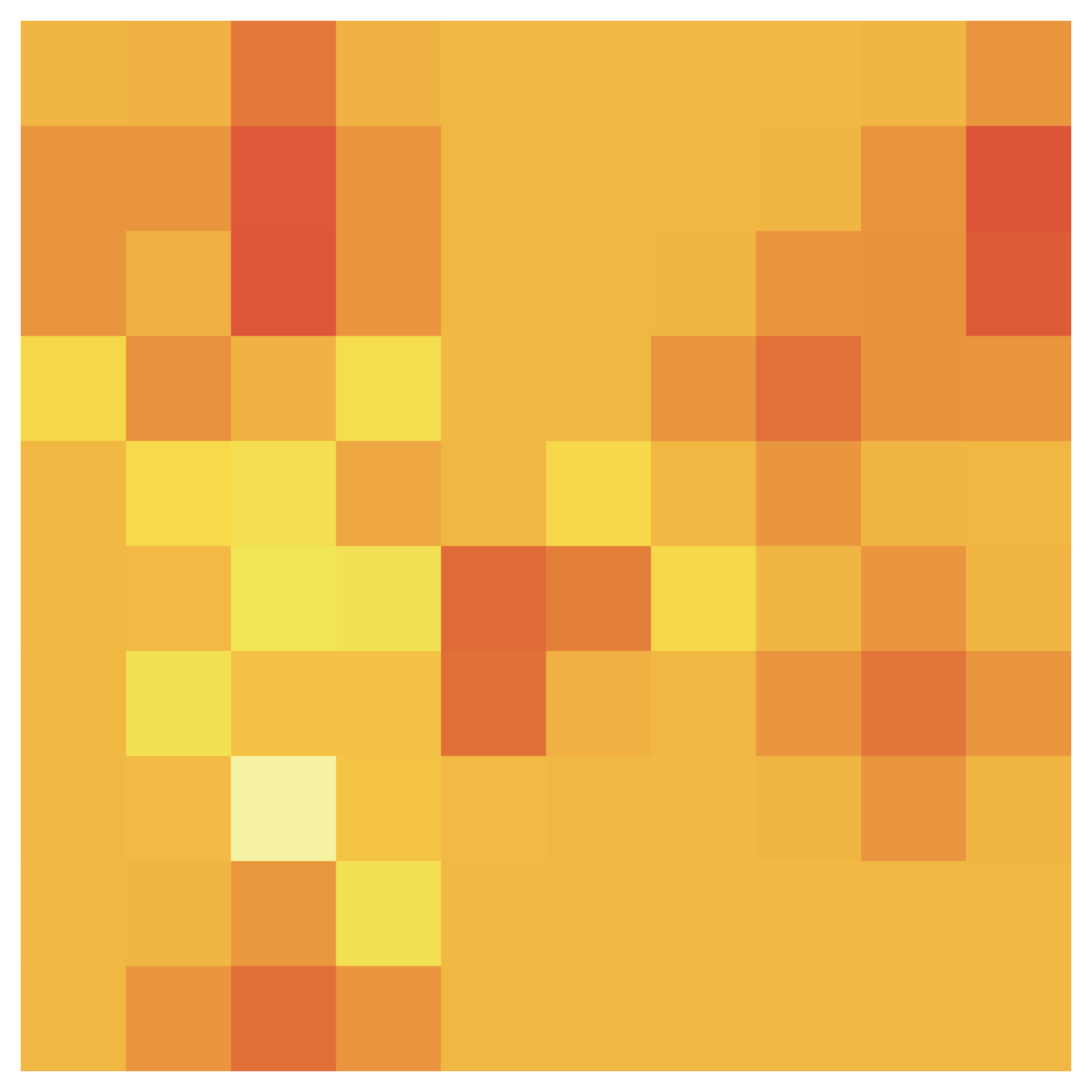}
     \end{subfigure}
        \begin{subfigure}[]{0.15\columnwidth}
        \caption{$v,\lambda=\tfrac{1}{2},s=\tfrac{4}{5}$}
        \includegraphics[width=\textwidth]{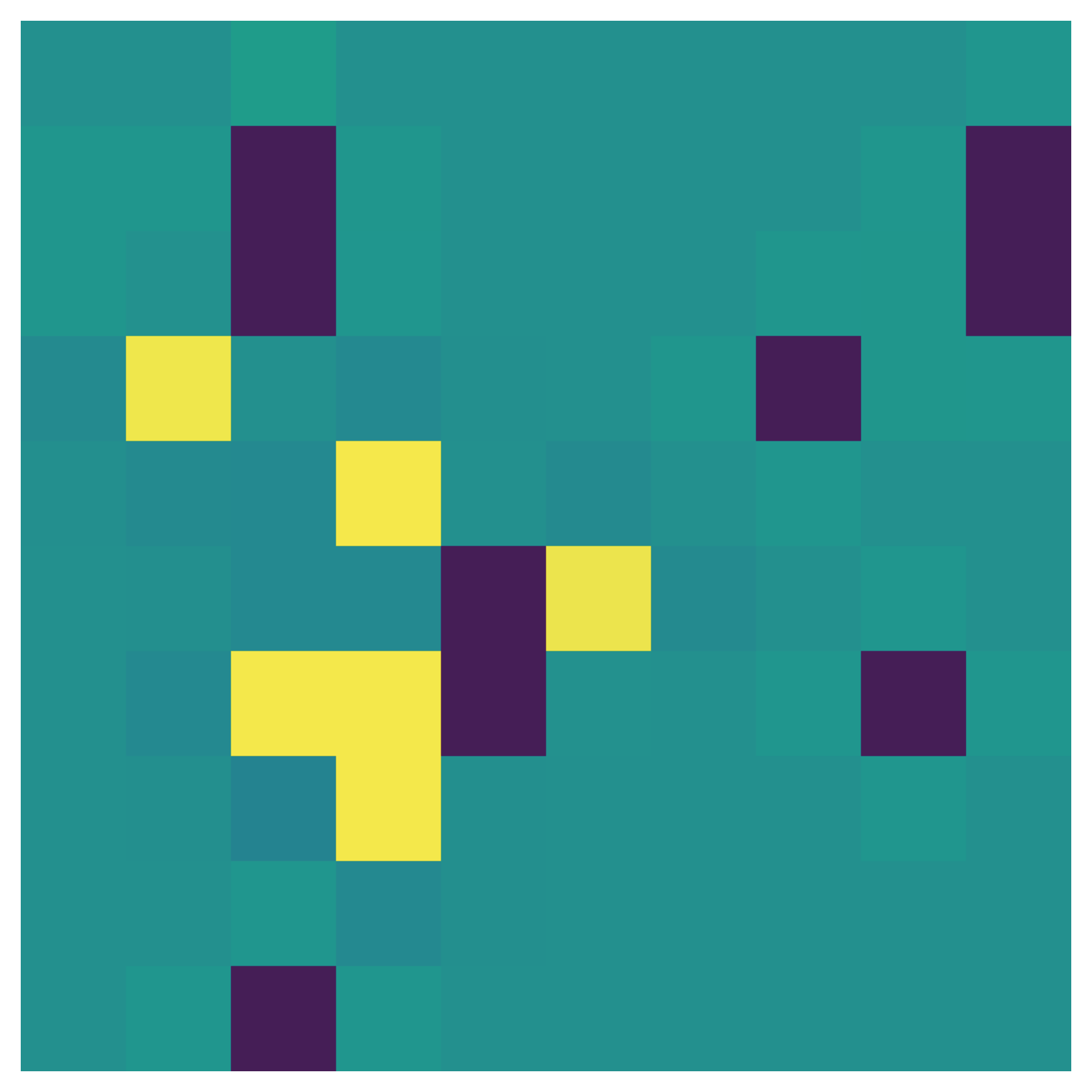}
    \end{subfigure}\\
     \begin{subfigure}[]{0.15\columnwidth}
        \caption{$c,\lambda=\tfrac{1}{2},s=1$}
        \includegraphics[width=\textwidth]{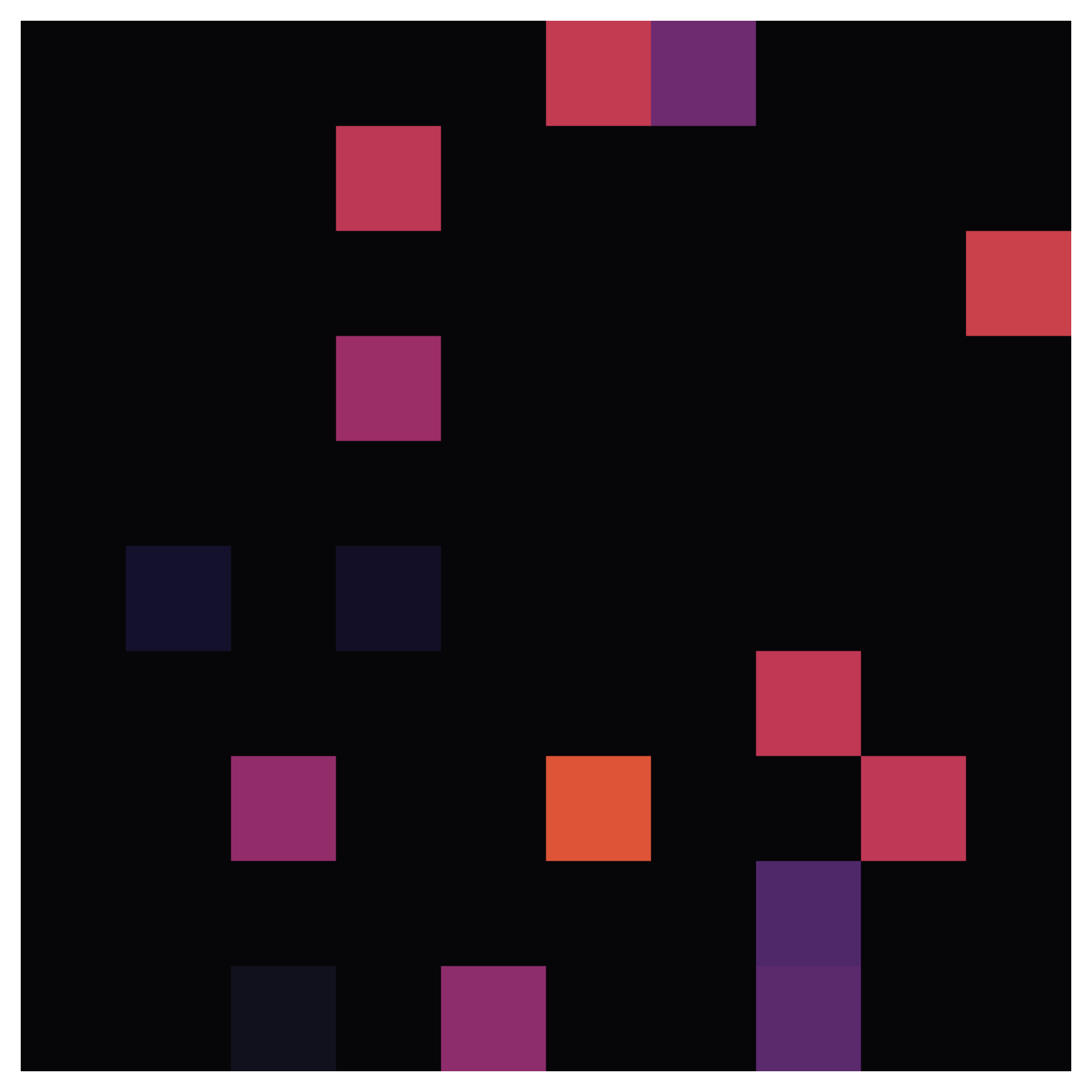}
    \end{subfigure}
    \begin{subfigure}[]{0.15\columnwidth}
        \caption{$g,\lambda=\tfrac{1}{2},s=1$}
        \includegraphics[width=\textwidth]{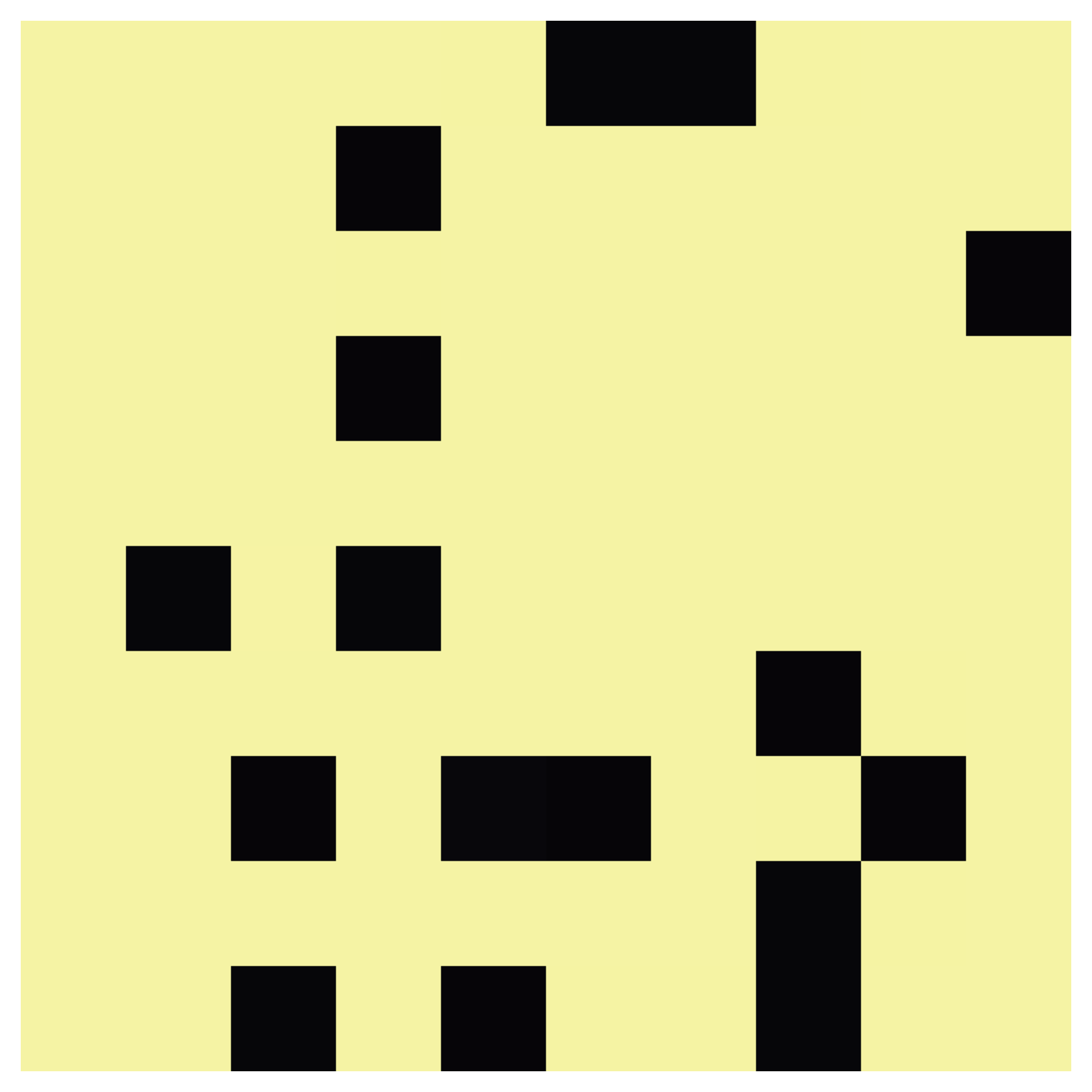}
    \end{subfigure}
        \begin{subfigure}[]{0.15\columnwidth}
        \caption{$z_1,\lambda=\tfrac{1}{2},s=1$}
        \includegraphics[width=\textwidth]{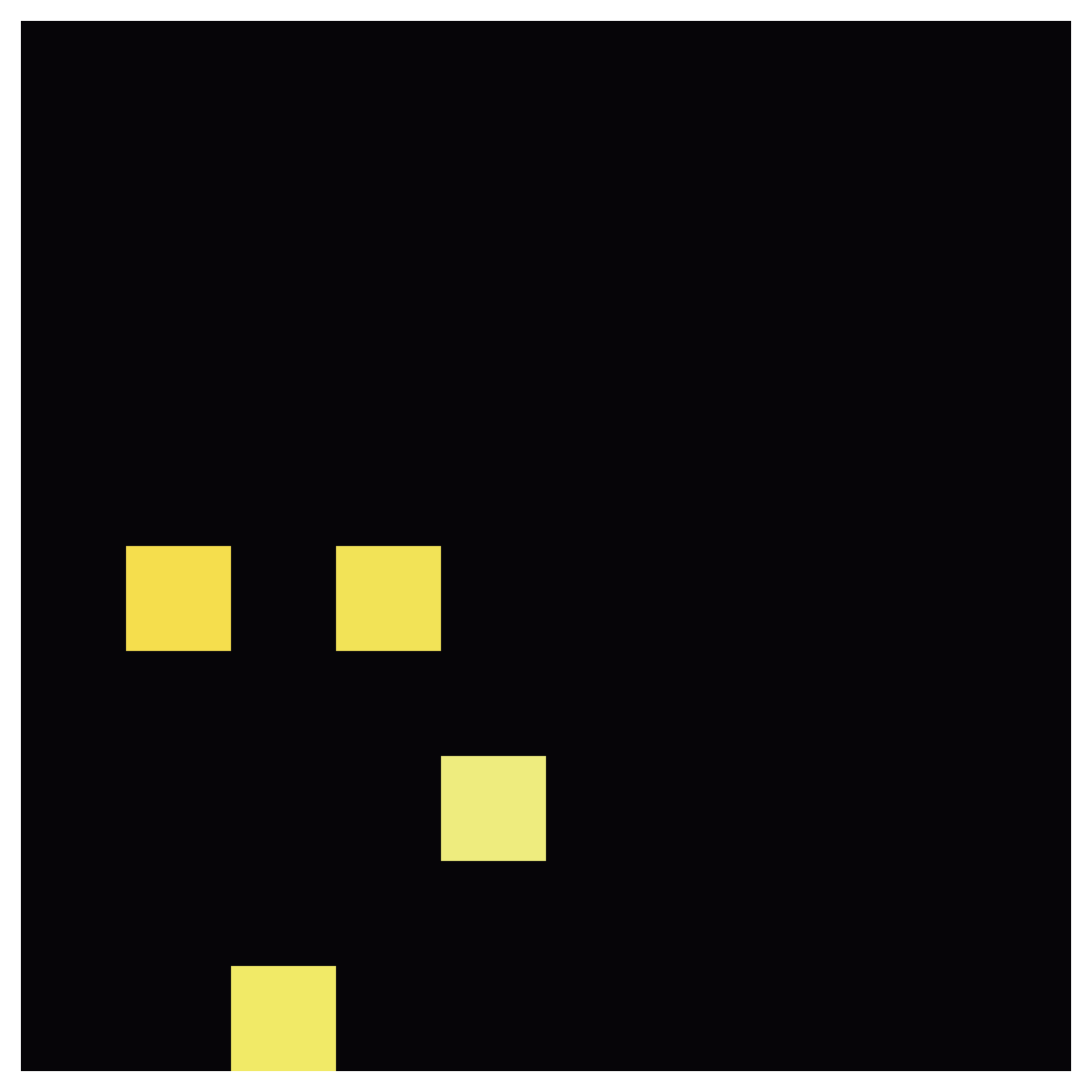}
    \end{subfigure}
        \begin{subfigure}[]{0.15\columnwidth}
        \caption{$z_2,\lambda=\tfrac{1}{2},s=1$}
        \includegraphics[width=\textwidth]{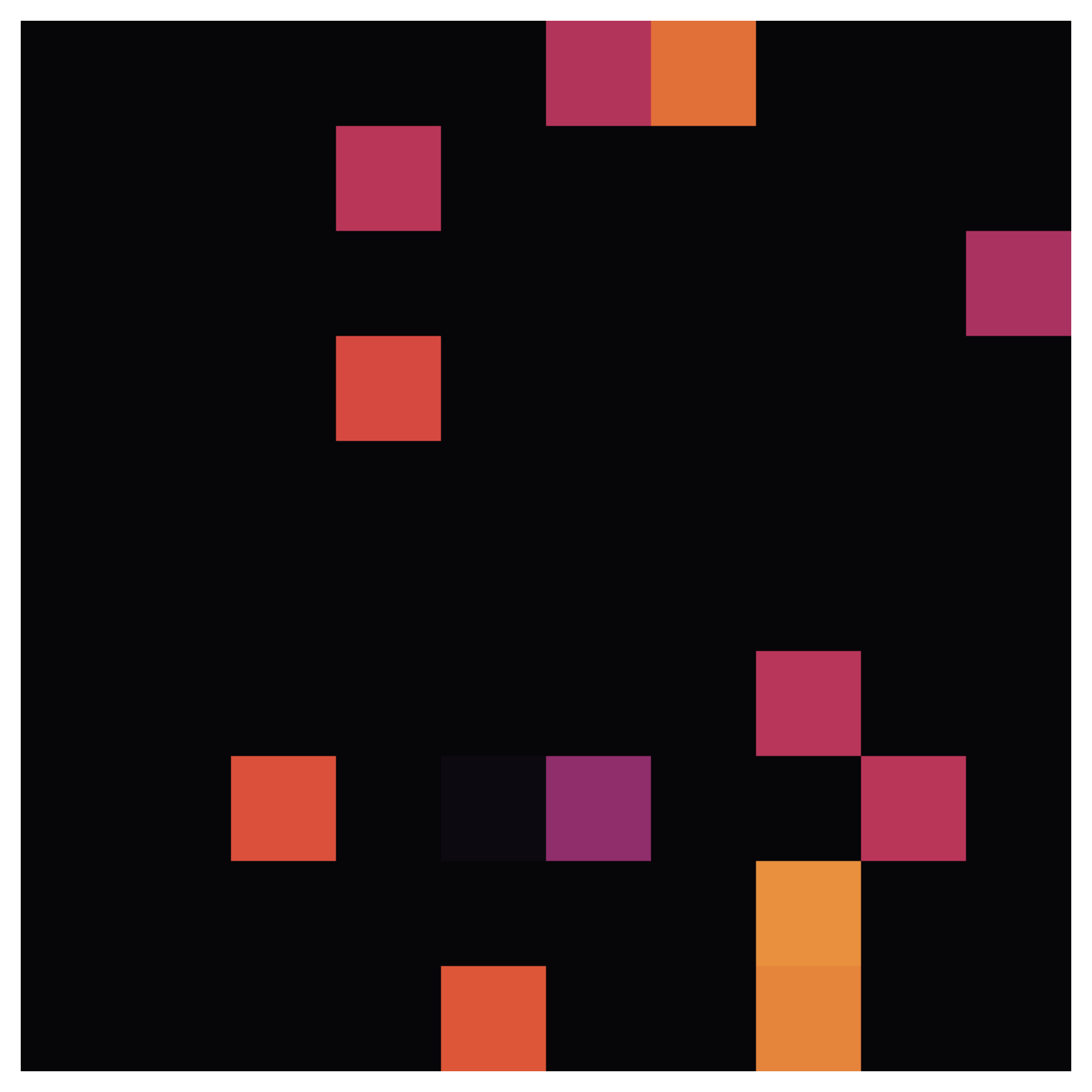}
    \end{subfigure}
         \begin{subfigure}[]{0.15\columnwidth}
         \caption{$\tilde{p},\lambda=\tfrac{1}{2},s=1$}
         \includegraphics[width=\textwidth]{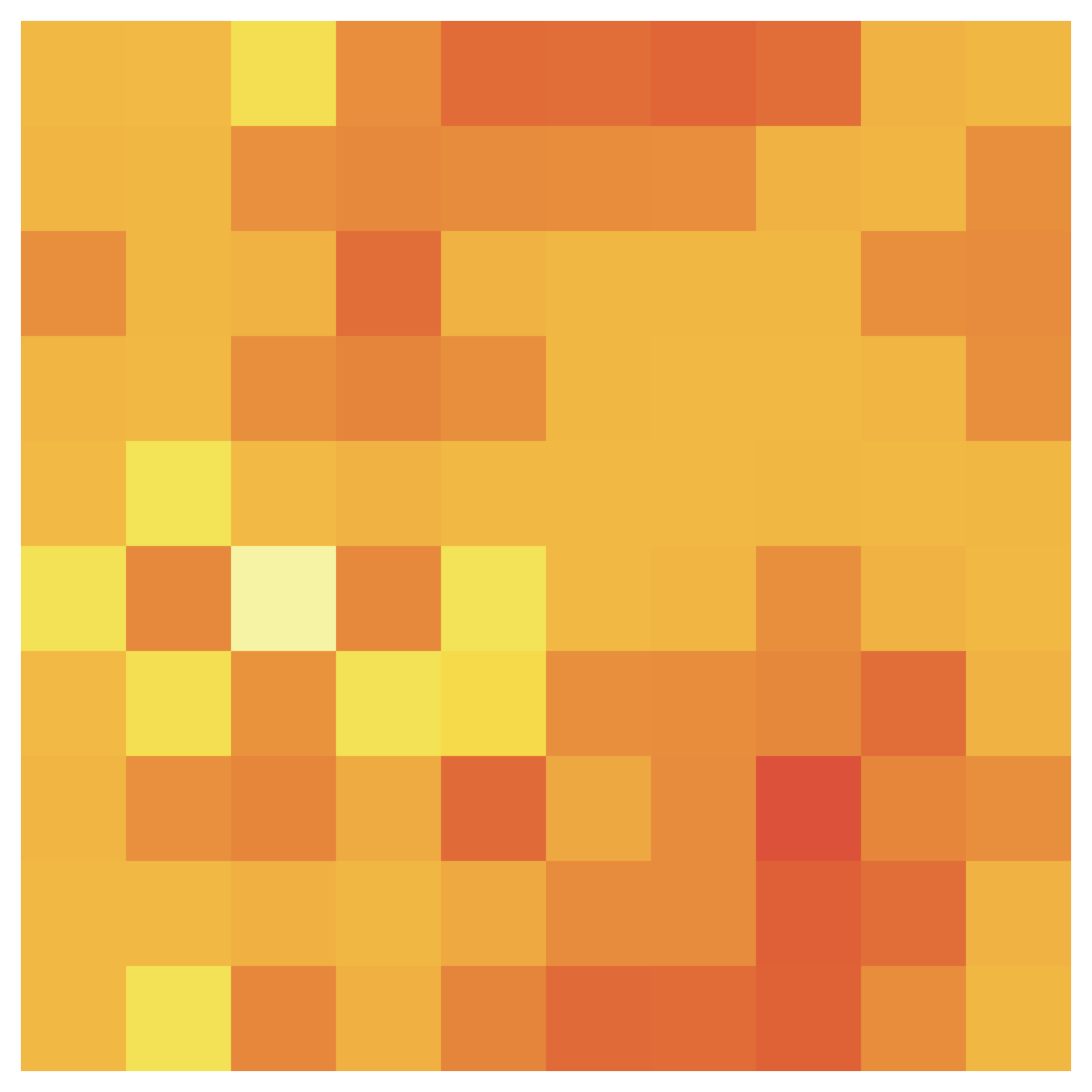}
     \end{subfigure}
        \begin{subfigure}[]{0.15\columnwidth}
        \caption{$v,\lambda=\tfrac{1}{2},s=1$}
        \includegraphics[width=\textwidth]{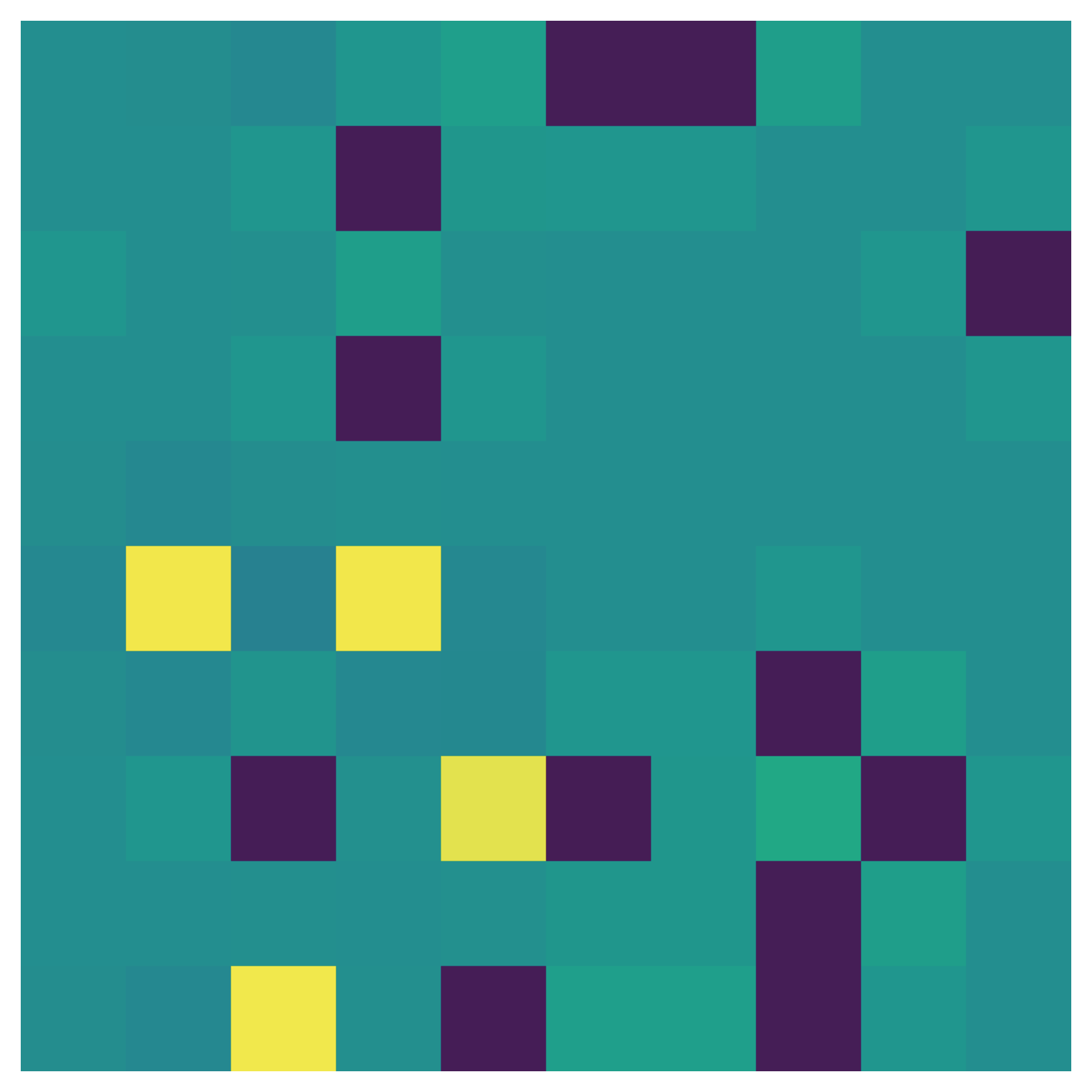}
    \end{subfigure}\\
        \begin{subfigure}[]{0.15\columnwidth}
        \caption{$c,\lambda=1,s=\tfrac{4}{5}$}
        \includegraphics[width=\textwidth]{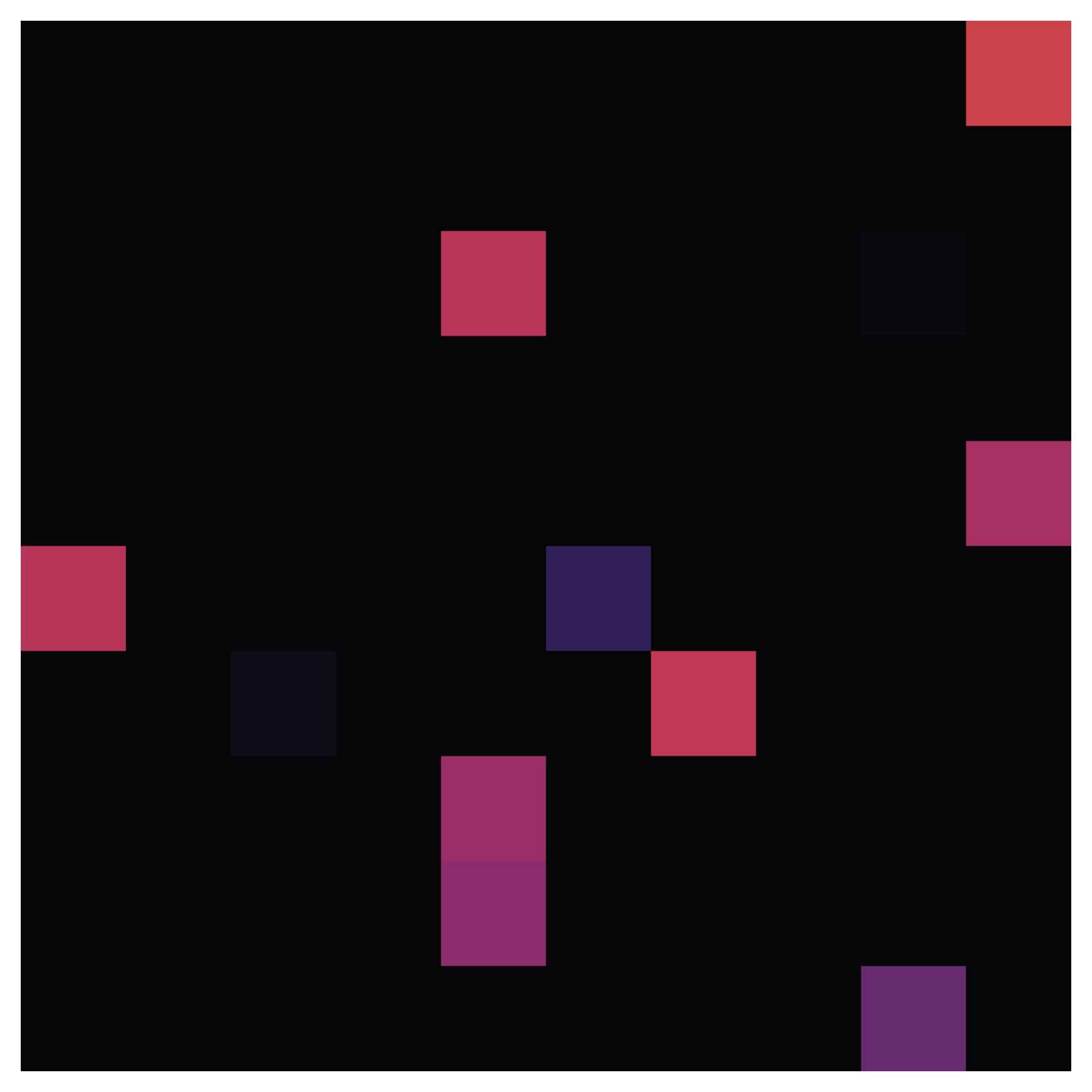}
    \end{subfigure}
    \begin{subfigure}[]{0.15\columnwidth}
        \caption{$g,\lambda=1,s=\tfrac{4}{5}$}
        \includegraphics[width=\textwidth]{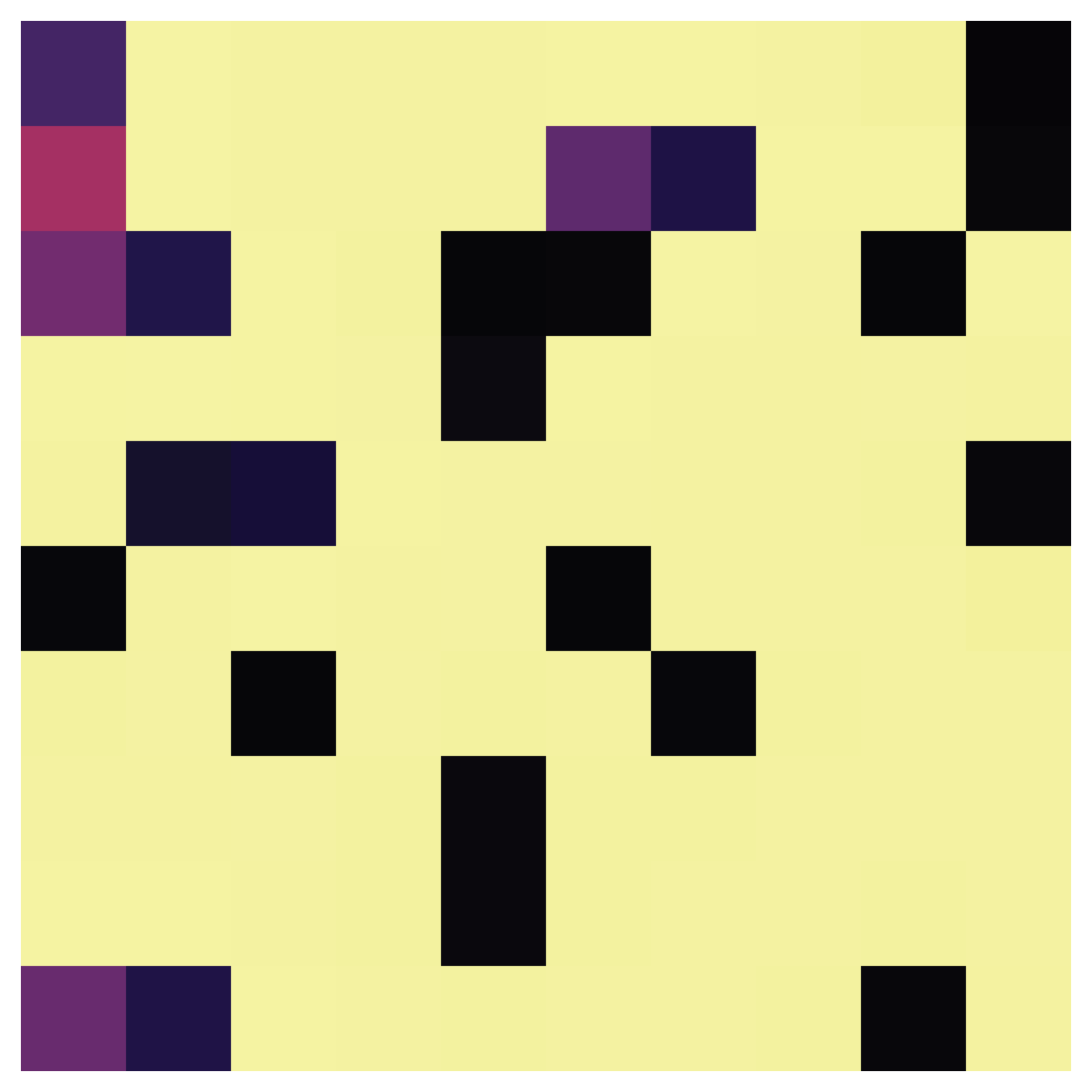}
    \end{subfigure}
        \begin{subfigure}[]{0.15\columnwidth}
        \caption{$z_1,\lambda=1,s=\tfrac{4}{5}$}
        \includegraphics[width=\textwidth]{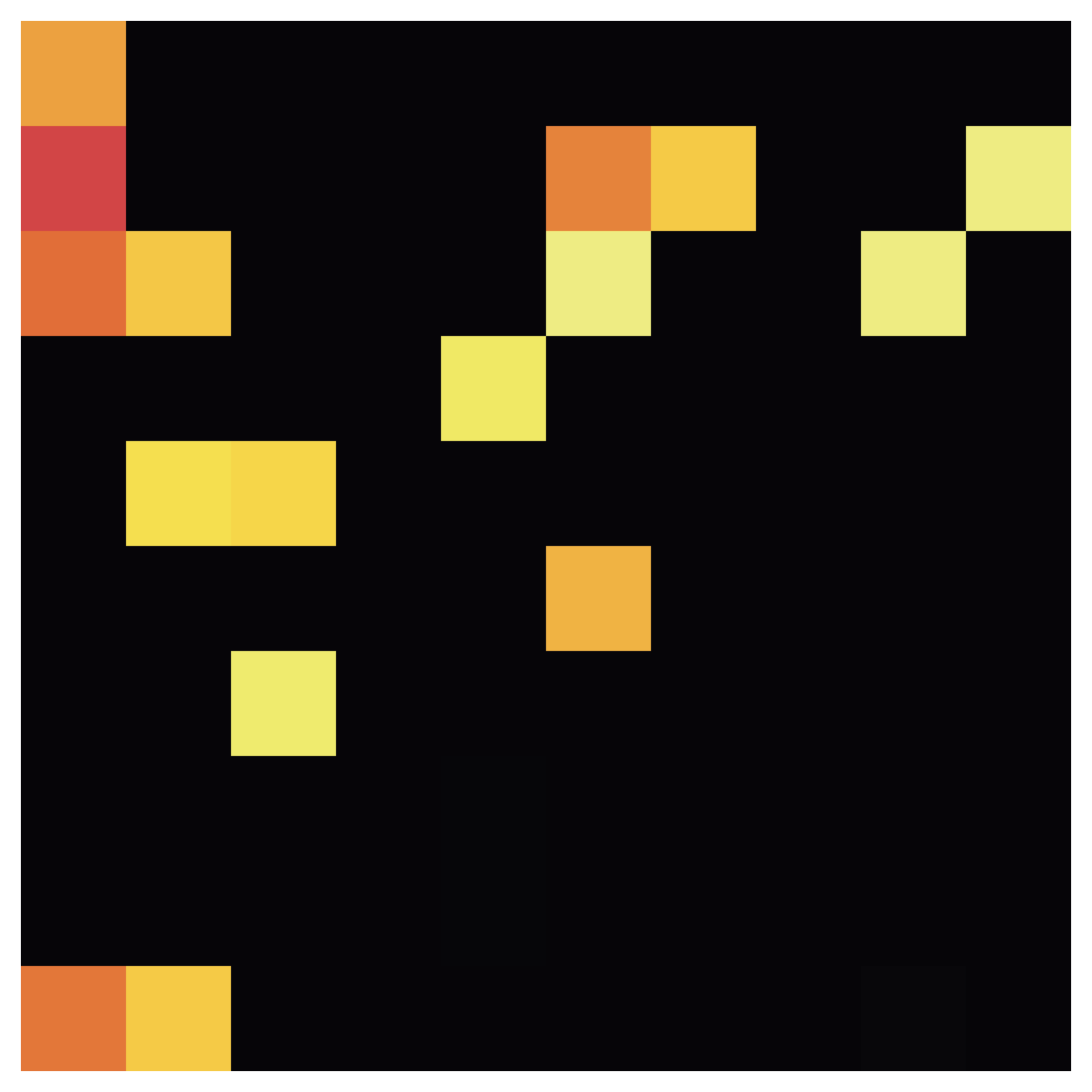}
    \end{subfigure}
        \begin{subfigure}[]{0.15\columnwidth}
         \caption{$z_2,\lambda=1,s=\tfrac{4}{5}$}
        \includegraphics[width=\textwidth]{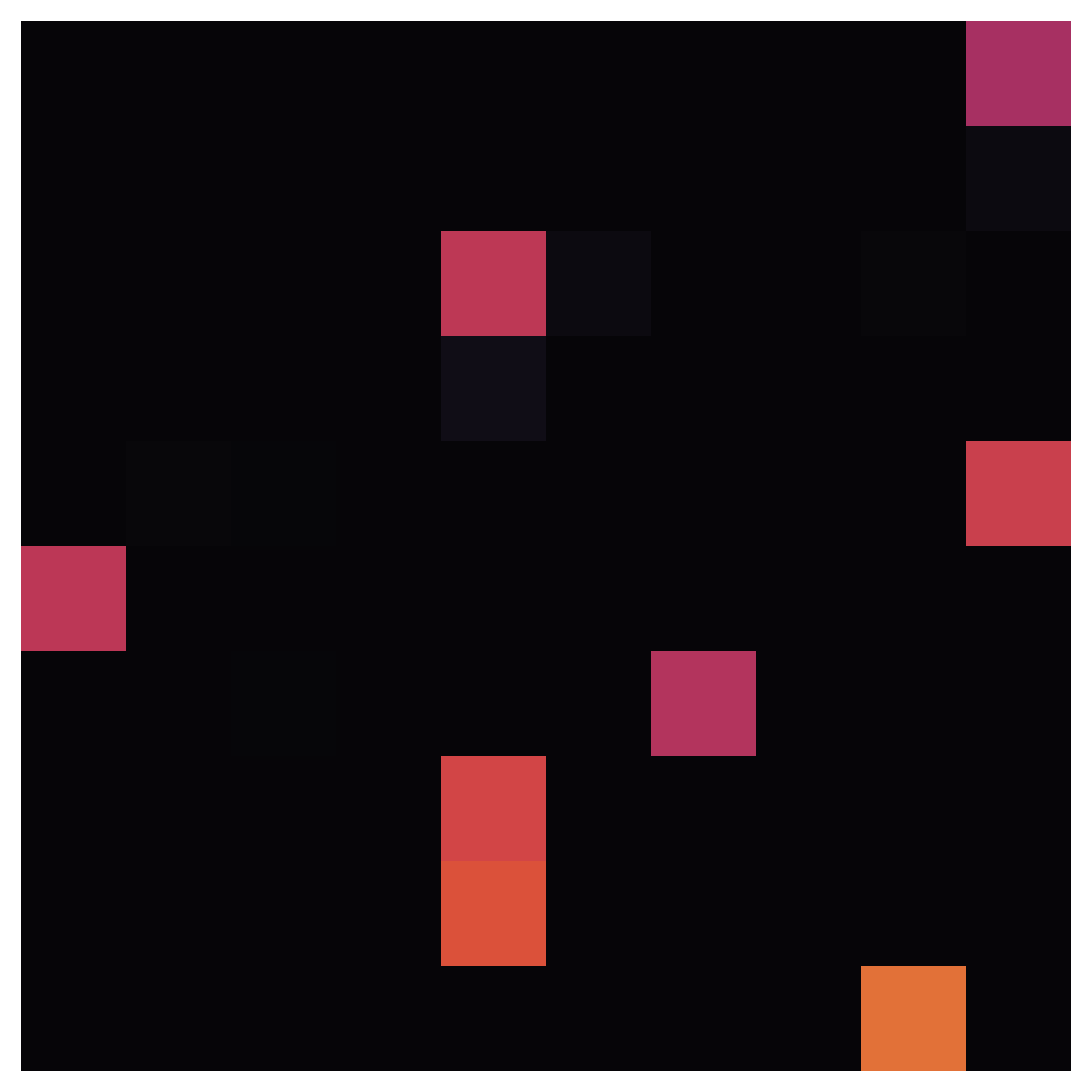}
    \end{subfigure}
         \begin{subfigure}[]{0.15\columnwidth}
         \caption{$\tilde{p},\lambda=1,s=\tfrac{4}{5}$}
         \includegraphics[width=\textwidth]{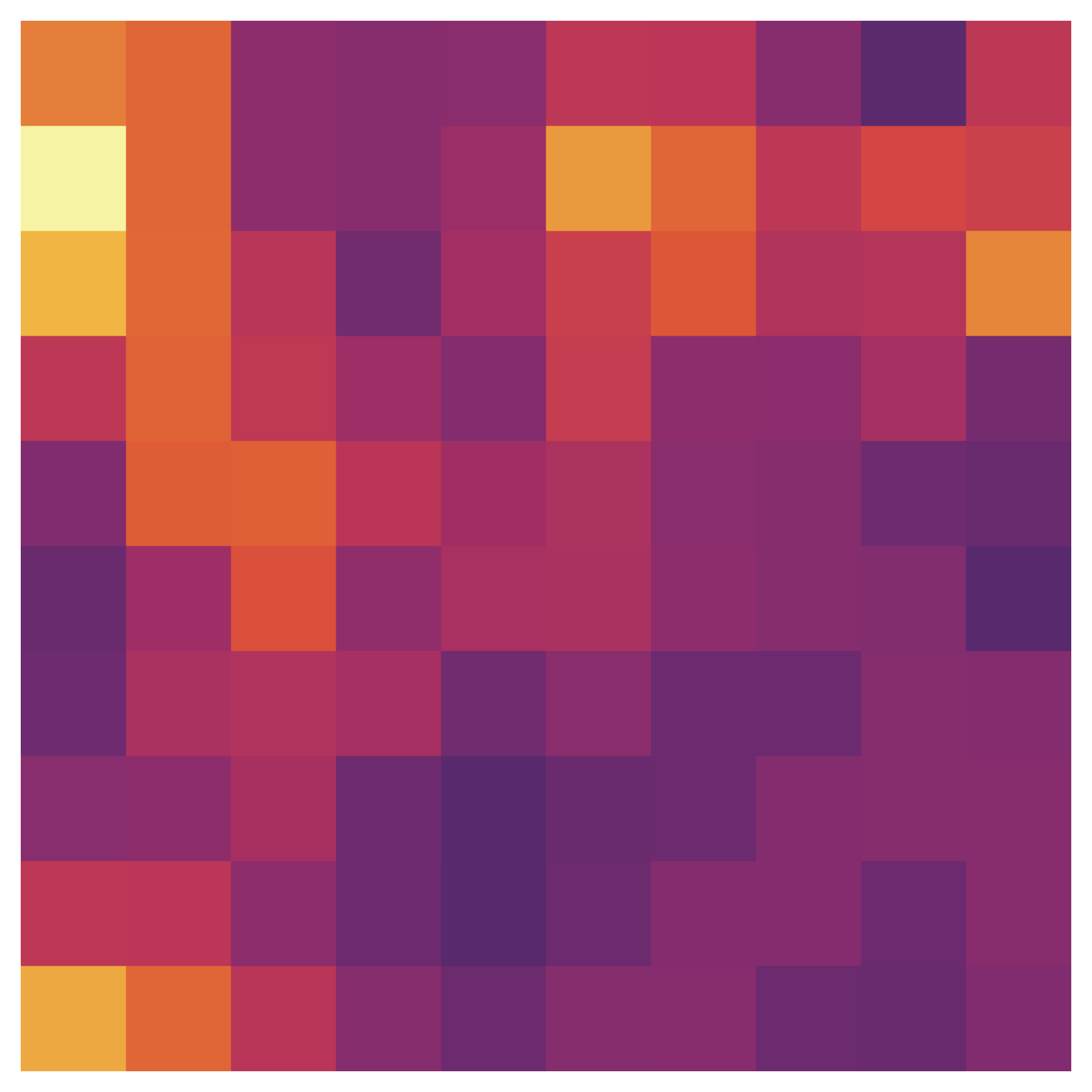}
     \end{subfigure}
        \begin{subfigure}[]{0.15\columnwidth}
        \caption{$v,\lambda=1,s=\tfrac{4}{5}$}
        \includegraphics[width=\textwidth]{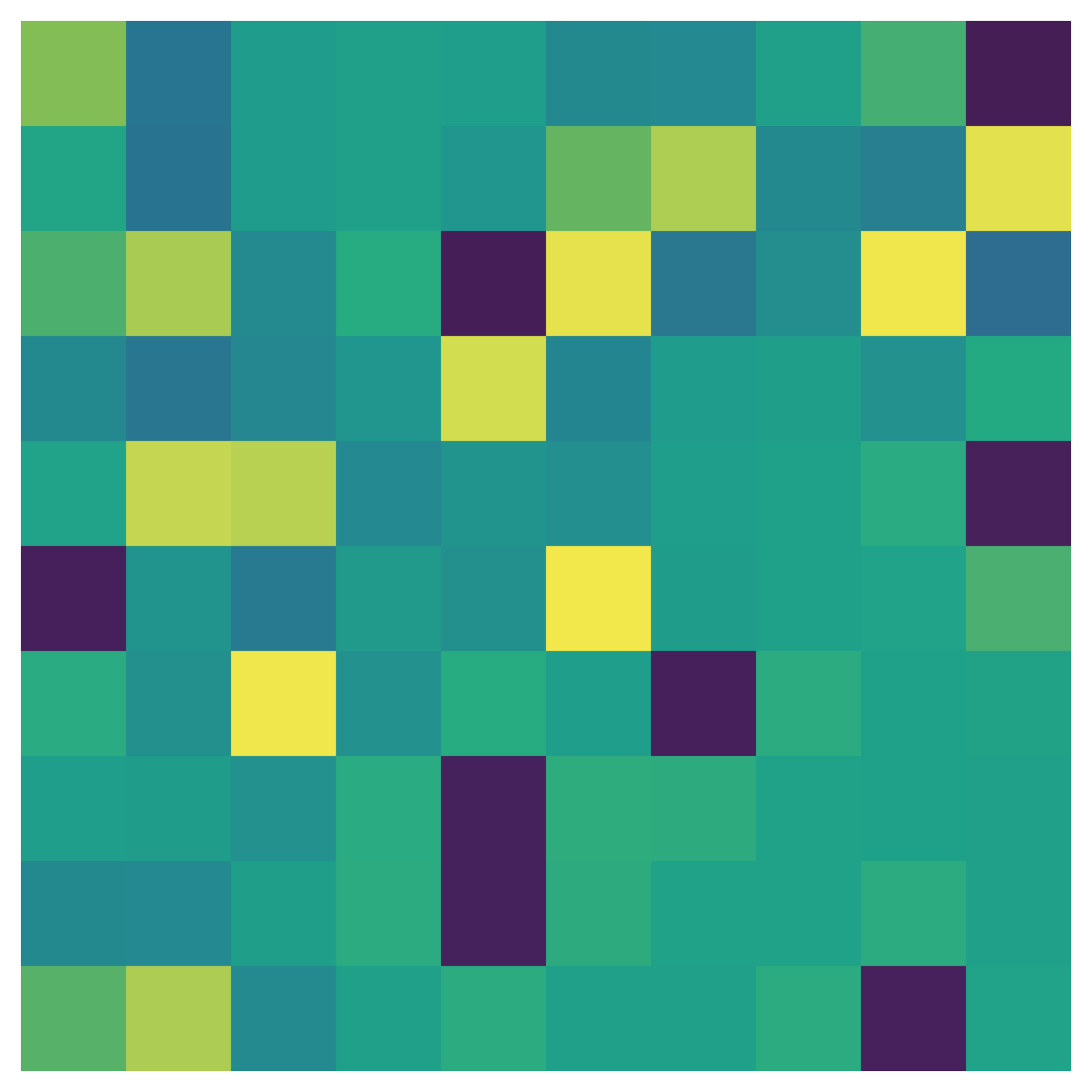}
    \end{subfigure}\\
     \begin{subfigure}[]{0.15\columnwidth}
        \caption{$c,\lambda=1,s=1$}
        \includegraphics[width=\textwidth]{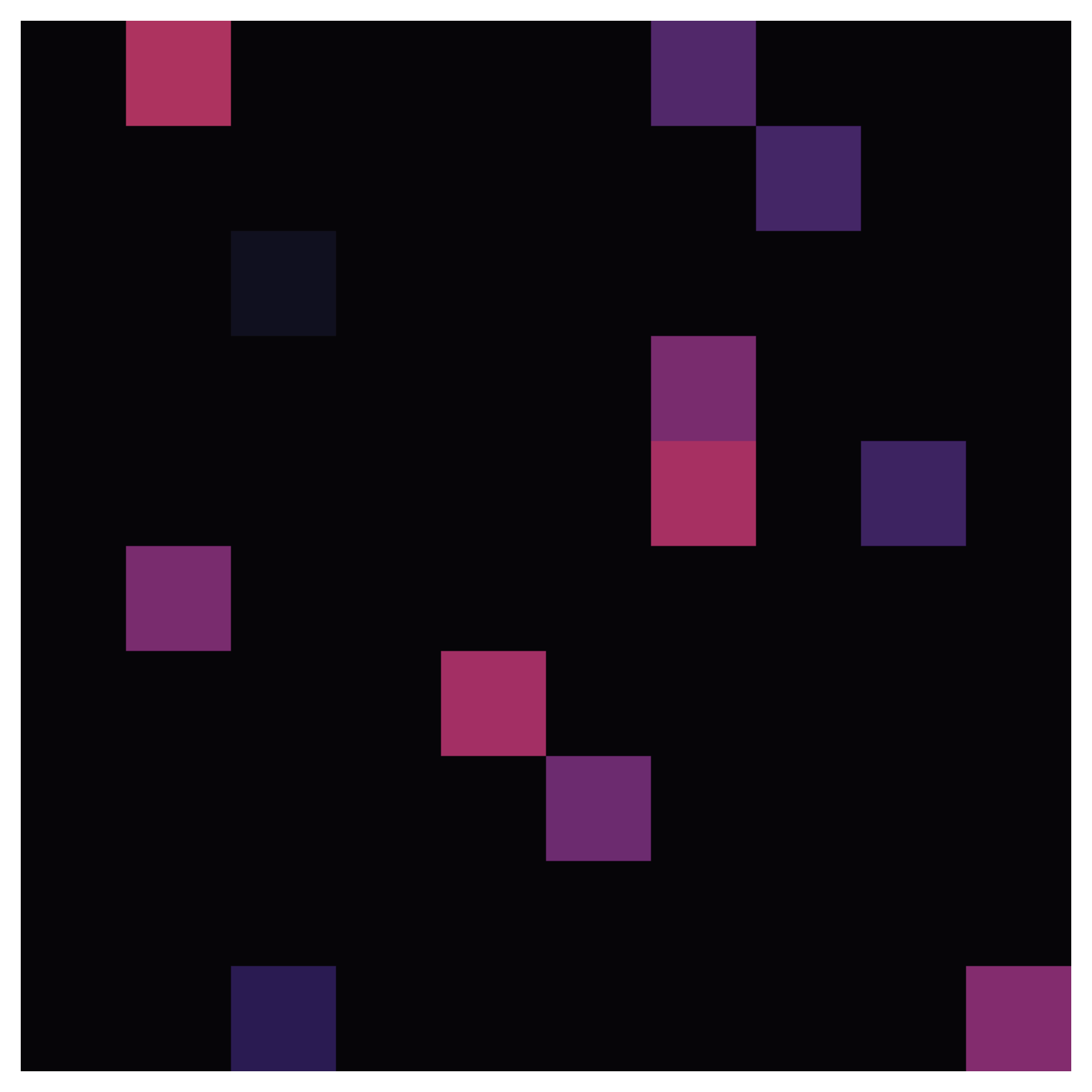}
    \end{subfigure}
    \begin{subfigure}[]{0.15\columnwidth}
        \caption{$g,\lambda=1,s=1$}
        \includegraphics[width=\textwidth]{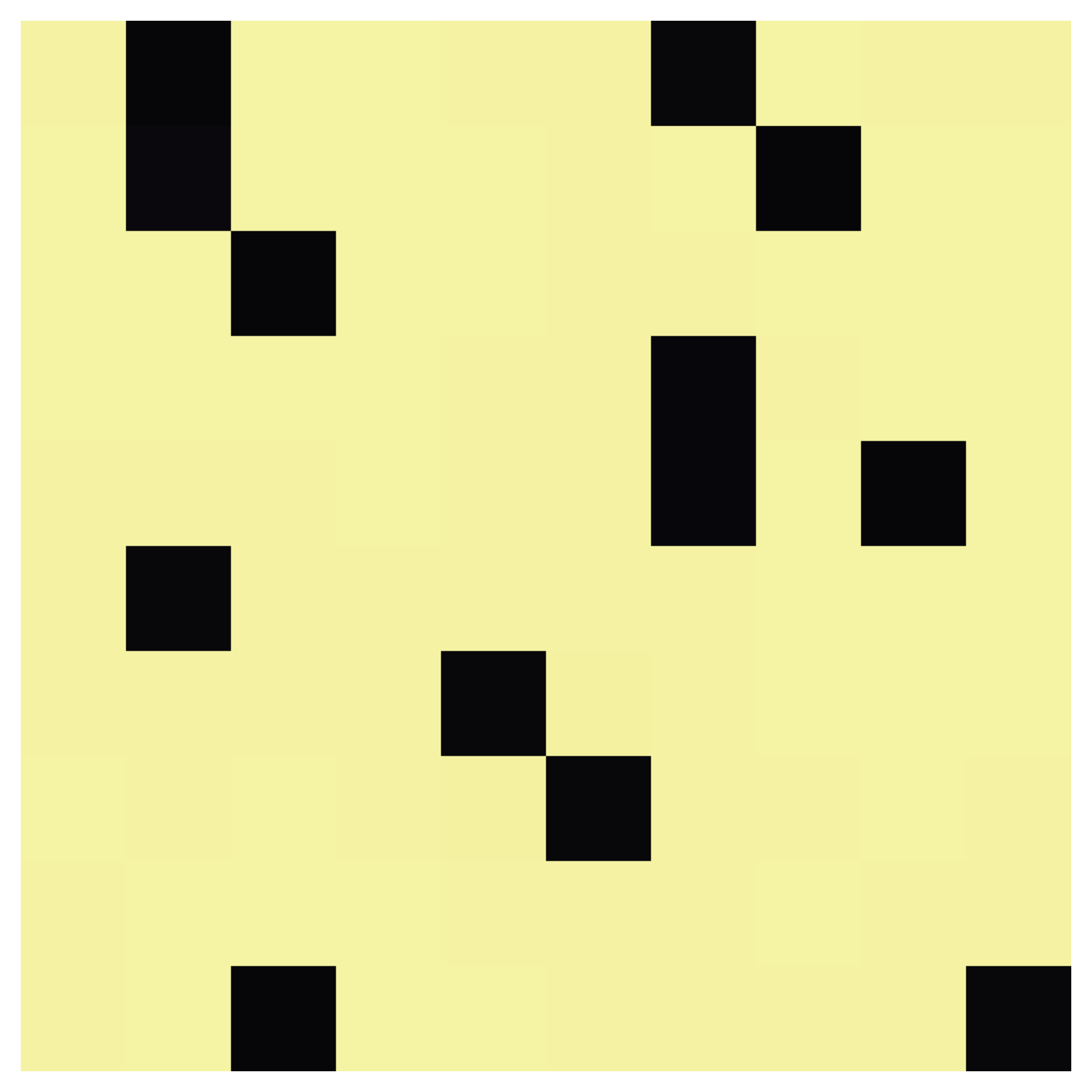}
    \end{subfigure}
        \begin{subfigure}[]{0.15\columnwidth}
        \caption{$z_1,\lambda=1,s=1$}
        \includegraphics[width=\textwidth]{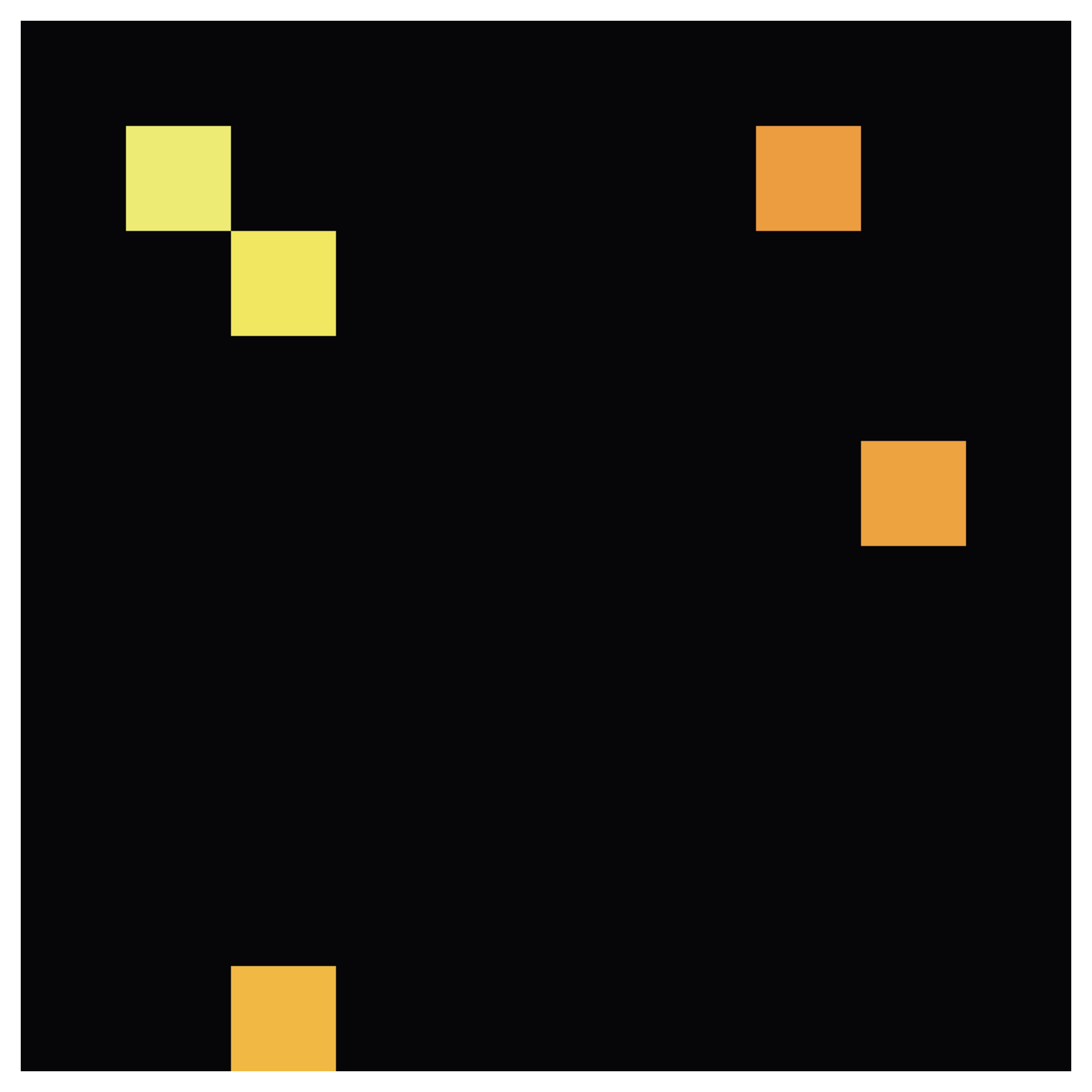}
    \end{subfigure}
        \begin{subfigure}[]{0.15\columnwidth}
        \caption{$z_2,\lambda=1,s=1$}
        \includegraphics[width=\textwidth]{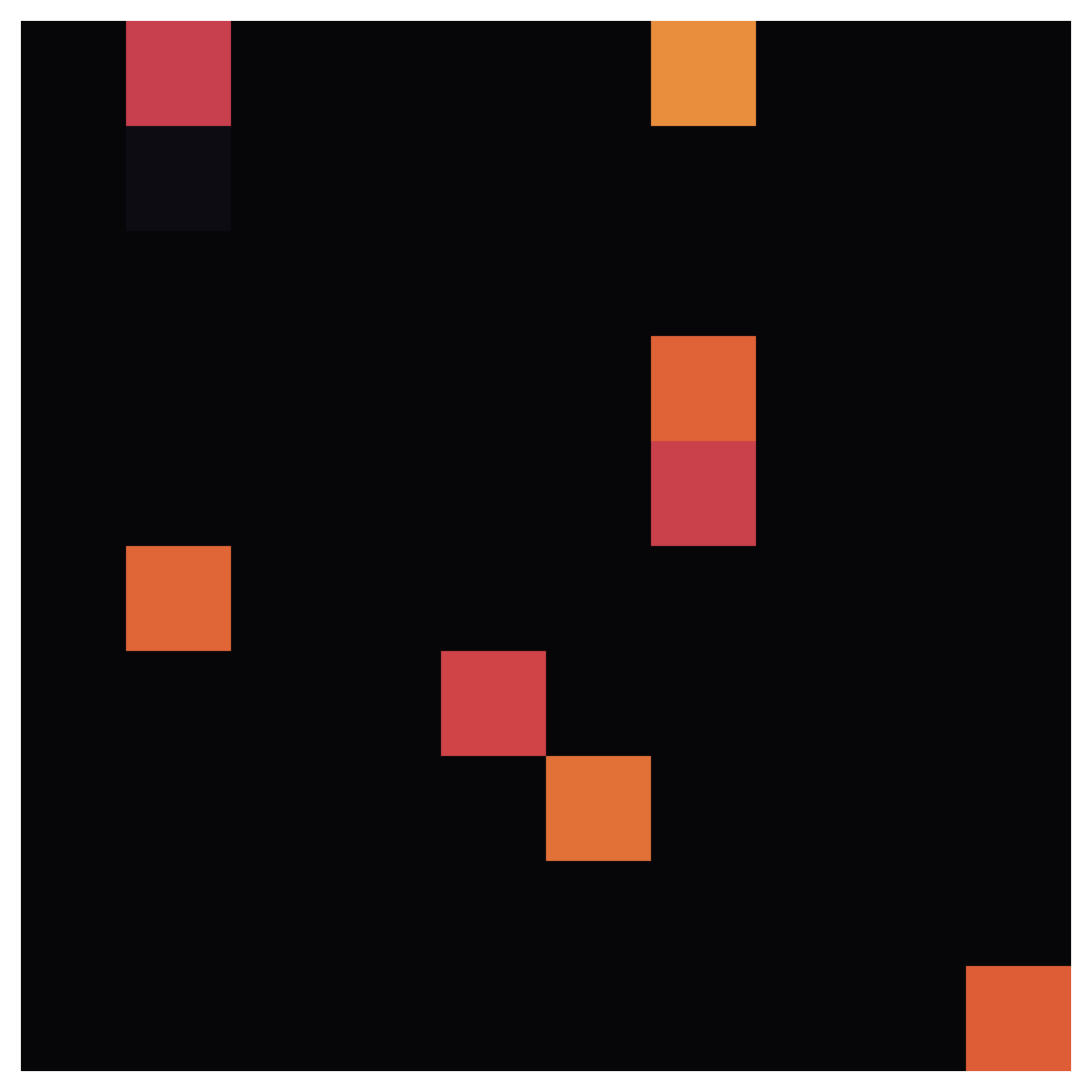}
    \end{subfigure}
         \begin{subfigure}[]{0.15\columnwidth}
         \caption{$\tilde{p},\lambda=1,s=1$}
         \includegraphics[width=\textwidth]{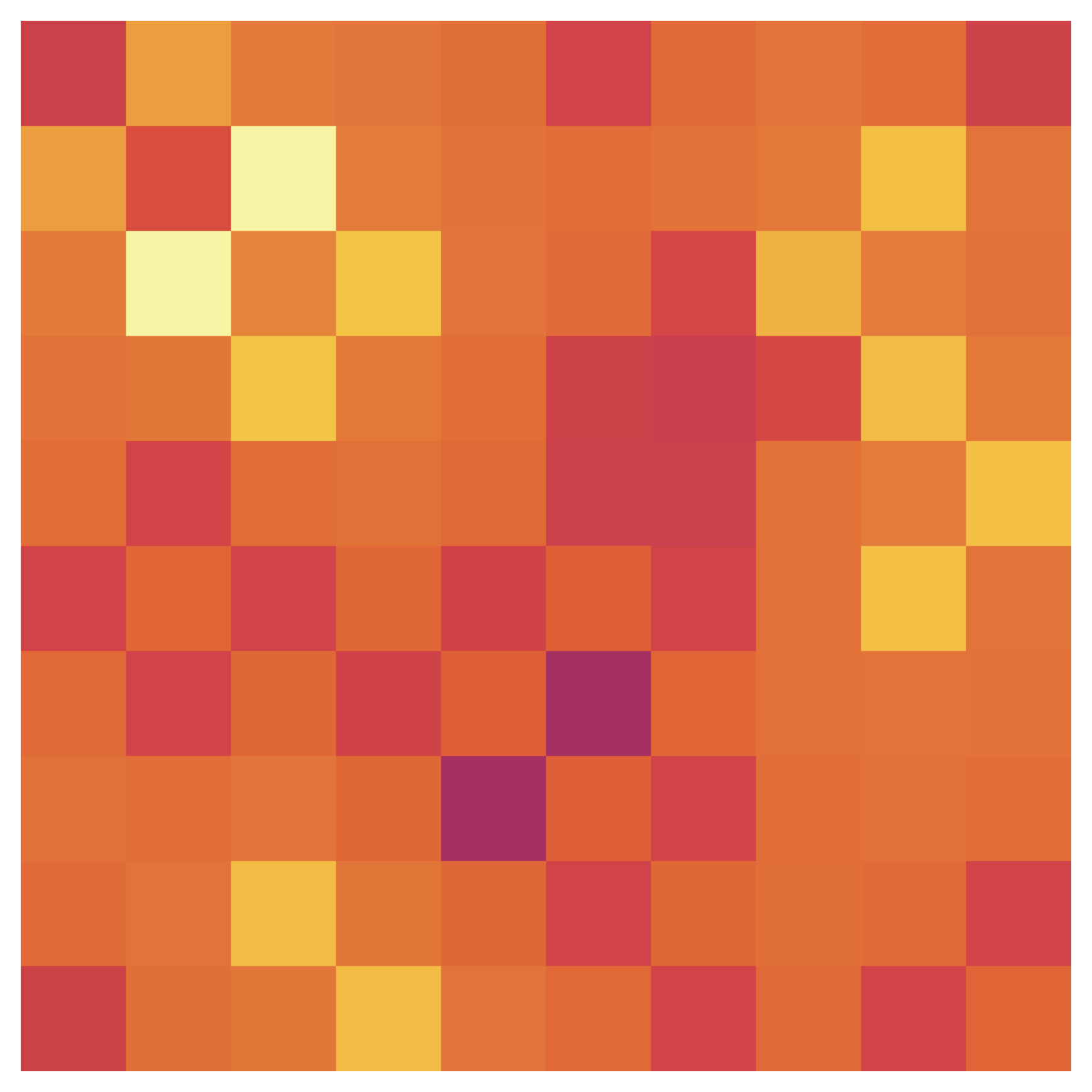}
     \end{subfigure}
        \begin{subfigure}[]{0.15\columnwidth}
        \caption{$v,\lambda=1,s=1$}
        \includegraphics[width=\textwidth]{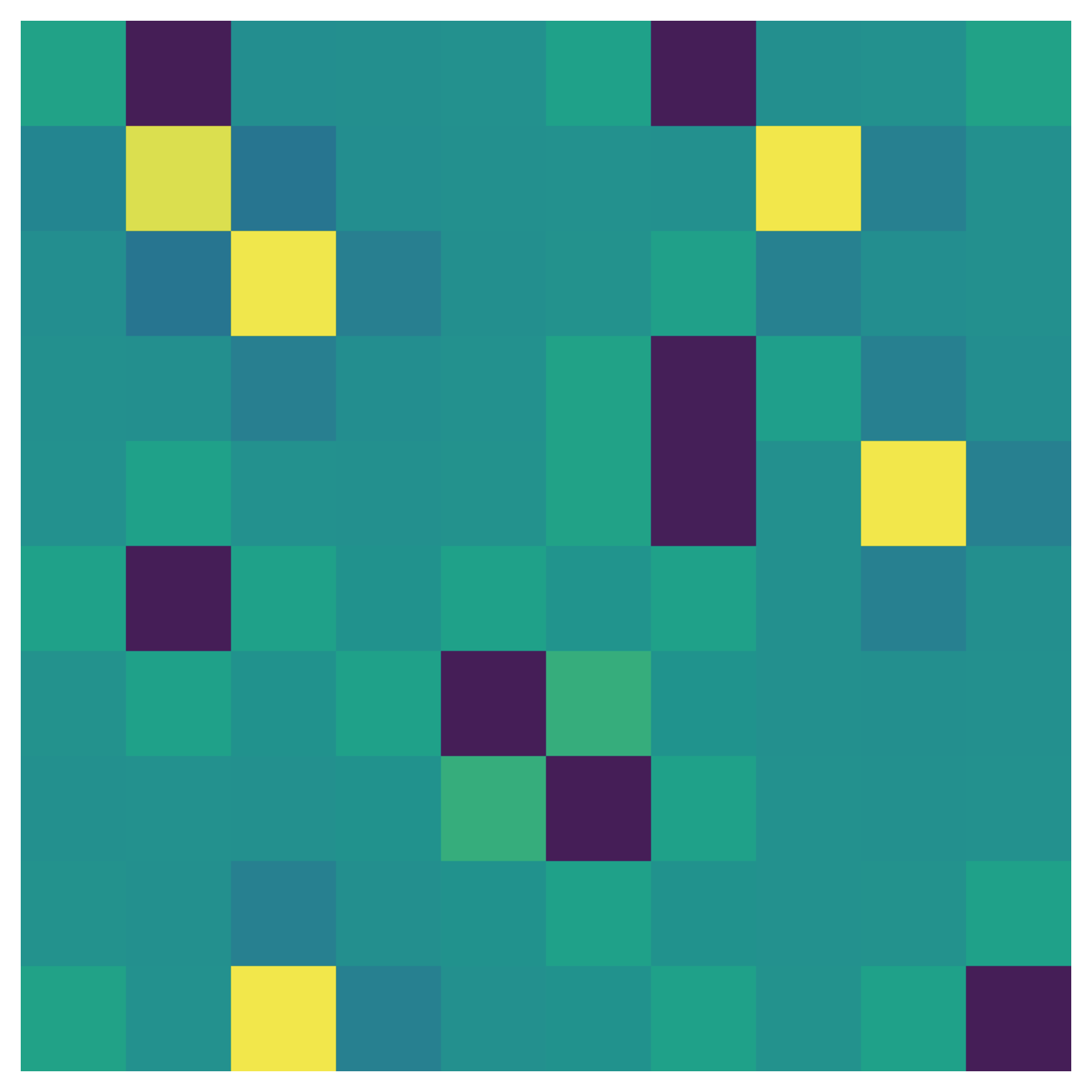}
    \end{subfigure}\\
    \caption{Heatmaps for the directed movement scenario with spillovers at time $t=200$. Here, we vary the preference for economic utility and spillovers.}
    \label{fig:HM_spillovers}
\end{figure}

Figure \ref{fig:HM_spillovers} depicts the heatmaps at time $t=200$. These results are for the same simulations that produced the time series in Figure \ref{fig:TS_spillovers}. Consider first the case where both economic and social utility drive movement ($\lambda=1/2$). Comparing these figures to the no spillover case (Figures \ref{fig:HM_directed_c_lambda0.5}-\ref{fig:HM_directed_v_lambda0.5}), we observe many similarities as well as distinct differences. For example, Gridlockers are prevalent across space in both sets of figures and there are pockets of the remaining strategies. However, votes are more polarized in these pockets and population size across space is more even when spillovers are present. Without spillover effects, population size falls into approximately three categories: high, intermediate, and low. Spillovers generate more variety in these sizes, though also less extremes. In particular, no node has a very low population size, unlike the no spillover case. In comparing the $s=4/5$ and $s=1$ spillover cases, we observe little variation. Juxtaposed with the $\lambda=1$ cases, we can see that social factors in determining migration act as a moderating effect.

Lastly, we consider the extreme case where economic utility solely drives movement ($\lambda=1$). Recall that the no spillover case, depicted in Figures \ref{fig:HM_directed_c_lambda1}-\ref{fig:HM_directed_v_lambda1}, exhibits large differences in population sizes. For $s=4/5$, these differences are dampened (and to a greater extent for $s=1$). Voting outcomes $v$ across space are also more varied, particularly for $s=4/5$. These results show that the free-rider effect disrupts the concentration of individuals in relatively high $v$ nodes.

\section{Discussion}

Here, we have developed a model of an electorate voting on the funding of a public good in a two-party system. We considered four voting strategies: Consensus-makers, Gridlockers, Party $1$ Zealots, and Party $2$ Zealots. In the ODE model, we observe bistability at both time scales. For the fast voting dynamics and a sufficient prevalence of Consensus-makers, consensus on one or the other party can be reached. Zealots enlarge the basins of attraction to consensus on their party. With respect to the slow imitation dynamics, stable outcomes are either gridlock or consensus. However, coexistence of gridlock and consensus outcomes can be maintained in the spatial model, though gridlock is most common.

For the spatial model, we considered two different types of movement, undirected and directed. Undirected movement generally results in consensus for one party or the other at non-gridlocked nodes. Directed movement, in contrast, can result in intermediate local voting outcomes. This effect is most intense when economic utility is the driver of movement. Further, disparities in population sizes across space are most pronounced when individuals move solely to increase economic utility. And the lower-populated areas are generally vote for gridlock. We find that preferences for social utility --- where voters sort by their voting strategies --- moderate these effects, reducing spatial variations in population size. Additionally, free-riding caused by spillovers disrupts large population centers, which produce high levels of the public good.

Political drivers of migration have been well documented such as the phenomenon of ``voting with your feet'' in the United States \citep{Liu20}. The recent Sunbelt migration also shows the effect of migration and political influence. Individuals moving to the South and Southwest areas of the United States have had a substantial impact on reinforcing and diluting party strength \citep{Jurjevich12}. However, political migration is not the only major driver in movement of voters. Policies on local funding of public goods can drive migration. For example, after the overturning of \textit{Dobbs v. Jackson}, the regulation of abortion to state legislatures, \cite{Blanchar25} found evidence of the ideological migration hypothesis. Individuals were more likely to consider relocating to a state with abortion polices that aligned with their political views.

Lastly, the effect of real-world spillovers can influence the outcome of political action. For example, after the COVID-19 pandemic, the European Union aimed to take aggressive economic action to counteract the economic impact of the pandemic through a coordinated fiscal expansion. Spillover effects from individual country measures, on average, composed a third of the European Union's GDP \citep{Pfeiffer23}. Furthermore, for smaller countries with fewer resources, spillover effects accounted for the majority of the GDP impact. On a more micro level, local governments in Spain have seen the impact of both beneficial spillovers, coming from the provision of a public good, and crowding spillovers, coming from the overcrowding in neighbourhoods where free-riding is prevalent \citep{Sole-olle06}.

There are multiple limitations to the model presented here, stemming from assumptions that could be relaxed or altered in future work. For one, we assume that the fitness functions are smooth. In reality, they may be discontinuous: voters may have thresholds at which their fitnesses change. Varying the fitness functions could alter the outcomes of the voting dynamics and thereby the funding of the public good. Our model could also be extended to include different voting strategies and more political parties. Spatially, we assumed that each node was intrinsically identical. However, variations in local exogenous conditions could have an impact on outcomes. Our model also only considered a single public good supported by one party and not supported by the other. A possible model extension is to consider two public goods each supported by a different party. Finally, inter-regional migration is often driver by non-economic and non-politcal factors such as lifestyle preferences and nearness to family. Future work could develop a more sophisticated model of migration.

\subsection*{Statements and Declarations}
The authors have no relevant financial or non-financial interests to disclose.

\subsection*{Code and Data Availability}
Code to run the simulations is available at github.com/J3ngle/Voting-PublicGoodFunding.

\bibliography{Spatial}
\bibliographystyle{plainnat}

\end{document}